\documentclass[useAMS,usenatbib]{mn2e}

\usepackage{graphicx}
\usepackage{amsmath}
\usepackage{amssymb}

\numberwithin{equation}{section}

\title[The Origin of Planetary System Architectures. I.]
{The Origin of Planetary System Architectures. I. Multiple Planet Traps in Gaseous Discs}
\author[Y. Hasegawa and R. E. Pudritz]{Yasuhiro Hasegawa$^{1}$\thanks{E-mail:
hasegay@physics.mcmaster.ca (YH); pudritz@physics.mcmaster.ca (REP)} and Ralph E. Pudritz$^{1,2}$\footnotemark[1]\\
$^{1}$Department of Physics and Astronomy, McMaster University, Hamilton, ON L8S 4M1, Canada\\
$^{2}$Origins Institute, McMaster University, Hamilton, ON L8S 4M1, Canada}

\begin{document}

\date{}

\pagerange{\pageref{firstpage}--\pageref{lastpage}} \pubyear{2009}

\maketitle

\label{firstpage}

\begin{abstract}

The structure of planetary systems around their host stars depends on their initial formation conditions. 
Massive planets will likely be formed as a consequence of rapid migration of planetesimals and low mass cores into 
specific trapping sites in protoplanetary discs. We present analytical modeling of inhomogeneities in protoplanetary 
discs around a variety of young stars, - from Herbig Ae/Be to classical T Tauri and down to M stars, - and show 
how they give rise to planet traps. The positions of these traps define the initial orbital 
distribution of multiple protoplanets. We investigate both corotation and Lindblad torques, and show that a new trap 
arises from the (entropy-related) corotation torque. This arises at that disc radius where disc heating changes from 
viscous to stellar irradiation dominated processes. We demonstrate that up to three traps (heat transitions, 
ice lines and dead zones) can exist in a single disc, and that they move differently as the disc accretion 
rate $\dot{M}$ decreases with time. The interaction between the giant planets which grow in such traps may be a 
crucial ingredient for establishing planetary systems. We also demonstrate that the position of planet traps 
strongly depends on stellar masses and disc accretion rates. This indicates that host stars establish preferred scales 
of planetary systems formed around them. We discuss the potential of planet traps induced by ice 
lines of various molecules such as water and CO, and estimate the maximum and minimum mass of planets which undergo 
type I migration. We finally apply our analyses to accounting for the initial conditions proposed in the Nice model 
for the origin of our Solar system. 

\end{abstract}


\section{Introduction}

Observations of exoplanets (nearly $\sim$ 1700 if candidates are included) show that the formation of 
multiple planets around their host stars is relatively common.\footnote{See the website http://exoplanet.eu/} This 
trend is confirmed by both the radial velocity and transit techniques such as the Kepler mission \citep[e.g.][]{h11}. 
Many previous studies based on N-body simulations, investigated the dynamics of the planetary systems 
\citep[e.g.][]{rf96}. It is well known that these simulations can reproduce the observed distribution of 
eccentricities of exoplanets very well, if they adopt a specific initial condition that planets are closely packed 
\citep[e.g.][]{fr08}. For our Solar system, the Nice model which requires a specific initial arrangement of Jupiter 
and Saturn explains the dynamics very well \citep[references herein]{m10}. The obvious question is whether or not 
such conditions are reasonable. In order to answer the question, one needs to consider the early stage of planet 
formation in which physical processes are controlled by gas dynamics in discs \citep[e.g.][]{tmr08}. The 
presence of gas is necessary for gas giants to be formed. However, it also causes rapid inward planetary migration 
\citep{ward97}. Since this radial motion strongly depends on the disc properties such as the 
gas surface density and the disc temperature, it becomes a huge challenge to systematically investigate what are the 
realistic initial conditions for the later evolution of planetary systems.  

Recently, planet traps have received a lot of attention \citep{mmcf06,mcmn08}. Barriers to planetary migration arise 
because of inhomogeneities in discs where the direction of planetary migration switches from inwards to outwards, 
so that migrating planets are halted. 
\citep[Equivalently, the net torque exerted on planets is zero there,][hereafter HP10, references herein]{hp10}. 
Planets may acquire 
most of their mass as they accrete material at these barriers - which we call planet traps in globally evolving discs. 
We use the term barriers for our local analyses ($\S$ \ref{wo_dz} and \ref{w_dz}) while the term of planet traps 
are used for our global, unified analyses ($\S$ \ref{synthesis}).

Planet traps were originally proposed by \citet{mmcf06} in order to solve the well known rapid migration problem 
wherein planets can be lost to discs within $10^{5}$ years. \citet[hereafter MPT07]{mpt07} first addressed a 
(possible) link between the planet traps and the diversity of exoplanets by showing that planet traps can move due 
to the time dependent, viscous evolution of discs \citep[also see][]{mpt09}. Although they focused on 
dead zones in discs \citep[which are the high density, inner regions, so that turbulence there induced by 
magnetorotational instability (MRI) is quenched,][]{g96}, their basic idea is valid for any type of planet trap. 
More recently, it has been pointed out that a single disc can have a couple of planet traps \citep{il08v,lpm10}. 
Population synthesis models confirm that the planet traps and their movements can be important for the diversity of 
observed exoplanetary systems \citep{mda11}. As already noted, planet traps have the possibility of strongly 
enhancing the growth rate of planetary cores \citep{sld11} and the formation of giant planets. Thus, planet traps 
have considerable potential for understanding the formation of planetary systems.

In a series of papers, we systematically investigate how inhomogeneities in discs can trap migrating planets in 
overdense regions where they undergo most of their growth, and how trapped planets establish their planetary systems in 
viscously evolving discs. In this paper, we undertake a comprehensive study of the various mechanisms that produce 
planet traps in discs around a variety of young stars, - from high to low mass. One of our major findings is 
that a single disc can have up to three planet trap regions. We show that the position of any trap depends upon the 
disc's accretion rate, which decreases with time as the disc is used up and star formation is terminated. The 
decreasing accretion rate forces the planet traps to move inwards at varying rates. This ultimately sets up the 
condition for the mutual interaction of planets in the traps, which can provide the realistic initial conditions 
for the evolution of planetary systems. 

The plan of this paper is as follows. We describe physical processes that create inhomogeneities in discs and 
summarise our analytical approach for them in $\S$ \ref{disc_model}. The general reader may then proceed to 
$\S$ \ref{synthesis} for discussion of the general results (also see Fig. \ref{fig7}). Armed with physical 
understanding of the inhomogeneities, we investigate corotation and Lindblad torques which are the driving force of 
planetary migration in $\S$ \ref{wo_dz} and \ref{w_dz}, respectively. For the former case, we show that a new barrier 
arises due to the change of the main heat source for gas disc from viscosity to stellar irradiation. For both torques, 
ice lines play some role in trapping planets (\citealt{il08v}, hereafter IL08; \citealt{lpm10}). In $\S$ \ref{discus}, 
we discuss other possible barriers and estimate the maximum and minimum mass with which planets undergo type 
I migration. Also, we investigate the possibility of the presence of ice lines due to various molecules. In $\S$ 
\ref{synthesis}, we integrate our analyses and discuss the roles of these planet traps in the formation of planetary 
systems. We apply our results to an explanation of the Nice model for the architecture of the Solar system in $\S$ 
\ref{nice}. In $\S$ \ref{conc}, we present our conclusions. We summarise important quantities which often appear in 
this paper in Table \ref{table1}.

\begin{table*}
\begin{minipage}{17cm}
\begin{center}
\caption{Important quantities}
\label{table1}
\begin{tabular}{cc}
\hline
Symbols    &  Meaning                        \\ \hline
$M_*$      &  Stellar mass                   \\
$R_*$      &  Stellar radius                 \\
$T_*$      &  Stellar effective temperature  \\
$\dot{M}$  &  Accretion rate (see equation (\ref{mdot})) \\
$G$        &  Gravitational constant          \\
$\Omega$    &  Angular frequency (see equation (\ref{omega})) \\
$\Omega_{Kep}$ & Keplerian frequency ($=\sqrt{GM_*/r^3}$)  \\
$M_p$      &  Planetary mass                 \\
$r_p$      &  Planetary orbital radius       \\
$\mu$       &  $M_p/M_*$                      \\
$\rho$     &  Gas volume density             \\
$\Sigma^1$   &  Gas surface density         \\
$\Sigma_0$ &  Gas surface density at $r=r_0$ \\
$s^1$        &  Power-law index of $\Sigma$ if $\Sigma \propto r^s$ \\
$T$        &  Disc temperature ($\propto r^{t}$) \\
$t$        &  Power-law index of $T$ \\
$T_e$      &  Disc effective temperature \\
$T_m$      &  Disc temperature of the mid-plane \\
$T_{m,k}(r_{il})$  & Condensation temperature for species $k$ at the ice line\\
$T_s$      &  Disc temperature of the surface \\
$c_s$      &  Sound speed ($\propto \sqrt{T}$)   \\
$H$        &  Disc scale height ($=c_s/\Omega$)  \\
$\bar{H}$  &  Disc photosphere height \\
$h$        &  Disc aspect ratio ($=H/r$)   \\
$r_{trans}$   &  Disc radius of disc inhomogeneities \\
$F_g$      &  Gas density modification at $r_{trans}$   \\ 
$\omega$   &  Transition width at $r_{trans}$ ($=cH$) \\
$r_{ht}$   &  Disc radius of the heat transition \\
$r_{il}$   &  Disc radius of ice lines \\
$r_{edge}$ &  Disc radius of the outer edge of dead zones \\
$p$        &  Gas pressure ($=\rho c_s^2$)   \\
$\kappa$   &  Epicyclic frequency (see equation (\ref{kappa})) \\
$m$        &  Wavenumber (see equation (\ref{wave_num})) \\
$\alpha_r$ &  Lindblad resonant position ($=r / r_p$) \\
$\psi$     &  Forcing function (see equations (\ref{psi1}) and (\ref{psi2})) \\
$X_p$      &  Value of a quantity $X$ at $r_p$ \\
$\Sigma_A$ &  Surface density of active regions (see equation (\ref{Sigma_A}))    \\
$s_A$      &  Power-law index of $\Sigma_A$ (generally $>0$) \\
$r_0$      &  Characteristic disc radius \\
$f_{ice}$  &  Modification factor of $\Sigma_A$ due to ice lines ($=f(r_{il}, f_{d1}, f_{d2})$)\\
$f_{d1}$   &  Parameter representing density jumps due to ice lines \\
$f_{d2}$   &  Parameter representing dust traps due to ice lines \\
$g_d$      &  Parameter representing density bumps due to ice lines \\ 
$\alpha$   &  Mean strength of turbulence (see equation (\ref{mean_alpha})) \\
$\alpha_{A}$ &  Strength of turbulence in the active zone ($=10^{-2}$) \\
$\alpha_{D}$ &  Strength of turbulence in the dead zone ($=10^{-5}$) \\
$\nu$      &  Kinematic viscosity ($=\alpha c_s H$) \\
$\gamma$   &  Adiabatic index (=1.4) \\
$F_z$      &  Heat flux in the vertical direction \\
$D_{vis}$  &  Viscous dissipation rate per unit mass \\
$\sigma_{SB}$ & the Stefan-Boltzmann constant \\
$k_{B}$    &  the Boltzmann constant \\
$\alpha_{GA}$ &  Grazing angle (see equation (\ref{alpha_geo})) \\
$\bar{\kappa}$ &  Opacity  \\
$\tau$         &  Optical depth ($= \bar{\kappa} \Sigma$)  \\
\hline
\end{tabular}
 
$^1$ The simplest assumption that $\Sigma \propto r^{s}$ is only adopted in $\S$ \ref{ht_il} and 
\ref{simple_power-law}. 
\end{center}
\end{minipage}
\end{table*}

\section{Disc inhomogeneities} \label{disc_model}

We describe physical processes governing the structure of protoplanetary discs and discuss how disc inhomogeneities 
arises from these processes. We discuss the basic features of our analytical modeling of the resultant disc structures 
- which will be presented in $\S$ \ref{wo_dz} and \ref{w_dz}. Table \ref{table2} summarises the disc inhomogeneities, 
related nomenclature that often appears in the literature, the dominant torque that transports angular momentum in 
that region of the disc, and the section of the paper that treats the analysis.

\begin{table*}
\begin{minipage}{17cm}
\begin{center}
\caption{Summary of disc inhomogeneities}
\label{table2}
\begin{tabular}{cccc}
\hline
Disc inhomogeneity               & Nomenclature       & Torque      & Section                  \\ \hline
Opacity transitions               & e.g. Ice lines$^1$ & Corotation  & \ref{ice_line}           \\
Heat transitions                   & N/A                & Corotation  & \ref{steirra}            \\
Turbulence transitions            & Dead zones         & Lindblad    & \ref{dead_zone}          \\
Opacity and turbulent transitions & Ice lines$^2$      & Lindbald    & \ref{layered_structures} \\
\hline
\end{tabular} 

$^1$ Disc turbulence is assumed to originate from non-MRI based processes

$^2$ Disc turbulence is assumed to originate from MRI based processes
\end{center}
\end{minipage}
\end{table*}

\subsection{Physical processes}

Heating of protoplanetary discs is one of the most important processes in order to understand their geometrical 
structure \citep[e.g.][]{dhkd07}. Since discs are accreted onto the central stars, release of gravitational 
energy through disc accretion becomes one of the main heat sources. This energy can be dissipated by viscous 
stresses, leading to viscous heating. 
Once discs are heated up, then the absorption efficiency of discs which is regulated by their optical depth establishes 
the thermal structure of discs. In protoplanetary discs, dust gives the main contribution to opacity for photons with 
low to intermediate energy while gas is the main absorber of high energy photons. In the inner region of discs, 
viscous heating is very efficient and leads to high disc temperatures 
\citep[$\gtrsim$ 1000 K for classical T Tauri stars (CTTSs),][]{dccl98}. As a result, both metal dust grains and 
molecules are destroyed there \citep[e.g.][]{bl94}. Thus, this high disc temperature and resultant destruction of 
opacity sources produces opacity transitions. It is interesting that opacity transitions are also produced in a 
completely different situation, that is, of low disc temperatures wherein opacity is enhanced by "freeze-out" 
processes. Ice lines are one of the most famous opacity transitions created by this process. Ice lines are defined 
such that the number density of ice-coated dust grains suddenly increases, which is a consequence of low disc temperatures there. In $\S$ 
\ref{ice_line}, we treat ice lines as an example of opacity transitions. 

It is well known that protoplanetary discs are also heated up by stellar irradiation \citep{cg97,hp09b}. Since viscous 
heating becomes less efficient for larger disc radii, stellar irradiation plays a dominant role in regulating the 
thermal structure of discs there. Thus, a transition exists for discs wherein the dominant heating mechanism transits 
from viscous heating to stellar irradiation. We call this a heat transition. 
(see $\S$ \ref{steirra}).

We have focused so far on heating processes which involve with photons that have relatively low energy. High energy 
photons such as X-rays from the central stars and cosmic rays, are important for understanding the ionization of 
protoplanetary discs. Since the MRI is the most favoured process to excite turbulence in discs, it is 
crucial to evaluate the ionization structure of discs \citep{g96}. In the inner region of discs, the column density 
is very high, so that high energy photons cannot penetrate the entire region. In this "layered" structure, 
the mid-plane is effectively screened from being ionized while the surface region is efficiently ionized. Thus, 
the inner region is less turbulent, and characterised by a dead zone. On the other hand, the outer region has lower 
column density and hence is fully ionized for the entire region including the mid-plane. As a result, the outer region 
is fully turbulent. Thus, transitions in the amplitude of turbulence exist in discs which we denote as turbulent 
transitions (see $\S$ \ref{dead_zone}). Dead zones are the most famous product of the turbulent transitions. 

Finally, we introduce interesting transitions which arise from the combination of opacity and turbulent transitions. 
If the MRI is the mechanism that drives disc turbulence, then ice lines can also be regarded as this type of 
transition (IL08, see $\S$ \ref{ls_il} for the complete discussion).  At the ice lines, 
the sudden increment of ice-coated dust grains is likely to result in the sudden decrement of free electrons 
there, since such sticky grains can efficiently absorb them \citep{smun00}. This process, initiated by the 
opacity transitions, reduces the ionization level there, and therefore removes the coupling between the magnetic field 
and the gas - which kills the MRI instability. Hence ice lines create turbulent transitions. In summary, ice lines 
are regarded as an opacity transition if disc turbulence is excited by non MRI processes while they act as an 
opacity and turbulent transition if turbulence is excited by the MRI. In $\S$ \ref{layered_structures}, we investigate 
ice lines again, but as an example of the opacity and turbulent transitions.    

\subsection{Our analytical approach}

We present analytical modeling of the disc structures affected by the disc inhomogeneities in $\S$ \ref{wo_dz} and 
$\ref{w_dz}$. In order to make analytical treatments possible, we adopt expressions that are simple enough, but well 
capture the physics arising from the inhomogeneities. As an example, we adopt a $\tanh$ function for representing an 
inhomogeneity in the surface density (see equation (\ref{sigma_density_jump})). This profile is more general than 
a power-law and is likely to be applicable for the cases of opacity, heat, and turbulent transitions. For the opacity 
and heat transitions, the results of \citet[hereafter MG04, see their fig1]{mg04} validate the usage of this function 
as do the results of MPT07 for the turbulent transitions. Disc structures are approximated as power-laws for 
regions far away from the inhomogeneities (see equation (\ref{sigma_density_jump})). Thus, we adopt more general 
profiles for the disc structures. In addition, we make use of the simplest expressions in order to reduce mathematical 
complexity, and hence different functions are adopted for different transitions. 

Bearing these in mind, we will demonstrate that the disc inhomogeneities are the most plausible sites to produce 
barriers to type I migration by undertaking a comprehensive study of both corotation and Lindblad torques in 
$\S$ \ref{wo_dz} and $\ref{w_dz}$, respectively.  

\section{Barriers arising from corotation torques} \label{wo_dz}

Corotation torques play an important role in planetary migration in regions with high viscosity 
($10^{-1} \lesssim \alpha \lesssim 10^{-3}$), because the corotation torque there is (partially) unsaturated 
(i.e. effective, see Appendix B in \citealt{hp10c}, references herein). The corotation torque is known to act as a 
barrier for two situations. The first situation is when $\Sigma\propto r^s$ with $s\gtrsim 1$ \citep{mmcf06,pp09a}. 
This case can be generally established only at the inner edge of discs. In typical discs, the inner edge is located 
around a few hundredths of au, which corresponds to the semi-major axis of Hot Jupiters. Thus, the positive gradients 
of surface density unlikely play a dominant role in explaining the diversity of the observed planetary systems. 
The second situation arises for adiabatic discs that have inhomogeneities \citep{lpm10}. This is more promising 
since the main sites of planetary formation in protoplanetary discs are generally optically thick and reasonably 
approximated to be adiabatic. In this section, we investigate only the second case for the above reason.

\subsection{Conditions for outward migration} \label{cc}

We derive conditions for the disc structures that are required for planets to migrate outwards. These conditions 
are crucial for the subsequent subsection where we discuss how disc inhomogeneities work as a barrier.

\subsubsection{Disc models}

We discuss our disc models (also see Table \ref{table1}). For the surface density affected by disc inhomogeneities, 
we adopt the following equation;
\begin{equation}
 \Sigma = \Sigma_{int} \left[ 1 + 
        \frac{F_g-1}{2} \left( 1- \tanh \left( \frac{r - r_{trans}}{\omega} \right) \right) \right],
\label{sigma_density_jump}
\end{equation}
where $\Sigma_{int} \propto r^s$ is the initial density profile, $F_g$ characterise the density distortion which 
is a consequence of disc inhomogeneities, $r_{trans}$ is a orbital radius of the disc inhomogeneities, and 
$\omega=cH$ is the width of the transition (also see Table \ref{table1}). This analytical modeling is motivated 
by the results of MG04 who first found the significant effects of opacity transitions on planetary migration 
by solving the detailed, 1D disc structure equations. Based on their results (see their fig. 1), any temperature 
distortion created by opacity or heat transitions, is likely to result in a surface density structure  
which is well expressed by equation (\ref{sigma_density_jump}). For the initial profile ($s$), we examine two cases 
both of which are well discussed in the literature: $s=-3/2$ and $s=-1$. The former power-law index is known as minimum 
mass solar nebula (MMSN) while the latter one is a steady state solution to disc accretion. For the disc temperature, 
we adopt simple, power-law structures ($T\propto r^t$) and derive the critical value of $t$ that results in outward 
migration below.

\subsubsection{Torque formula}

We adopt the torque formula derived by \citet{pbck09} in which discs are assumed to be 2D. In the formula, 
the total torque is comprised of the linear Lindblad torque and non-linear corotation torques, known as horseshoe 
drags. If discs are (locally) isothermal, both Lindblad and corotation torques dictate that standard rapid inward 
migration will occur. More specifically, the vortensity-related horseshoe drag that is only an active (corotation) 
torque in this case, cannot exceed the Lindblad torque for power-law discs \citep{pp09a}. For discs with more general 
profiles, the vortensity-related horseshoe drag can exceed the Lindblad torque, but still results in inward migration. 
Therefore, we assume discs to be adiabatic, which is reasonable for the inner part of discs ($r \lesssim $ 80 au) 
around CTTSs \citep{cg97}. Adopting the torque formula in adiabatic discs \citep[see][]{pbck09}, the direction 
of migration may be given as 
\begin{equation}
 \mbox{sgn} \left[ \left( 2.5 -1.7t +0.1 \bar{s} \right) - 1.1 \phi_{v} \left( \frac{3}{2} + \bar{s} \right) + 
                  \frac{7.9}{\gamma} \left(t-(\gamma-1) \bar{s} \right) \right],
 \label{corotation_barrier}
\end{equation}
where $\bar{s}=d \ln \Sigma / d \ln r$ is an "effective" power-law index for $\Sigma$ which can have a more general 
profile, $\gamma$ is 
the adiabatic index and we took the softening length $b=0.4h$ for simplicity (also see Table \ref{table1}). 
The terms in the first brackets arise from the Lindblad torque, the terms in the second from the vortensity-related 
horseshoe drag, and the terms in the third from the entropy-related horseshoe drag. As \citet{pm06} first discovered 
and subsequent works interpreted it, the entropy-related horseshoe drag which occurs only in adiabatic discs, 
is scaled by radial, entropy gradients and can drive planets into outward migration (see Appendix B in \citealt{hp10c} 
for a summary, references herein). 

The original formula derived by \citet{pbck09} is valid exclusively in power-law discs. In order to take into 
account discs with more general profiles such as equation (\ref{sigma_density_jump}), we add a vortensity correction 
factor $\phi_v$ in the second term which is defined as
\begin{equation}
 \phi_v= \frac{1}{\bar{s}+3/2} \frac{d \ln (\Sigma / B)}{d \ln r},
 \label{phi_v}
\end{equation}
where 
\begin{equation}
B=\frac{1}{2r}\frac{d}{dr}\left( r^2 \Omega \right)
\label{Boort}
\end{equation}
is the vorticity of gas, also known as one of the Oort constants. This is because gradients of vortensities 
($d \ln (\Sigma/B) /d \ln r$) are very sensitive to disc structure.\footnote{Lindblad torques also need a 
similar correction factor for discs with general profiles. Indeed, we derive an analytical relation for Lindblad 
torques using equation (\ref{sigma_density_jump}) in $\S$ \ref{dead_zone} (see equation (\ref{density_barrier})). 
As shown in Appendix \ref{app1}, however, they switch the direction of migration only if $F_g> 1.5$ for $c=1$. 
In general, the density distortion created by opacity and heat transitions is likely to be less than that. 
In addition, it is more consistent to use the above terms than equation (\ref{density_barrier}) in this torque 
formulation. Therefore, we adopt the original form for the Lindblad torques.} We note that $\phi_v=1$ for 
pure power-law behaviours. In fact, gradients of vortensities are the core of the vortensity-related corotation 
torque and regulate the transfer of angular momentum there (see Appendix B in \citealt{hp10c}). Therefore, inclusion 
of the factor $\phi_v$ is likely to be crucial for properly evaluating the importance of the vortensity-related 
corotation torque relative to the others.

The condition for outward migration is, therefore, 
\begin{equation}
 t < \frac{[1.1 \phi_v (3/2+\bar{s})) - 2.5 + 7.8\bar{s}] \gamma -7.9s}{-1.7 \gamma+7.9}.
 \label{required_t}
\end{equation}
It is fruitful to first consider pure power-law discs where $\phi_v=1$. Setting $\gamma=1.4$, the required 
temperature profile is $t\lesssim - 1.4$ for MMSN discs ($s=-3/2$) while for $s=-1$, the temperature profile 
$t\lesssim - 1.1$ is needed. We stress that such steep temperature profiles can be only achieved by viscous 
heating in optically thick discs, which is discussed more in the next subsection.

\begin{figure}
\begin{center}
\includegraphics[width=8.5cm]{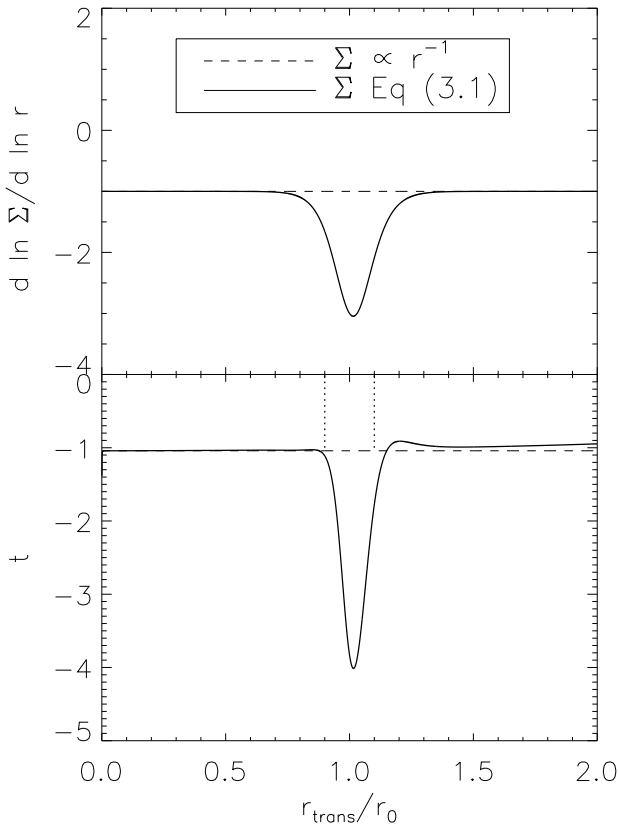}
\caption{The effects of the vortensity correction factor $\phi_v$ on $\bar{s}$ and $t$ on the top and bottom panel, 
respectively. For both panels, the solid lines denote the results for discs with general profiles (see equation 
(\ref{sigma_density_jump})) while the dashed lines are for pure power-law discs. Only the results of 
$\Sigma_{int}\propto r^{-1}$ are shown because those of the MMSN case are qualitatively similar. We set the 
transition width parameter $c=1$. For both panels, the effects of $\phi_v$ are well confined in the 
transition region ($\sim \omega$) (see the vertical dotted lines on the bottom panel). In addition, inclusion of 
$\phi_v$ reduces the possibility of outward migration there.}
\label{fig1}
\end{center}
\end{figure}

Let us now investigate how the required $t$ deviates from the power-law predictions due to the factor $\phi_v$. In 
general, we find that the structure of $\phi_v$ is very complicated. Therefore, we present the detail discussion of 
$\phi_v$ in Appendix \ref{app1} and briefly summarise the two most important effects on $t$ here. The first is that 
the effects of $\phi_v$ are very local and are likely to be confined within the transition region 
$\vartriangle r \sim \omega=cH$. This is clearly shown in Fig. \ref{fig1} (see the vertical dotted line on the bottom 
panel for representing the transition region). In this figure, we set $c=1$ that is the most likely value for 
the opacity and heat transitions. For the bottom panel, we plot the critical value of $t$ derived from 
equation (\ref{required_t}) as a function of $r_{trans}/r_0$, where the characteristic disc radius is set $r_0=1$ au 
unless otherwise stated. The top panel shows the behaviour of $\bar{s}$. For both panels, the solid line denotes the 
results for the disc with the general profiles (see equation (\ref{sigma_density_jump})) while the dashed line is for 
the pure power-law discs. We show the results only for the case that $\Sigma_{int}\propto r^{-1}$, because the results 
of $\Sigma_{int}\propto r^{-3/2}$ are qualitatively similar. The second important point is that, in the local region 
where the effects of $\phi_v$ become crucial, a required $t$ reaches values that are unattainable by any physical 
process in protoplanetary discs. This means that outward migration cannot happen due to large vortensity-related 
corotation torques that results in inward migration. In summary, the inclusion of 
$\phi_v$ reduces the possible region wherein planets can migrate outwards, but this reduction is well confined in a 
local region ($\sim$ an order of $\omega$) centered at $r=r_{trans}$.

Finally, we note that this torque formula does not take into account any effect of saturation. The problem of 
saturation, which is central to the problem of corotation torque, is very complicated and strongly dependent on the 
flow pattern at the horseshoe orbit (\citealt{mc10}; also see Appendix B in \citealt{hp10c}). 
\citet{pbk10,mc10} attempted to derive analytical formulae in which the effects of saturation are included. 
However, both formulae depend on thermal diffusivity in discs that is totally unknown in protoplanetary discs. 
Hence, we adopted the unsaturated torque formula. This implies that the above required temperature profiles 
may be the minimum value. Steeper profiles may be needed if (partial) saturation effects are taken into account. 

\subsection{Heat and opacity transition barriers} \label{ht_il}

The entropy-related horseshoe drag can result in outward migration only in discs with steep temperature 
profiles, as discussed above. Only viscous heating can establish such profiles. On the other hand, stellar 
irradiation can heat up protoplanetary discs as well. By deriving temperature profiles for both heating 
processes, we show that the heating transition from viscosity to stellar irradiation activates a new barrier. 
In addition, we examine the effects of opacity transitions on the temperature structure. We exclusively 
focus on ice lines as an example of the opacity transitions. For both transitions, we identify the positions of 
barriers.

\subsubsection{Disc models}

We adopt simple, power-law profiles for both the surface density and disc temperature ($\Sigma \propto r^{s}$ and 
$T \propto r^{t}$). This is supported by the argument done in the above subsection. Since the effects of 
$\phi_v(\neq 1)$ occurs only in a local region centered at the position of a transition, none of our findings 
discussed below is affected. In addition, the resultant deviation for our estimate of the position of barriers is 
only $<$ 10 per cents. Thus, it is reasonable to use power-law discs here.

We take the value of surface density at $r_0$ (labeled by $\Sigma_0$) as 10 times larger than the standard values of 
discs for all cases. Many previous studies showed that gas giants can be formed only if the disc mass increases by 
this amount relative to the MMSN models \citep[e.g.][]{ambw05}. Table \ref{table3} tabulates the values of $\Sigma_0$ 
around stars with various masses.

In addition, we assume stationary accretion discs that have accretion rates which can be modeled as
\begin{equation}
 \dot{M}=3 \pi \nu \Sigma= 3 \pi \alpha c_s H \Sigma
 \label{mdot}
\end{equation}
using the famous $\alpha$-prescription \citep{ss73}. 

\begin{table}
\begin{center}
\caption{Typical quantities for stars with various masses}
\label{table3}
\begin{tabular}{cccc}
\hline
                                & Herbig Ae/Be stars  & CTTSs      & M stars     \\ \hline
$\dot{M}$ ($M_{\odot}$/ year)   & $10^{-6}$           & $10^{-8}$  & $10^{-10}$  \\
$M_*$ ($M_{\odot}$)             & 2.5                 & 0.5        & 0.1         \\
$R_*$ ($R_{\odot}$)             & 2                   & 2.5        & 0.4         \\
$T_*$ (K)                       & 10000               & 4000       & 2850        \\
$\Sigma_0^1$ (g cm$^{-3}$)      & $10^{6}$            & $10^{4}$   & $10^{2}$        \\
\hline
\end{tabular} 

$^1$ The value of $\Sigma_0$ is 10 times larger than the standard values.
\end{center}
\end{table}
 
\subsubsection{Viscous heating: Origin of outward migration} \label{vh} 

We first discuss viscous heating. Assuming local thermal equilibrium in 
geometrically thin discs, the energy equation can be written as
\begin{equation}
 \frac{1}{\rho} \frac{\partial F_z}{\partial z} = D_{vis},
 \label{energy_eq}
\end{equation}
where $F_z$ is the heat flux in the vertical direction and $D_{vis}$ is the viscous energy dissipation rate per 
unit mass \citep[e.g.][also see Table \ref{table1}]{rl86}. In Keplerian discs, the viscous dissipation rate becomes 
\begin{equation}
 D_{vis}=\left( r \frac{d \Omega}{ dr} \right)^2 \nu.
\end{equation}
Integrating equation (\ref{energy_eq}) gives
\begin{equation}
 F_z =\frac{9}{4} \Sigma \nu \Omega^2.
\end{equation}

If discs are assumed to be optically thin and isothermal, the flux $F_z=2 \sigma_{SB} T_e^4$ with the Stefan-Boltzmann 
constant $\sigma_{SB}$. Thus, the resultant effective temperature profile becomes
\begin{equation}
 T_e \propto r^{-3/4}
\end{equation}
for discs which are assumed to reach a steady state. In this case which is valid in the late stage of disc 
evolution, accretion rates ($\dot{M}$) is constant over the entire disc. This famous $3/4$ law was first 
derived by \citet{lbp74} and is exactly identical to the temperature profile for flat discs which are heated by 
stellar irradiation \citep{als87,cg97}. In addition, it is well known that the resultant spectral energy 
distributions (SEDs) do not reproduce the observed ones - flared discs are required. 

If the accretion rate ($\dot{M}$) changes with disc radius, then the self-consistent temperature profile is 
\begin{equation}
 T_e \propto r^{s/3-1/2},
\end{equation}
where $\Sigma\propto r^{s}$ is the surface density of discs. For the MMSN models ($s=-3/2$), $T_e\propto r^{-1}$ 
while for discs with $s=-1$, $T_e\propto r^{-5/6}$. It is obvious that the temperature profiles derived from the 
isothermal assumption are not steep enough for the corotation torque to provide a barrier. 
 
The isothermal assumption can break down, especially in the main site of planetary formation 
$(1 \mbox{au} \lesssim r \lesssim 20 \mbox{au})$. The region of discs is reasonably considered to be optically 
thick and the energy is mainly transported by radiation \citep{dccl98}. In this case, the flux is generally 
described by 
\begin{equation}
 F_z = - \frac{16 \sigma_{SB} T^3 }{3 \bar{\kappa} \rho} \frac{\partial T}{\partial z},
 \label{thick_eq}
\end{equation}
where $\bar{\kappa}$ is the opacity \citep[e.g.][also see Table \ref{table1}]{rl86}. Assuming the flux $F_z$ is 
constant over $z$, integration of equation (\ref{thick_eq}) gives the relation between the temperatures of surface 
and midplane regions;
\begin{equation}
 T_s^4 - T_m^4 = - \frac{3 \tau}{16 \sigma_{SB}}F_z,
\end{equation}
where $\tau= \bar{\kappa} \Sigma$ is the optical depth and $T_s$ and $T_m$ is the temperature of surface and midplane 
region, respectively \citep[also see][Table \ref{table1}]{nn94}. If only viscous heating is taken into account, 
the temperature of the surface region should be much smaller than that of the midplane. As a result, 
the temperature structure is governed by 
\begin{equation}
 T_m^4 \approx \frac{27 \tau}{64 \sigma_{SB}} \Sigma \nu \Omega^2.
 \label{t_vis}
\end{equation}
When radiative transfer equation is treated explicitly, $\tau$ is replaced by $\tau_{eff}$ \citep{h90,kc08};
\begin{equation}
 \tau_{eff}= \frac{3}{8} \tau + \frac{\sqrt{3}}{4} + \frac{1}{4 \tau}.
\end{equation}

If discs have a constant $\dot{M}$ (see equation (\ref{mdot})), then the temperature profile 
becomes
\begin{equation}
 T_m \propto r^{s/4-3/4}, 
 \label{temp_thick1}
\end{equation}
assuming $\bar{\kappa}$ to be independent of $\Sigma$ and $T$. 
For the MMSN models ($s=-3/2$), $T_m\propto r^{-9/8}$ while for discs with $s=-1$, $T_m\propto r^{-1}$. Again, these 
temperature profiles are not steep enough. 

However, \citet{kc08} performed numerical simulations by solving a more complicated energy equation, and showed that 
the resultant temperature profile goes to $r^{-1.6}$ for discs with $\Sigma \propto r^{-0.5}$ in steady state. We 
can gain a similar profile if we adopt $\bar{\kappa}\propto T^2$ \citep{bl94}. In this case, $T_m \propto r^{s/2-3/2}$. 
Therefore, the assumption that $\bar{\kappa}$ is independent of $\Sigma$ and $T$, likely underestimates the temperature 
structure. If $\dot{M}$ is not constant, then the self-consistent temperature profile becomes
\begin{equation}
 T_m \propto r^{2s/3-1/2}, 
 \label{temp_thick2}
\end{equation}
assuming $\bar{\kappa}$ to be independent of $\Sigma$ and $T$.
For the MMSN models ($s=-3/2$), $T_m\propto r^{-1.5}$ while for discs with $s=-1$, $T_m\propto r^{-1.2}$. These 
profiles are exactly what is required (equation (\ref{corotation_barrier})) in order for planets to migrate outwards. 
If $\bar{\kappa} \propto T^2$ is adopted, $T_m\propto r^{2s-3/2}$, which is more preferred for outward migration. 

Thus, outward migration due to the entropy-related horseshoe drag is expected for viscously 
heated, optically thick discs. If protoplanetary discs were homogeneous in opacity and viscosity were the only 
physical process heating them, the results would suggest that large region of discs ought to be devoid of planets. 
This does not occur, however. We examine two kinds of inhomogeneity of discs below. One of them is stellar 
irradiation, which gives a new barrier. The other is the ice line which arises as a consequence of an opacity 
transition \citep{lpm10}. 

\subsubsection{Stellar irradiation: Origin of heat transition barriers} \label{steirra}

Stellar irradiation is well known to be the main heat source for regions in the disc beyond $r \gtrsim 2-3$ au 
\citep[in CTTSs,][]{dccl98}. Inside of that radius, viscous heating dominates over stellar irradiation. Since the 
temperature slope controlled by stellar irradiation is much shallower than that of viscous heating, planets which 
migrate inwards can be halted at the turning point where viscous heating begins to take over. We call this 
barrier a heat transition barrier.

In order to calculate the position of the turning point, we adopt radiative disc models of \citet{cg97}. In this model, 
two kinds of disc temperature are calculated. One of them represents the surface layer which is directly heated by the 
central star, called the super-heated layer. The other is for the midplane layer which is heated by the super-heated 
layer. Assuming radiative equilibrium discs, the temperature of the super-heated layer is given as
\begin{equation}
 T_s\approx T_{s0} \left( \frac{r}{\mbox{au}} \right)^{-2/5},
\end{equation}
where $T_{s0}=550$ K for discs around the CTTSs. Since the super-heated layer radiates equal 
amounts of energy inwards and outwards, the temperature of the midplane layer which is approximated to be 
optically thick is written as
\begin{equation}
 T_m\approx \left( \frac{\alpha_{GA}}{4} \right)^{1/4} \left( \frac{R_*}{r} \right)^{1/2} T_*,
\end{equation}
where $\alpha_{GA}$ is the grazing angle at which photons emitted from the star strike the discs 
(also see Table \ref{table1}). In general, $\alpha_{GA}$ is written as 
\begin{equation}
 \alpha_{GA} \approx \frac{0.4 R_*}{r} + r \frac{d}{dr} \left( \frac{\bar{H}}{r} \right),
 \label{alpha_geo}
\end{equation}
where $\bar{H}$ is the height of the visible photosphere above the mid-plane. Thus, the temperature of the 
midplane is strongly dependent on the geometrical shape of discs if stellar irradiation is taken into account.

For flat discs in which the aspect ratio $h$ is the constant, the temperature 
of the midplane layer becomes 
\begin{equation}
 T_m \approx \left( \frac{1}{3 \pi} \right)^{1/4} \left( \frac{R_*}{r} \right)^{3/4} T_*.
\end{equation}
Again, this is the famous $3/4$ law. For flared discs which are needed for reproducing the observed SEDs 
\citep{kh87}, the second term of the right hand side in equation (\ref{alpha_geo}) becomes dominant. Assuming 
that dust is well mixed with the gas in discs,
\begin{equation}
 \frac{\bar{H}}{r} = \frac{\bar{H}}{H} \frac{H}{r} = \frac{\bar{H}}{H} \left( \frac{T_e}{T_c} \right)^{1/2} 
                \left( \frac{r}{R_*} \right)^{1/2},
\label{H_r}
\end{equation}
where 
\begin{equation}
 T_c \equiv \frac{GM_* \mu_g}{k_B R_*},
\end{equation}
$\mu_g$ is the mean molecular weight of the gas and $k_B$ is the Boltzmann constant (also see Table \ref{table1}). 
Assuming $\bar{H}/H$ is constant, the self-consistent temperature of the midplane is 
\begin{equation}
 T_m \approx  T_{m0} \left( \frac{R_*}{r} \right)^{3/7},
 \label{t_irr}
\end{equation}
where
\begin{equation}
 T_{m0} \equiv  \left( \frac{\bar{H}}{16H} \right)^{2/7} \left( \frac{T_*}{T_c} \right)^{1/7} T_*.
\end{equation}
It is obvious that this temperature profile is much shallower than that due to viscous heating. 

Thus, the radius at which temperature profile switches is (equating equations (\ref{t_vis}) and (\ref{t_irr}))
\begin{equation}
 \frac{r_{ht}}{r_0} = \left[ T_{m0} \left( \frac{R_*}{r_0} \right)^{3/7} 
                        \frac{64 \sigma_{SB} \mu_g }{27 \bar{\kappa}_0 \Sigma_0^2 \alpha \gamma k_B \Omega_0} 
                 \right]^{\frac{1}{2s-3/2+3/7}},
 \label{r_transit}
\end{equation}
where we adopt $\bar{\kappa}=\bar{\kappa}_0 T^2$ with $\bar{\kappa}_0= 2 \times 10^{-4}$ \citep{bl94}. The usage of 
this form of $\bar{\kappa}$ that is only valid for the region outside of ice lines is reasonable, which is discussed 
below.  

\begin{figure}
\begin{center}
\includegraphics[width=8.5cm]{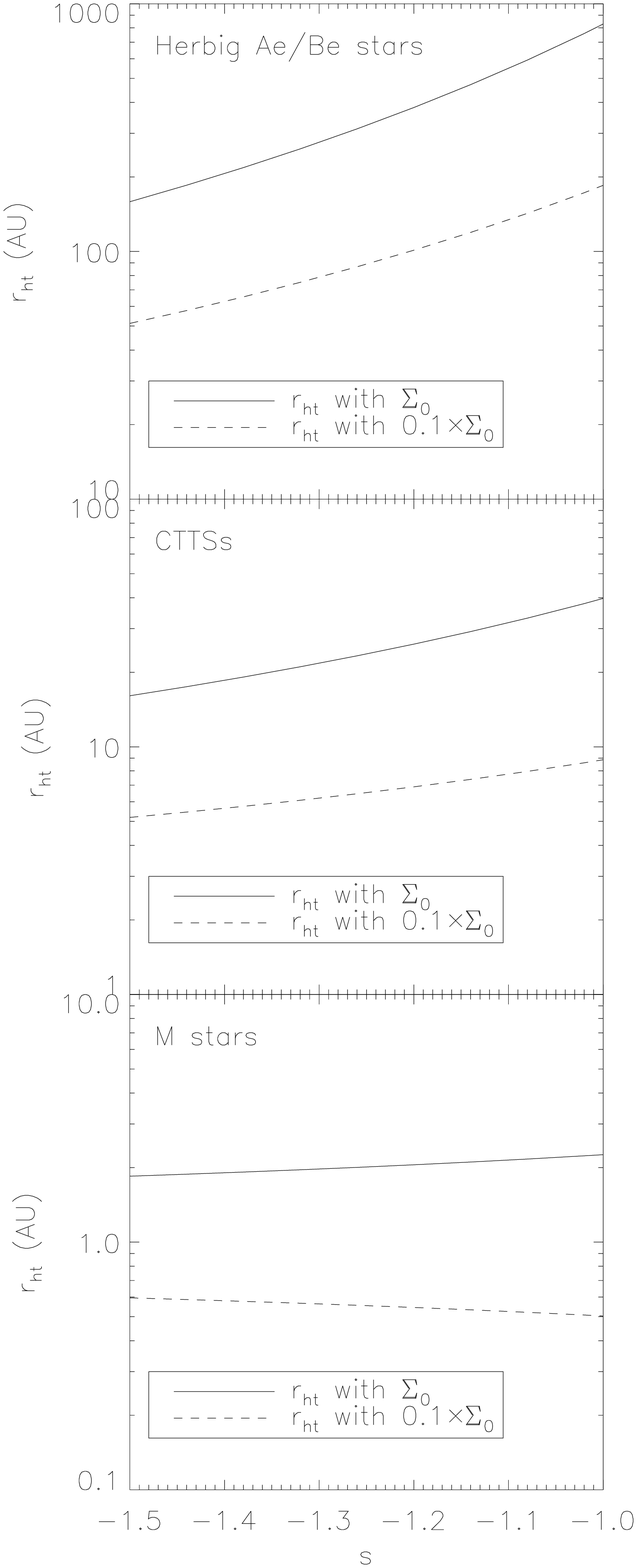}
\caption{The heat transition radius as a function of the power-law index $s$. For all panels, the solid line denotes 
the case using the values in Table \ref{table3} while only $\Sigma_0$ decreases by a factor of 10 for the dashed line, 
which is the value used more often in the literature. In general, the radius increase with increasing $s$, since 
the quantity in the brackets of equation (\ref{r_transit}) is larger than unity. If it is less than unity (see the 
dashed line for the case of M stars), the radius decreases.}
\label{fig2}
\end{center}
\end{figure}

Fig. \ref{fig2} shows the heat transition radius for discs around stars with various masses. For the solid lines, 
we adopt the values of $\Sigma_0$ in Table \ref{table3} with $\alpha=\alpha_A$ (also see Table \ref{table1}). For 
comparison purposes, the dashed lines denote the case with $0.1 \times \Sigma_{0}$ (which is the standard surface 
density in the literature). The transition radius generally increases with increasing $s$. Otherwise, it becomes a 
decreasing function of $s$ (see the dashed line for the case of M stars). Table \ref{table4} summarises the values 
of $r_{ht}$ for both $s=-3/2$ and $s=-1$ with $\Sigma_0$. It is obvious that massive discs around Herbig Ae/Be stars 
extend $r_{rh}$ to the order of several hundred au while lowest mass discs around M stars shrink $r_{ht}$ to the 
order of 1 au. For the intermediate disc mass around CTTSs, $r_{ht}$ is the order of a few ten au.

\begin{table}
\begin{center}
\caption{The typical values of $r_{ht}$ for $\Sigma_0$}
\label{table4}
\begin{tabular}{cccc}
\hline
                           & Herbig Ae/Be stars  & CTTSs  & M stars   \\ \hline
$r_{ht}$ (au) for $s=-3/2$ & 159                 & 16     & 1.8       \\
$r_{ht}$ (au) for $s=-1$   & 827                 & 40     & 2.3       \\
\hline
\end{tabular} 
\end{center}
\end{table}

\subsubsection{Ice lines: Origin of ice line barriers} \label{ice_line}

We now discuss the effects of ice lines on temperature profiles which provide another barrier \citep{lpm10}. 
We note that our following approach is applicable to any other opacity transitions. We discuss them more in 
$\S$ \ref{ot_all}. 
At the ice line, the number density of icy dust grains suddenly increases due to the low disc temperature. 
This sudden increment of dust provides an opacity transition. As a result, the temperature profile around the ice 
lines becomes shallow, resulting in inward migration while the profile well inside of the ice line becomes steep 
enough for planets to migrate outwards. Thus, planets migrating inwards are halted around the ice lines. This is 
known as the ice line barrier. This assumes that corotation torques are active -i.e. non-saturated, which is not 
necessarily clear at the ice lines (see $\S$ \ref{ls_il} and \ref{ot_all}).

At first, we examine which heat source controls the location of ice lines, $r_{il}$. In order to proceed, we adopt 
equation (\ref{t_vis}) for viscous heating and equation (\ref{t_irr}) for stellar irradiation. Equating $T_m$ to the 
condensation temperature for H$_2$O ice, $T_{m,\mbox{H}_2\mbox{O}}(r_{il})=170$ K \citep{js04}, equation (\ref{t_vis}) 
becomes
\begin{equation}
 \frac{r_{il}}{r_0} = \left[ T_{m,\mbox{H}_2\mbox{O}}^{10}(r_{il})
                        \frac{64 \sigma_{SB} \mu_g }{27 \bar{\kappa}_0 \Sigma_0^2 \alpha \gamma k_B \Omega_0} 
                 \right]^{\frac{1}{2s-3/2}},
 \label{ril_vis}
\end{equation}
where $\bar{\kappa}=\bar{\kappa}_0 T^{-7}$ with $\bar{\kappa}_0= 2\times 10^{16}$ \citep{bl94}, and equation 
(\ref{t_irr}) becomes
\begin{equation}
 \frac{r_{il}}{r_0} = \left( \frac{T_{m0}}{T_{m,\mbox{H}_2\mbox{O}}(r_{il})} \right)^{7/3} \left( \frac{R_*}{r_0} \right). 
 \label{ril_irr}
\end{equation}
In general, the exponent index $s$ is negative, so that $r_{il}\propto \Sigma_0^{2/(3/2-2|s|)}$. This shows that 
as the surface density ($\Sigma_0$) decreases, the position of the ice line barriers moves inwards. 
We note that we examine the effects of a water-ice line here although the above argument is applicable to ice lines 
of any material by changing the relevant condensation temperature for species $k$, $T_{m,k}(r_{il})$. We will discuss 
them more in $\S$ \ref{il_all}.

\begin{figure}
\begin{center}
\includegraphics[width=8.5cm]{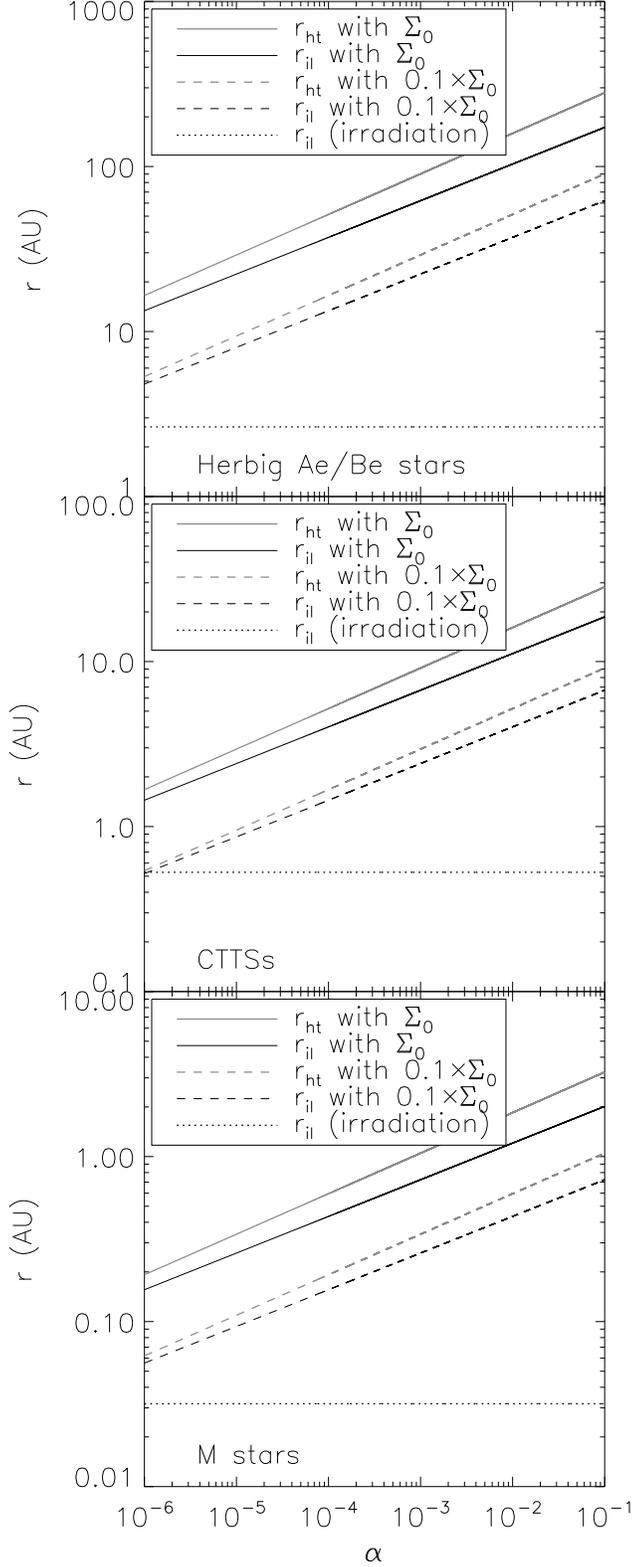}
\caption{The location of the water-ice line as a function of the turbulence parameter, $\alpha$. For every panel, 
the solid and dashed, black lines 
denote the radius derived from viscous heating using $\Sigma_0$ and $0.1 \times \Sigma_0$, respectively 
(see equation (\ref{ril_vis}) and Table \ref{table3}). The dotted lines are from stellar irradiation 
(see equation (\ref{ril_irr})). For comparison purposes, the heat transition radius is denoted by the two gray 
lines (see equation \ref{r_transit}). For all panels, viscous heating defines the location of the water-ice lines. In 
addition, the heat transition radius is larger than that of the water-ice lines.}
\label{fig3}
\end{center}
\end{figure}

Fig. \ref{fig3} shows the above two equations as a function of $\alpha$ for discs around stars with various masses. 
We set $s=-1.5$, since this choice of $s$ minimises the importance of viscous heating in discs 
(see Fig. \ref{fig2}). The black, solid and dashed lines denote equation (\ref{ril_vis}) with $\Sigma_0$ and 
$0.1 \times \Sigma_0$, respectively while the dotted line is for equation (\ref{ril_irr}). In all cases, $r_{il}$ 
defined by viscous heating is located at greater distances from the star than that by stellar irradiation. Furthermore, 
the transition radius defined by equation (\ref{r_transit}) is larger than any of these two radii (see the gray, 
thick lines). Thus, we can conclude that $r_{il}$ is determined by viscous heating for discs with a wide range of 
$\alpha$. This agrees with numerical work by \citet{mdk11} who showed the same results by numerically solving the full 
wavelength dependent, radiative transfer equation by means of a Monte Carlo method in 3D discs. In their simulations, 
viscous heating is explicitly included as well as stellar irradiation. In addition, this supports the findings of 
\citet{lpm10} who adopt disc models which are only heated by viscous heating. Furthermore, Fig. \ref{fig3} shows 
that the radius of the heat transition is located outside of the ice line. This validates our choice of 
$\bar{\kappa}$ in equation (\ref{r_transit}).

We now investigate temperature profiles created by viscous heating around ice lines by adopting appropriate 
$\bar{\kappa}$ for each region \citep{bl94}. For the region inside of the ice lines, 
$\bar{\kappa}=\bar{\kappa}_0 T^{1/2}$ with $\bar{\kappa}_0=0.1$. As a result, equation (\ref{t_vis}) becomes
\begin{equation}
 T_m \propto r^{4s/5-3/5}
\end{equation}
for discs with general $\dot{M}(r)$ (see equation (\ref{mdot})). $T_m \propto r^{-1.8}$ for the MMSN disc models 
($s=-3/2$) while $T_m \propto r^{-1.4}$ for the disc models with $s=-1$. These steep profiles obviously indicate 
that planets migrate outwards. For the region around the ice lines, $\bar{\kappa}=\bar{\kappa}_0 T^{-7}$ with 
$\bar{\kappa}_0=2 \times 10^{16}$. Consequently, equation (\ref{t_vis}) becomes
\begin{equation}
 T_m \propto r^{s/5-3/20}
\end{equation}
for discs with $\dot{M}(r)$. $T_m \propto r^{-0.5}$ for the MMSN disc models ($s=-3/2$) while 
$T_m \propto r^{-0.4}$ for the disc models with $s=-1$. Under these shallow profiles, the entropy-related horseshoe 
drag cannot exceed the Linblad torque, and hence planets migrate inwards. Thus, a planet that migrates inwards beyond 
the ice line will be halted near the ice line due to this opacity transition. 

\subsection{Comparison}

We compare the heat transition barriers to the ice line barriers. We especially focus on the relative location of each 
barrier. As discussed above, Fig. \ref{fig3} shows the relative location of these two barriers (see the black and 
gray lines). The ice line barrier (equation (\ref{ril_vis})) is located inside of the heat transition barrier 
(equation (\ref{r_transit})) for discs with a wide range of $\alpha$. However, we again emphasise that the above 
analyses for these two barriers can be valid only for the disc region with ($10^{-1} \lesssim \alpha \lesssim 10^{-3}$). 
In the region with a low value of $\alpha$ such as dead zones ($\alpha\sim 10^{-5}$), any corotation torque will be 
saturated (i.e. zero), so that these two barriers are never activated there. We also note that for such low values 
of $\alpha$, ice lines can become barriers as a consequence of the Lindblad torque rather than the corotation torque. 
In addition, if the adiabatic approximation breaks down, which can arise for the late stage of disc evolution or the outer 
part of discs, these two mechanisms cannot be effective. We now examine the mechanisms of barriers activated by 
Lindblad torque.

\section{Barriers arising from Linblad torques} \label{w_dz}

Let us now examine regions wherein any torques arising from corotation resonances are negligible, so that 
the direction of migration is determined only by the Lindblad torque. Such regions occur for discs with a shallow 
temperature profile or a low value of $\alpha$ in dead zones \citep[$\sim 10^{-5}$, HP10;][]{pbk10,mc10,hp10c}. 

\subsection{Lindblad torque}

\subsubsection{Basic equations}

We adopt analytical formulae of the Lindblad torques described in 
\citet[see their Appendix A for the complete expression]{hp10c}. In the formulae, the standard Lindblad torque 
derived by \citet{ward97} is modified, so that the vertical thickness of discs is taken into account without 
rigorously solving the 3D Euler equations. Thus, we treat the Lindblad torque density $d^2 \Gamma^L/dzdr$ of 
a layer at the height $z$ from the mid-plane which is written as 
\begin{equation}
\frac{d}{dz} \left( \frac{d\Gamma^L}{dr}(r,z) \right) = \epsilon  
                                \frac{2\mu ^2 \rho r_p^4 \Omega_p^4}{r(1+4\xi ^2)\kappa^2}m^4\psi ^2,
\label{torque_density}
\end{equation} 
where $\epsilon = +(-)$ for the outer(inner) resonances, the wavenumber $m$ is treated as a continuous variable;
\begin{equation}
m(r,z)=\sqrt{\frac{\kappa^2}{\left( \Omega-\Omega_p \right)^2 -c_s^2/r^2}},
\label{wave_num}
\end{equation}
$\Omega_p=\sqrt{GM_*/r_p^3}$ is the angular velocity of a planet with $r=r_p$, 
the angular velocity $\Omega$ is 
\begin{equation}
\Omega^2(r,z)=\Omega_{Kep}^2{\left( 1 + \frac{z^2}{r^2}\right)^{-3/2}} 
              + \frac{1}{r\rho}\frac{\partial p}{\partial r},
\label{omega}
\end{equation}
the epicyclic frequency $\kappa$ is
\begin{equation}
\kappa^2(r,z)=\frac{1}{r^3}\frac{\partial}{\partial r} \left( r^4 \Omega^2 \right),
\label{kappa}
\end{equation}
$\xi = mc_s/r\kappa $, and the forcing function $\psi$ is
\begin{equation}
\psi \approx \frac{\epsilon (\alpha_r^2-\zeta ^2-1)}{2\alpha_r\sqrt{(\alpha_r-1)^2+\zeta ^2}}K_1(\Lambda )
      + \left( \frac{\epsilon }{2m}+2\frac{\Omega}{\kappa} \sqrt{1+\xi^2}\right) \frac{K_0(\Lambda )}{\sqrt{\alpha_r}},
\label{psi1}
\end{equation}
where 
\begin{equation}
\Lambda(\alpha_r,\zeta) =m\sqrt{\frac{(\alpha_r -1)^2+\zeta ^2}{\alpha_r}},
\end{equation}
$\alpha_r=r/r_p$ is the resonant position, and $\zeta=z/r$ (also see Table \ref{table1}). 
We have adopted the standard approximation for $\psi$ \citep{gt80,js05}. In addition, we have added 
a factor $\Omega/\kappa$ in the second term of $\psi$ following MG04.

Finally, the total torque exerted by the planets on discs is found by summing over all layers; 
\begin{equation}
\Gamma^L=\int^{\infty}_{-\infty}dz \int^{\infty}_{0}dr \frac{d}{dz} \left( \frac{d\Gamma^L}{dr}(r,z) \right).
\label{total_torque}
\end{equation}
Note that the integration of equation (\ref{torque_density}) in terms of $z$ with $\zeta=0$ leads to the famous 
analytical formulae of the Lindblad torque density in 2D discs \citep{ward97}. Also, note that the tidal torques 
shown above are exerted by the planets on discs. Thus, the positive sign of $\epsilon$ in equation 
(\ref{torque_density}) means that the planets exert the tidal torques on discs, resulting in the loss of angular 
momentum from the planets and vice versa.

The validity of the above approach for mimicking the reduction of the tidal torques in 3D discs \citep{ttw02} is 
confirmed by \citet{lo98} who showed that, in thermally stratified discs, 2D modes carry off more than 95 per cent 
of angular momentum and vertical modes are not dominant for transferring angular momentum between planets and their 
disc \citep[also see][]{ward88,a93,js05}.

\subsubsection{Basic assumptions}

We assume discs to be geometrically thin ($z=0$) and Keplerian. Under this assumption, we have
\begin{eqnarray}
\rho  & \approx & \frac{\Sigma}{H} \\
\label{rho_1}
\Omega^2  & \approx  &  \kappa^2 \approx  \Omega_{Kep}^2, \\
\alpha^{3/2}_r & \approx  & \left( 1+\frac{\epsilon }{m}\sqrt{1+\xi ^2} \right).
\label{alpha_r} 
\end{eqnarray}
Thus, the torque density becomes
\begin{equation}
\frac{d}{dz} \left( \frac{d \Gamma^L}{dr}(r,z=0) \right) \approx  
                           \epsilon 2\mu ^2 GM_* \frac{m^4}{(1+4\xi ^2)} \frac{\Sigma \alpha_r^2 \psi ^2}{H},
\label{td_approx}
\end{equation}
where
\begin{equation}
\psi \approx \frac{1}{2}\left(1+\frac{1}{\alpha_r} \right) K_1(\Lambda )+ 
                  \left( \frac{\epsilon }{2m}+2\sqrt{1+\xi^2} \right) \frac{K_0(\Lambda )}{\sqrt{\alpha_r}},
\label{psi2}
\end{equation}
and 
\begin{equation}
\Lambda =m\frac{|\alpha_r -1|}{\sqrt{\alpha_r}}.
\end{equation}
 
Thus, the behavior of the torque density depends on $\Sigma \alpha_r^2 \psi^2/H$. In the following subsections, 
we expand every quantity ($\alpha_r, \Sigma, H,$ and $\psi$) as a series in $1/m (\ll 1)$, by assuming 
simple power-law disc structures in $\S$ \ref{simple_power-law} and non-power-law disc structures in $\S$ 
\ref{dead_zone} and \ref{layered_structures}. This expansion is assured, since the torque takes the maximum value 
at $m\approx 10$ \citep{ward97}, and consequently, allows us to derive simple relations governing the 
direction of migration. Thus, we perform local analyses as done in $\S$ 2.

\subsection{Simple power-law structures} \label{simple_power-law}

We derive an analytical relation for the direction of planetary migration. Here, we assume simple power-law structures 
for the background density and temperature in discs, that is, $\Sigma \propto r^s$ and $T \propto r^t$. As a result, 
the relation depends only on the exponent of $\Sigma$ and $T$. 

We can further simplify the mathematics by taking the limit $\xi\rightarrow 0$ (since the effect of gas pressure 
becomes negligible in the Keplerian discs; see Appendix \ref{app2}). Equation (\ref{td_approx}) then becomes, to 
first order in $1/m$,    
\begin{equation}
\frac{d}{dz} \left( \frac{d \Gamma^L}{dr}(r,z=0) \right) \approx  
                      \epsilon \Gamma_0(r_p) \left[ 1+ \frac{2\epsilon }{3m} 
                            \left( s - \frac{t}{2} + \frac{1}{2} + \frac{5}{4} \right) \right],
\end{equation}
where
\begin{equation}
\Gamma_0(r_p)=2\mu ^2 GM_* \frac{\Sigma_p }{H_p}m^4\psi_0 ^2
\end{equation}

Thus, the total torque is
\begin{eqnarray}
\Gamma^L & \approx & \int^{\bar{H}_p}_{-\bar{H}_p}dz \Gamma_0(r_p)\frac{4 }{3m} 
                                         \left( s - \frac{t}{2} + \frac{7}{4}  \right) \nonumber \\
    & =       & \frac{8}{3m} \Gamma_0(r_p)\bar{H}_p \left( s - \frac{t}{2} + \frac{7}{4} \right),
\end{eqnarray}
where the integration range for $z$ is approximated to extend over the height of the photosphere 
$\bar{H}_p \sim 2H_p$ at the location of the planet since the density above the photosphere is about two orders of 
magnitude less than that at the mid-plane. 

Consequently, the sign of the net torque depends on the sign expression;
\begin{equation}
 \mbox{sgn} \left( s - \frac{t}{2} + \frac{7}{4}\right).
 \label{thermal_barrier}
\end{equation}
The torque becomes positive (inward migration) when $s -t/2 +7/4>0$ and negative (outward migration) when 
$s -t/2 +7/4<0$. When the MMSN disc models ($s=-3/2$) are adopted, $t>1/2$ is needed for outward migration while, 
when disc models have $s=-1$, $t>3/2$ is required. In general, $T$ is a decreasing function of $r$. Therefore, 
some specific features in discs are required for the Lindblad torque to excite barriers. In the following subsections, 
we focus on dead zones and ice lines.
 
\subsection{Dead zone barriers} \label{dead_zone}

Dead zones are known to play very interesting roles in planetary formation and migration \citep[e.g.][]{mp06}. 
They can excite two barriers: density barriers (MPT07) and thermal barriers (HP10). These two barriers are 
interesting products of turbulent transitions. In this subsection, we derive 
the analytical relation which predicts when the direction of migration is reversed from inwards to outwards for 
these cases. 

\subsubsection{Thermal barriers}

Thermal barriers were uncovered by HP10 who first investigated numerically the effects of dead zones on the temperature 
structure of discs \citep[also see][]{hp09b}. In their paper, the temperature structure of discs with dust settling and 
a dead zone was simulated by solving the full wavelength dependent, radiative transfer equation by means of a Monte 
Carlo method. Stellar irradiation is assumed to be the main heat source. HP10 demonstrated that planets migrate 
outward in a region where a positive temperature gradient is established (see their figs 1 and 3). This 
temperature gradient is a result of the back heating of the dead zone by a thermally hot dusty wall, and is well 
represented by $T\propto r^{t'}$ with $t'\gtrsim 3/2$. The hot dusty wall is produced by the enhanced dust settling 
in the dead zone and the resultant enhanced absorption of stellar irradiation by the wall. Since they adopted 
$\Sigma \propto r^s$ with $s=-1$, the profile of the positive temperature gradient is identical to the profile 
required by equation (\ref{thermal_barrier}) in order for planets to migrate outwards. Thus, our relation predicts 
the results of the detail numerical simulations very well.

\subsubsection{Density jumps} \label{dz_dj}

Density barriers were found by MPT07 who undertook a pioneering study on the effects of dead zones on planetary 
migration. They are created by the formation of a density jump at the outer edge of dead zones which arises as a 
consequence of time-dependent, viscous evolution of discs \citep[also see][for a more complete study]{mpt09}. 
This density jump develops because of the huge difference in the strength of turbulence ($\alpha$) between the active 
and dead zones of the disc.

Here, we analytically derive an analogous relation for density barriers. We adopt equation (\ref{sigma_density_jump}) 
as the background surface density. In this case, $F_g$ is the density difference between the active and dead zones, 
$r_{trans}= r_{edge}$ is a orbital radius of the outer boundary of the dead zone, and $\omega=cH$ is 
the width of the transition (also see Table \ref{table1}). Except for this, we adopt the same approximation as above. 
As a result, the relation depends on the structure of the dead zones as well as the exponent of $\Sigma$ and $T$. 
Since planets are slowed down or stopped near $r_{edge}$, we can approximate $r_{edge}\approx r_p$. 

We note that a slightly more accurate shape of the boundary in equation (\ref{sigma_density_jump}) can be expressed 
in terms of error functions. However, the difference between the error and $\tanh$ functions is less than 1 per cent 
with the proper choice of coefficients \citep{b98}. Furthermore, $\tanh$ functions are fully analytical. For these 
reasons, we adopt $\tanh$ functions. 

Adopting the same approximation as above, we find 
\begin{equation}
 \Sigma / \Sigma_{int} \approx 1 + \frac{F_g-1}{2} \left[ 1- \tanh \left( \frac{\epsilon}{c} \right)
               + h_p\frac{t+3}{2c}\frac{1}{\cosh ^2 (\epsilon / c)} \right].
\end{equation}
Note that we expanded $\Sigma$ in terms of $1/m$ and finally expressed it in terms of $h_p$ 
assuming 
\begin{equation}
 h_p\approx \frac{2}{3m}.
\end{equation}
This is reasonable because the resonant positions are pushed away from the planet by a distance $\sim H(=rh)$ due to 
the gas pressure \citep[also see equation (\ref{alpha_r})]{a93,ward97}.

Inserting the above surface density into equation (\ref{td_approx}), the sign of the net torque consequently 
becomes
\begin{equation}
 \mbox{sgn} \left[ -(F_g-1) \tanh \left( \frac{1}{c} \right) 
    + (F_g+1)  h_p  \left( s - \frac{t}{2} + \frac{7}{4}  \right) \right].
\label{density_barrier}
\end{equation}

Again, we can check the validity of equation (\ref{density_barrier}) by comparing it with the numerical simulations.  
In this case, we compare the results of MPT07. Changing parameters $F_g$ and $c$ in equation (\ref{sigma_density_jump}), 
they investigated what values of $F_g$ and $\omega$ result in outward migration (see their Appendix B). 
Table \ref{table5} summarises their experiments with our predictions (given in the brackets). One observes immediately 
that equation (\ref{density_barrier}) well explains the results of their numerical simulations. One might wonder if 
the corotation torque is significant around the density jump, but this is not the case. MPT07 clarified that planets 
migrate outwards due to the reversal of the balance of Lindblad torques, not corotation torques. This arises because 
the Lindblad resonant positions are located further away from planets than the corotation ones. Therefore, migrating 
planets first encounter the inner Lindblad torque at the density jump and this is sufficient to reverse the direction 
of migration.  

In summary, both our simple analytical relations (equations (\ref{thermal_barrier}) and (\ref{density_barrier})) well 
reproduce the results of the detailed numerical simulations of dead zones as planet traps. This indicates that our 
assumptions and treatments are reasonable to capture the physics arising in the more complicated numerical simulations.

\begin{table}
\begin{center}
\caption{Summary of parameter study}
\label{table5}
\begin{tabular}{cccc}
\hline
         & $F_g$=5       & $F_g$=10       & $F_g$=100      \\ \hline \hline
$c= 4$   & in (0.07)     & in (-0.28)     & out (-6.6)   \\
$c= 2$   & out (-0.8)    & out (-2.2)     & out (-28)    \\
$c= 1$   & out (-2)      & out (-4.9)     & out (-57)    \\
\hline
\end{tabular}

$h_p$=0.35, $s$=-3/2, and $t$=-1/2.
in(out) means the inward(outward) migration calculated by MPT07.
The number in the brackets is calculated from equation (\ref{density_barrier}).
\end{center}
\end{table}

\subsection{Ice line barriers} \label{layered_structures}

We finally examine ice line barriers. As discussed in $\S$ \ref{ice_line}, the disc radius of the ice lines 
is determined by viscous heating for a wide range of $\alpha$ (see Fig. \ref{fig3}). In this subsection, 
we discuss how ice lines establish layered structures, and derive analytical relations for the resultant barrier. 
Thus, we investigate ice lines as an example of opacity and turbulent transitions. 
Again, we specify a ice line due to water, although all our analyses are applicable to ice lines of any material. 
The discussion is presented in $\S$ \ref{il_all}.

\subsubsection{Layered structures at ice lines} \label{ls_il}

As mentioned before, the number density of icy grains suddenly increases at the ice lines. If discs are considered to 
be turbulent due to the MRI, this sudden increment of dust can result in layered structures: the MRI active, surface 
layer and the MRI dead, inner layer \citep{g96}. This can be understood as follows. At the ice lines, the number 
density of free electrons in the disc also suddenly drops because they are absorbed by such dust grains. Since free 
electrons are the main contributor of coupling with magnetic fields threading the disc \citep{smun00}, the surface 
density of the MRI active region is strongly diminished, and consequently the surface density of the MRI dead region 
is enhanced. As a result, a density bump appears which acts as a barrier (see Fig. \ref{fig4}). Thus, the mean value 
of $\alpha$ around the ice lines becomes $\alpha \simeq \alpha_D$ (see equation (\ref{mean_alpha})), and hence, the 
ice lines can be regarded as a self-regulated, localised dead zone. In summary, the subsequent process initiated by 
icy grains can reduce the value of $\alpha$ at the ice lines. The disc radius at which it appears is determined by 
viscous heating (see equation (\ref{ril_vis})).

We recall from $\S$ \ref{ice_line} that ice lines become a barrier due to the corotation torque (also see IL08). 
However, any corotation torque may saturate at the ice lines because the turbulence dies out there - a conclusion 
which is valid for any disc in which turbulence is excited by the MRI. If the MRI is the most general source of 
disc turbulence, we expect that the ice lines may be natural barriers arising from the Lindblad torques rather than 
corotation torques.   

\subsubsection{Disc models}

We adopt a parameterised treatment of dead zones in a disc described by IL08. In the region with layered structures, 
the effective $\alpha$ can be generally written as 
\begin{equation}
 \alpha = \frac{\Sigma_A \alpha_A+(\Sigma-\Sigma_A) \alpha_D}{\Sigma},
 \label{mean_alpha}
\end{equation}
where $\Sigma_A$ is the surface density of the active layer, and $\alpha_A$ and $\alpha_D$ are 
the strength of turbulence in the active and dead layers, respectively 
\citep[also see Table \ref{table1}]{kl07,mpt09}. Assuming stationary accretion discs (see equation (\ref{mdot})) with 
equation (\ref{mean_alpha}), the surface density is given as
\begin{equation}
 \Sigma= \frac{\dot{M}}{3 \pi c_s H \alpha_D} - \Sigma_A \frac{\alpha_A - \alpha_D}{\alpha_D},
\label{sigma_density_jump2}
\end{equation}
where $\Sigma_A$ cannot exceed $\Sigma$. Thus, equation (\ref{sigma_density_jump2}) is useful for representing 
the surface density within dead zones where $\Sigma_A < \Sigma$. Also, this equation implies that the structure of 
$\Sigma$ in the dead zones strongly depends on $\Sigma_A$. However, the structure of $\Sigma_A$ is not well 
constrained by theory, since it is very sensitive to the distribution of dust grains as well as the chemical models 
\citep{smun00,in06}. Therefore, we adopt simple prescriptions proposed by \citet{kl07} and IL08 as follows. 
For the active region where $\Sigma_A \approx \Sigma$, the surface density simply becomes 
\begin{equation}
 \Sigma= \frac{\dot{M}}{3 \pi c_s H \alpha_A}, 
 \label{sigma_active}
\end{equation}
if there is no ice line. The effects of the ice lines on $\Sigma$ are described below.

\subsubsection{Density jumps without ice lines}

At first, we examine density barriers produced by dead zones (see equation (\ref{sigma_density_jump2})) to compare 
the our previous analysis done in $\S$ \ref{dead_zone}. For the surface density of the active layer, we assume 
\begin{equation}
 \Sigma_A=\Sigma_{A0} f_{ice }\left( \frac{r}{r_0} \right)^{s_A}
 \label{Sigma_A}
\end{equation}
following \citet{kl07}. We note that $\Sigma_{A0}$ and $f_{ice}$ are very sensitive to the dust distribution 
(also see Table \ref{table1}). However, we set $\Sigma_{A0}$ constant and $f_{ice}=1$ here. A more detailed 
analysis is presented below where the effects of ice lines on $f_{ice}$ are included. In discs with layered 
structures, it is generally considered that $s_A\geq 0$.

Using the above equation, equation (\ref{sigma_density_jump2}) is approximately written as
\begin{eqnarray}
 \Sigma & \simeq & \frac{\dot{M}}{3 \pi \alpha_D H_p^2 \Omega_p} 
                \left( 1 -\epsilon h_p \left( t + \frac{3}{2} \right) \right) \\  \nonumber
  &      &- \Sigma_A(r_p)\frac{\alpha_A - \alpha_D}{\alpha_D} \left( 1+ \epsilon s_A h_p \right),
\end{eqnarray}
where we performed Taylor expansion of $\Sigma$ in terms of $1/m$, and finally expressed it by $h_p$, assuming $
h_p \approx 2/3m$. Thus, the direction of migration is determined by 
\begin{eqnarray}
 &   & \mbox{sgn} \left[ \frac{\dot{M}}{3 \pi \alpha_D H_p^2 \Omega_p} h_p
                \left( -\frac{3}{2} t + \frac{1}{4} \right)  \right.       \\ \nonumber
 & &    \left.  - \Sigma_A(r_p)\frac{\alpha_A - \alpha_D}{\alpha_D} 
                h_p \left(  s_A -\frac{t}{2} + \frac{7}{4} \right) \right].
 \label{thermal_barrier2}
\end{eqnarray}
Since the outer edge of dead zones where a density jump develops is established around the region with 
$\Sigma_A \sim \Sigma/2$ , we can assume
\begin{equation}
 \Sigma_A(r_p)\frac{\alpha_A - \alpha_D}{\alpha_D} \approx 
      \frac{\dot{M}}{3 \pi \alpha_D H_p^2 \Omega_p} \frac{\alpha_A- \alpha_D}{\alpha_A + \alpha_D}.
 \label{eq_rp}
\end{equation}
We emphasise that equation (\ref{eq_rp}) gives the position of planets halted by the density barrier which defines 
the outer edge of the dead zones ($r_p \approx r_{edge}$). In addition, this equation is another expression that 
the necessary condition ($\Sigma_A < \Sigma$) is safely satisfied. As a result, the sign of the net torque is 
written as
\begin{equation}
 \mbox{sgn}\left[s_A -\frac{t}{2} + \frac{7}{4} - 
           \frac{\alpha_A+\alpha_D}{2 \alpha_D} \left(s_A + t + \frac{3}{2} \right) \right].
 \label{density_barrier2}
\end{equation}
Compared with equation (\ref{thermal_barrier}), the layered structure with a density jump gives an additional 
term which controls the direction of migration (since $\alpha_A \gg \alpha_D$). Thus, planets migrate inwards when 
sgn$(s_A+t+3/2)<0$ while they migrate outwards when $s_A+t+3/2>0$. In standard disc models, $s_A>0$ and $t\simeq -1/2$. 
Our relation therefore dictates outward migration, which is also consistent with our previous analysis done in 
$\S$ \ref{dz_dj}.

\subsubsection{Density jumps with ice lines} \label{iceline_dz}

\begin{figure}
\begin{center}
\includegraphics[height=5.5cm]{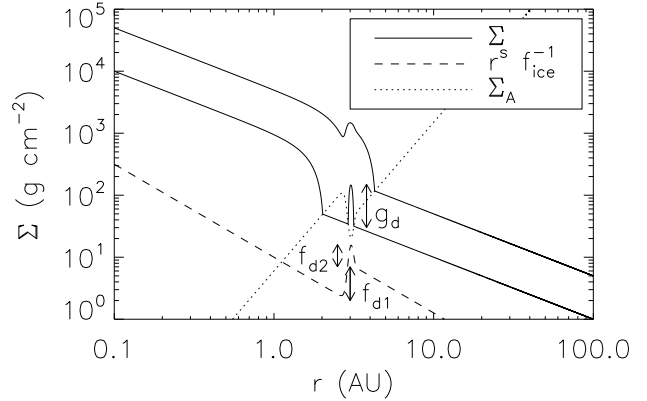}
\caption{The typical structures of $\Sigma$ and $\Sigma_A$ controlled by $f_{ice}$. The upper solid line denotes 
the case that an ice line is located within a dead zone while the lower solid line is for an ice line beyond the dead 
zone. The distribution of $\Sigma_A$ is denoted by the dotted line. The definitions of $f_{d1}$ and $f_{d2}$ in 
$f_{ice}$ and of $g_{d}$ in $\Sigma$ are labeled in the figure.}
\label{fig4}
\end{center}
\end{figure}

We incorporate the effects of ice lines on the surface density of the active layer, following IL08 
(see Fig. \ref{fig4}). In this case, we need to consider two separate cases, depending on the ratio of 
$r_{il}$ to $r_{edge}$, where $r_{il}$ is the disc radius of an ice line. For the case that $r_{il}/r_{edge} < 1 $, 
we can use equations (\ref{sigma_density_jump2}) and (\ref{Sigma_A}) as the surface density while we need to modify 
equation (\ref{sigma_active}) for the case that $r_{il}/r_{edge} > 1 $.

We first examine the case that $r_{il} < r_{edge}$, that is, an ice line is located inside of a dead zone. 
In this case, $f_{ice}$ in equation (\ref{Sigma_A}) is generally given as 
\begin{equation}
 f_{ice}=f_d\left[ 1+ \frac{f_{d1}}{2} \left(1+ \tanh\left( \frac{r-r_{il}}{w} \right) \right) \right]^{-1}, 
 \label{eq_f_ice}
\end{equation}
where $f_d$ and $f_{d1}$ parameterise the effects of an ice line at which the surface density of the active layer 
suddenly drops. As noted by \citet{kl07}, the surface density distortion induced by ice lines produces a radial, 
positive pressure gradient there, and consequently dust can be trapped there. Otherwise, it migrates inwards due to 
the so-called head winds arising from the gas motion \citep{w77}. Since the maximum migration speed of dust is 
$\sim 1/\alpha$ times faster than the viscous evolution of gas, dust is quickly piled up at the ice lines. 
This effect can be included in $f_d$ which is generally written as
\begin{equation}
 f_d=  \left[ 1+f_{d2} \exp \left(- \left( \frac{r-r_{il}}{w} \right)^2 \right) \right]^{-1}.
 \label{eq_f_d}
\end{equation}
We note that inclusion of the dust effects may upgrade the ice line barriers to planet traps although we keep 
calling them barriers. 

Fig. \ref{fig4} shows the typical $\Sigma$ (as controlled by equations (\ref{eq_f_ice}) and (\ref{eq_f_d})) and 
the role of the parameters $f_{d1}$ and $f_{f2}$ in $f_{ice}$ .

Consequently, we can approximate the surface density (see equations (\ref{sigma_density_jump2}), (\ref{Sigma_A}), 
(\ref{eq_f_ice}), and (\ref{eq_f_d})) as
\begin{eqnarray}
 \Sigma & \simeq & \frac{\dot{M}}{3 \pi H_p^2 \Omega_p \alpha_D} 
               \left(1 - \epsilon h_p \left( t + \frac{3}{2} \right) \right)  \\ \nonumber
        &      &   - \Sigma_A(r_p,f_{ice}=1) \frac{\alpha_A - \alpha_D}{\alpha_D} D,
\end{eqnarray}
where
\begin{eqnarray}
 D & = & A B + \epsilon h_p \\ \nonumber
   & \times &  \left( B \left(  A s_A  + \epsilon \frac{t+3}{2c} \frac{f_{d1}}{2}\frac{1}{\cosh^2(1/c)}  
                        \right) - A(B-1) \frac{t+3}{c^2} \right), 
\end{eqnarray}
\begin{equation}
 A= 1 + \frac{f_{d1}}{2} \left(1 + \epsilon \tanh \left( \frac{1}{c} \right) \right) 
  \equiv A_1 + \epsilon A_2 ,
\end{equation}
and
\begin{equation}
 B=  1 + f_{d2} \exp \left(- \frac{1}{c^2} \right).
\end{equation}
We emphasise that the location of the ice line barrier which can halt planets migrating inward ($r_p\approx r_{il}$) 
is determined by viscous heating (see equation (\ref{ril_vis})) while the location of the density barrier produced by 
dead zones ($r_p\approx r_{edge}$) is determined by equation (\ref{eq_rp}). 

As a result, the sign of the net torque is given as
\begin{equation}
 \mbox{sgn} \left[ \frac{\dot{M}}{3 \pi H_p^2 \Omega_p \alpha_D} 
               h_p \left(- \frac{3}{2}t + \frac{1}{4} \right) 
           - \Sigma_A(r_p,f_{ice}=1) \frac{\alpha_A - \alpha_D}{\alpha_D} E \right],
 \label{iceline_barrier1}
\end{equation}
where 
\begin{eqnarray}
 E & = & A_2 \left( B - (B-1)\frac{t+3}{c^2} \right)   \\ \nonumber
   & + & h_p A_1 \left[ B
                     \left( s_A -\frac{t}{2} + \frac{7}{4} \right) 
                     - (B-1)\frac{t+3}{c^2} \left( -\frac{t}{2} + \frac{7}{4}  \right)
                                          \right].
\end{eqnarray}
The above equation is identical to equation (\ref{thermal_barrier2}) if $A_1$ and $B$ $\rightarrow 1$, and 
$A_2\rightarrow 0$ (equivalently, $f_{d1}$ and $f_{d2}$ $\rightarrow 0$). Before fully examining equation 
(\ref{iceline_barrier1}) which looks very complicated, it is very useful to understand a simplified case. 
Taking the leading terms, outward migration arises when 
\begin{equation}
 \Sigma_A(r_p,f_{ice}=1) > \frac{\dot{M}}{3 \pi (\alpha_A - \alpha_D)} \frac{1}{r_p c_s(r_p)}.
 \label{iceline_lt}
\end{equation}
The relation shows that the ice line barrier is readily active for discs with low accretion rate, high disc 
temperatures, high surface density of the active region, and large disc radii of the ice lines ($r_p \approx r_{il}$). 
In addition, this relation is very useful for constraining the effectiveness of the ice line barrier, which is 
discussed more in the next subsection. 

\begin{figure}
\begin{center}
\includegraphics[width=7.7cm]{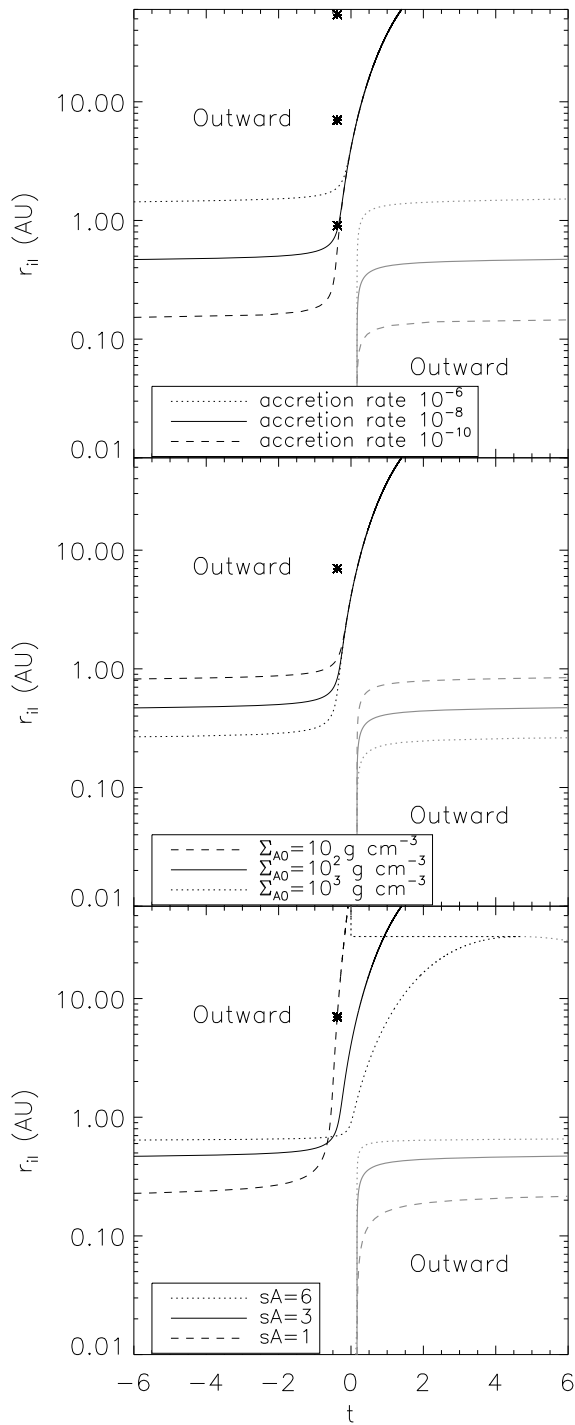}
\caption{The location of a water-ice line as a function of the temperature exponent $t$. The black lines denote 
the negative solution to equation (\ref{iceline_barrier1}) while the gray, thick lines are for the positive solution. 
The regions where planets migrate outwards are surrounded by one of the solutions and $x-$ and $y-$axes 
(labeled by Outward). For the top, middle, and bottom panels, $\dot{M}$, $\Sigma_{A0}$, and $s_A$ are varied, 
respectively. Otherwise, we adopt the values of the fiducial model. For the case of $t<0$ with higher $\dot{M}$ and $s_A$ and lower $\Sigma_{A0}$, the 
outward migration region shrinks. For the case of $t>0$ the situation is the opposite. For comparison purposes, 
we label $r_{il}$ by asterisks (see equation (\ref{ril_vis2})). For the top panel, the value of $r_{il}$ increases 
with increasing $\dot{M}$. This indicates that the conditions for outward migration are determined by the complex 
dependence on $\dot{M}$ (see equation (\ref{iceline_lt})).}
\label{fig5}
\end{center}
\end{figure}

We can readily solve the equation (\ref{iceline_barrier1}), since it is quadratic in $t$. Fig. \ref{fig5} shows 
$r_{il}$ as a function of $t$. 
The black lines denote the negative solution while the gray, thick lines are for the positive solution. The regions 
where planets migrate outwards are encompassed by one of the solutions, $x-$ and, $y-$axes (labeled by Outward). 
In this figure, we set $f_{d1}=6$, $f_{d2}=2$, $s_A=3$, and $\Sigma_{A0}/\Sigma_0 =0.01$ at $r_0=1$ au following IL08 
(also see Table \ref{table3}). We consider a water-ice line, so that $T_{m,\mbox{H}_2\mbox{O}}(r_{il})= 170$ K 
\citep{js04}. We use $c=1$, because any sharp transition in discs is smoothed out by disc viscosity \citep{ym10}. For 
stellar parameters, we adopt the values of CTTSs (see Table \ref{table3}). This disc setup is our fiducial model 
(see the solid lines). Based on equation (\ref{iceline_lt}), the choice of stellar parameters ($R_*$ and $T_*$) 
does not matter if $r_{il}$ is parameterised (see equations (\ref{t_irr}) and (\ref{ril_vis})). Although the complete 
treatment in which $r_{il}$ depends on stellar parameters is presented in $\S$ \ref{synthesis}, we simply label 
$r_{il}$ by asterisks for comparison purposes here (see equation (\ref{ril_vis2})). 

In Fig. \ref{fig5}, we explore parameter space by changing $\dot{M}$, $\Sigma_{A0}$, and $s_A$ 
(on the top, middle, and bottom panels). For $t<$0 which is general for any disc, higher $\dot{M}$ and $s_A$, 
and lower $\Sigma_{A0}$ reduce the outward migration region. This is consistent with the above argument 
(see equation (\ref{iceline_lt})). The situation is the opposite for $t>$0 which can happen due to 
hot dusty walls. If the dependence of $r_{il}$ on $\dot{M}$ is taken into account (see the top panel), 
higher accretion rates do not necessarily reduce the region where the water-ice line barrier is active. This 
arises because higher accretion rates result in larger water-ice line radii which are preferred for 
outward migration (see equation (\ref{iceline_lt})). Thus, the conditions for outward migration are controlled by the 
complex dependence on $\dot{M}$. Another interesting feature is shown in the bottom panel. Around $r_{il} \approx$ 
30 au with $t>0$, both negative and positive solutions merge with each other (see the black and gray dotted lines on 
the bottom panel). This happens because there is no solution to equation (\ref{iceline_barrier1}). 

Now, we examine the case that $r_{il}/r_{edge} > 1 $, that is, an ice line is located beyond the dead zone in the 
active region of the disc. In this case, we adopt the following functional form for $\Sigma$;
\begin{equation}
 \Sigma =  \frac{\dot{M}}{3 \pi c_s H \alpha_A} 
      \left[ 1+  g_d \exp \left(- \left( \frac{r-r_{il}}{w} \right)^2 \right)  \right].
\end{equation}
Compared with equation (\ref{sigma_active}), the ice line produces a density bump at $r=r_{il}$ which is a 
consequence of a localised layered structure. This function well represents the results of IL08 with a proper choice 
of $g_d$ (see the lower solid line in Fig. \ref{fig4}). Here, we only examine the power-law index of $\Sigma$ rather 
than expanding it in terms of $h$. This is because one of our approximations, $r_p\approx r_{il}$, limits the 
applicability to Gaussian-functions. In fact, planets migrating towards the density bump can be halted around 
$r_p=r_{il} + \vartriangle r$ where a negative steep surface density profile plays the critical role. However, 
the assumption that $r_p \approx r_{il}$ washes out this effect. This also indicates that equation 
(\ref{iceline_barrier1}) may underestimate the required condition. 

The power-law index of $\Sigma$ is given as
\begin{equation}
 \bar{s} \equiv \frac{d \ln \Sigma}{d \ln r} = - t - \frac{3}{2} - 
          \frac{ \frac{2r_p(r_p-r_{il})}{w^2} g_d \exp \left(- \left( \frac{r_p-r_{il}}{w} \right)^2 \right)}{1+  g_d \exp \left(- \left( \frac{r_p-r_{il}}{w} \right)^2 \right)}.
\end{equation}
We recall that $\bar{s}$ becomes equivalent to $s$ for pure power-law discs (see equation (\ref{corotation_barrier})). 
Assuming planets to be halted around $r_p \approx r_{il}+w\sqrt{\ln 2}$ where the Gaussian function takes half of the 
maximum value, the resultant value of $\bar{s}$ becomes
\begin{equation}
 \bar{s} = -t - \frac{3}{2} - \frac{ \frac{g_d}{ch_p} \sqrt{\ln 2}}{1+ g_d / 2}.
\end{equation}
As a result, we find that the direction of migration is controlled by 
\begin{equation}
 \mbox{sgn}\left( \bar{s} - \frac{t}{2} + \frac{7}{4} \right),
 \label{iceline_barrier2}
\end{equation}
where we have assumed any disc quantity to behave as power-law in a local region centered at $r\approx r_p$ 
(see $\S$ \ref{simple_power-law}). For discs with $h_p=0.1$, $c=1$, and $g_d=3$, $t>-6.5$ is required for outward 
migration while for discs with $h_p=0.1$, $c=1$, and $g_d=6$, $t>-8.2$ is needed. Thus, the direction of migration 
is readily reversed by the density bump produced by the ice lines. Furthermore, larger density bump $g_d$ expands 
the outward migration region, as expected.

\subsection{Locations of the barriers}

We can now compare the dead zone and ice line barriers in order to examine the relative importance of each. 
We adopt the analysis done for the case of $r_{il}/r_{edge}$. We especially focus on the relative location of each 
barrier.  Combining two equations (\ref{eq_rp}) and (\ref{iceline_lt}), we find the condition which is given as
\begin{equation}
 \frac{r_{il}}{r_{edge}} > 
 \left( h(r_{edge}) \frac{\alpha_A + \alpha_D}{\alpha_A - \alpha_D} \right)^{\frac{1}{s_A+t/2+1}},
 \label{condition_il}
\end{equation}
where we set $r_0=r_{edge}$. If this condition is satisfied, the ice line and density barriers are both active 
while if not, only the density barrier is active. Thus, the separation between the ice line and dead zone barriers is 
required to be relatively small in order for the ice line barrier to be effective. Table \ref{table6} summarises 
the threshold values of $r_{il}/r_{edge}$ for various $s_A$, $t$, and $h(r_{edge})$. As all three quantities increase, 
the threshold values of $r_{il}/r_{edge}$ increase. In other words, the parameter space where the ice line barrier is 
effective is larger for lower values of $s_A$, $t$, and $h(r_{edge})$. More complete treatments in which the ice line 
radius is determined by viscous heating are presented in $\S$ \ref{synthesis}. 
    
\begin{table}
\begin{center}
\caption{The condition for the ice line barrier to be active}
\label{table6}
\begin{tabular}{cccc}
\hline
($s_A$,$t$,$h(r_{edge})$) & (3,-1/2,0.01)  & (3,-1/2,0.1)  & (3,-1/2,1)     \\ 
$r_{il}/r_{edge}$         & 0.2            & 0.5           & 1              \\ \hline
($s_A$,$t$,$h(r_{edge})$) & (3,-1.5,0.1)   & (3,-1/2,0.1)  & (3,1.5,0.1)    \\ 
$r_{il}/r_{edge}$         & 0.4            & 0.5           & 0.7            \\ \hline
($s_A$,$t$,$h(r_{edge})$) & (1,-1/2,0.1)   & (3,-1/2,0.1)  & (6,1/2,0.1)    \\ 
$r_{il}/r_{edge}$         & 0.2            & 0.5           & 0.7            \\
\hline
\end{tabular} 
\end{center}
\end{table}

\section{Discussion} \label{discus}     

Before we integrate our analyses into a coherent picture, we discuss other possible barriers induced by other physical 
processes which are neglected in this paper. Also, we discuss opacity transitions that are neglected in our unified 
picture. 
In addition, we examine the possibility of molecules other than water to excite ice line barriers. Finally, we estimate 
the mass range in which planets are regarded as type I migrator.

\subsection{Other possible barriers}

We have so far assumed that stellar irradiation heats up only the dust in discs. However, photons emitted from 
stars, especially with very short wavelengths such as UV and X-rays can heat up the gas. This results in 
the photoevaporation of discs \citep{hjl94,jhb98} - which considerably affects planet formation and migration 
\citep[e.g.][]{mab09}. \citet{lpm10} showed that photoevaporation can activate another barrier. This 
arises by the reduction of the surface density of gas due to photoevaporation. Since viscous heating strongly 
depends on the surface density (see equation (\ref{t_vis})), this reduction results in shallower temperature profiles. 
Consequently, planets are halted at the region where the transition of the temperature slope occurs. 

In general, photoevaporation is considered to be important for discs surrounded by nearby OB associations which 
provide a huge input of high energy photons. In addition to photoevaporation, one expects that such 
massive stars can significantly affect the disc structure by changing the dust temperature. Recently, \citet{gh09} 
have shown that, without nearby massive stars, even low-mass central stars ($> 0.3 M_{\oplus}$) can drive 
photoevaporation of their surrounding discs ($r\gtrsim 20$ au). This arises from energetic stellar irradiation 
(far-UV, extreme-UV, and X-rays). We will address this in a future publication.
 
In addition, we neglect the effects of the inner edge of dead zones where a positive radial gradient of the surface 
density can form. Such a profile can produce a barrier due to the vortensity-related corotation torque. However, 
the disc radius of the inner edge of dead zones can be fixed around $r\approx 0.01$ au, because the radius is 
controlled by the thermal ionisation temperature of gas. Thus, the barrier is unlikely to play an important role in 
the diversity of the detected exoplanetary systems. For the reason, we disregard it in this paper. 

\subsection{Neglect of opacity transitions} \label{ot_all}

We have intensively investigated ice lines as an opacity transitions in $\S$ \ref{ice_line} while 
our approach is applicable to any other opacity transitions. In general, protoplanetary discs have several opacity 
transitions. MG04 showed that any opacity transition (including ice lines) affects the disc surface density and 
temperature in a similar fashion. Therefore, it is naturally expected that a single disc can have several barriers 
that are all excited by opacity transitions. Nonetheless, we will neglect all opacity transitions in 
$\S \ref{synthesis}$ for the following reasons.

Except for those produced by ice lines, all opacity transitions are a consequence of the high disc temperatures and 
subsequent destruction of opacity sources. As a result, their disc radii distribute well inside of 1 au (see fig. 1 
of MG04). MG04 found that these opacity transitions make the migration time significantly longer due to the Lindblad 
torque alone. (Equivalently, the net torque exerted on a planet becomes a very small (in magnitude), negative value.)
This indicates that, in order to actually stop the planet's inward migration, a small amount of outward-directed 
corotation torque would be needed. However, any corotation torque at the opacity transitions is likely to vanish 
due to the presence of dead zones. As discussed above, protoplanetary discs may have dead zones that are 1- 10 au in 
size from their central stars, depending on their column density. One of the key features of the dead zones is a 
low level of turbulence, which results in the saturation (ineffectiveness) of corotation torques there 
(see $\S$ \ref{w_dz}). Thus, opacity transitions within dead zones cannot be planet traps and therefore we disregard 
opacity transitions other than those created by ice lines. For the ice lines, it is more likely that the process 
initiated by the increment of icy dust grains reduces a level of turbulence there and hence ice lines are regarded 
as localised dead zones (see \ref{ls_il}). Thus, we include ice lines as an opacity and turbulent transition rather 
than an opacity transition in $\S$ \ref{synthesis}. In summary, any opacity transition is neglected below.

\subsection{Ice lines of other molecules} \label{il_all}

We have focused on a water-ice line in the above discussions, although our analyses 
apply to ice lines of any molecules in discs. It is well known that water is not the most abundant chemical species in 
protoplanetary discs \citep[e.g.][]{ah99}. In general, CO is the most abundant molecule in discs. \citet{d07} showed 
that the total amount of CO-ice is much larger than water-ice in irradiated accretion discs. Furthermore, different 
species have different condensation temperatures. \citet{d05} demonstrated that the location of ice lines strongly 
depends on molecules in consideration. Thus, it is very interesting to investigate which molecules can produce 
barriers in discs.

In order to proceed, we adopt the analyses done in $\S$ \ref{layered_structures}. We especially focus on the minimum 
abundance of molecules which is required for their ice lines to be a barrier. We present the detail analysis in 
Appendix \ref{app3} and summarise our findings here. We find that whether or not ice lines of other molecules act as 
barriers strongly depends on the disc parameters such as $s_A$ and $t$. Adopting reasonable values for them, we 
find that, for molecules that condense within a dead zone, only specific, abundant species such as CO have the 
possibility to work as a barrier. If molecules freeze out beyond the dead zone, then even a tiny fraction of the 
molecules can act as a barrier. In summary, other molecular species can produce ice lines but these will 
form near to the position of the outer edge of the dead zone ($r_{edge}$). We leave more comprehensive models 
for ice lines to a future publication. In the following discussions, we again assume water-ice to be abundant 
enough for creating a barrier.

\subsection{Mass limits for type I migration}

We discuss the mass limits that our analyses apply to. Type I migration is only applicable to low mass protoplanets. 
Massive protoplanets can open up a gap in their discs and undergo type II migration. This mode of migration is 
controlled by viscous evolution of discs and is much slower than the type I migration. The critical mass, also known 
as the gap-opening mass, is well discussed in the literature \citep[e.g.][]{mp06}, and given by
\begin{equation}
 \frac{M_{p,max}}{M_*} = \mbox{min}\left[ 3 C h_p^3, C\sqrt{40 \alpha h_p^5} \right],
 \label{M_max}
\end{equation}
where we have introduced a coefficient $C=3$ into the original condition by taking into account the reduction of 
the tidal torque due to the disc thickness \citep{hp10c}. The left term in the brackets of equation (\ref{M_max}) 
is derived by a Hill radius analysis while the right term for viscous ones. Table \ref{table7} summarises $M_{p,max}$ 
for stars with various masses at $r_p=1$ au. The stellar and disc parameters are given in Table \ref{table3}. For discs 
with $\alpha=10^{-2}$, the critical mass is established by the Hill radius analysis, and has the same 
order of Neptune to Saturn masses. On the other hand, the criterion derived for viscous discs provides the gap-opening 
mass for discs with $\alpha=10^{-5}$, and $M_{p,max}$ is on the order of Earth masses. 

What is the minimum mass for protoplanets to undergo type I migration? It is interesting that the dynamics of 
low mass bodies, from dust grains to planetesimals, can drastically differ from that of protoplanets 
\citep{ahn76,bs10,ng10}. This arises mainly from gas drag. One of the most famous consequences of the gas drag is the 
rapid inward migration of meter-sized particles discussed in $\S$ \ref{iceline_dz}. The size dependency of gas drag 
affects the formation of planetesimals \citep[see][for a recent review]{cy10}. However, we simply focus on the 
dynamics of planetesimals, since we are interested in minimum mass of type I migrator. In laminar discs, the rate of 
change of semi-major axis induced by gas drag is given by \citep[see][for derivation]{ahn76}
\begin{equation}
 \frac{d r_p}{dt} \simeq - K_{drag} \frac{\rho_p r_p^2 \Omega_p}{M_p^{1/3}},
\end{equation}
where $K_{drag}$ is a function of the eccentricity, the inclination, and the material density of a planetesimal, and 
the radial pressure gradient of discs. Following \citet{ki00}, we find $K_{drag} \sim 2.7 \times 10^{-5}$ in cgs 
units. As a result, the timescale of gas drag is written as 
\begin{equation}
 \tau_{drag} \simeq r_p \left| \frac{d r_p}{dt} \right|^{-1} = \frac{M_p^{1/3} h_p}{K_{drag} \Sigma_p \Omega_p},
 \label{tau_drag}
\end{equation}
where we have assumed $\rho_p \simeq \Sigma_p/H_p$. On the other hand, the migration timescale is generally 
written as
\begin{equation}
\tau_{mig} \simeq \frac{M_p r_p^2 \Omega_p}{2 \Gamma},
\label{tau_mig}
\end{equation}
where the tidal torque is scaled by 
\begin{equation}
 \Gamma = K_{mig} \left( \frac{M_p}{M_*} \right)^2 \frac{\Sigma_p r_p^4 \Omega_p^2}{h_p^2},
\end{equation}
and $K_{mig}$ is an order of 1 to 10, depending on disc models \citep{pbck09}. Equating equation (\ref{tau_drag}) with 
equation (\ref{tau_mig}), the critical mass is given as 
\begin{equation}
 M_{p,min}= \left( \frac{K_{drag}h_pM_*^2}{2K_{mig}r_p^2} \right)^{3/4}.
\end{equation}
Table \ref{table7} summarises the minimum mass of type I migrator for stars with various masses. We set $K_{mig}=5$. 
The values of $M_{p.min}$ are at least three orders of magnitude smaller than Earth masses. Thus, the dynamics of solids 
in laminar, gaseous discs derives from type I migration over four to six orders of magnitude in mass.

In turbulent discs, the situation becomes more complicated, because planetesimals are affected by gas drag as 
well as stochastic torque arising from density fluctuations. \citet{ng10} undertook numerical simulations of the 
dynamics of planetesimals in MHD turbulent discs and found that the stochastic torque becomes dominant over the gas 
drag for planetesimals which are $\simeq$ 25 meters in size. This implies that the critical mass $M_{p.min}$ is 
involved with the stochastic torque rather than the gas drag. Adopting the torque formula of \citet{ttw02}, 
\citet{ng10} estimated that $M_{p,min} \simeq 0.11 M_{\oplus}$ in discs with $\alpha \simeq 10^{-2}$. 

Our value for $M_{p,min}$, is estimated for laminar discs and is several orders of magnitude 
smaller than that in turbulent discs. However, equilibrium states for planetesimals larger than 100 m are not achieved 
in their simulations. Thus, more effort is required for accurately estimating the minimum mass of type I migrator.   

In summary, our analyses are likely applicable to protoplanets with the mass range from well below Earth to Saturn 
masses for either laminar or turbulent discs. It is important to emphasise that almost observed low-mass exoplanets 
are covered by our analyses.

\begin{table*}
\begin{minipage}{17cm}
\begin{center}
\caption{Mass limits of type I migrator at 1 au}
\label{table7}
\begin{tabular}{cccc}
\hline
                                              & Herbig Ae/Be stars  & CTTSs              & M stars     \\  \hline
$M_{p,max}$ ($M_{\oplus}$) ($\alpha=10^{-2}$) & 63.8                & 50.7               & 18.6        \\ 
$M_{p,max}$ ($M_{\oplus}$) ($\alpha=10^{-5}$) & 3.0                 & 1.9                & 0.6          \\ 
$M_{p,min}$ ($M_{\oplus}$)                    & $3.7\times10^{-3}$  & $4.6\times10^{-4}$ &  $4.8\times10^{-5}$    \\
\hline
\end{tabular} 
\end{center}
\end{minipage}
\end{table*}

\section{A unified picture of planet traps} \label{synthesis}

\begin{table*}
\begin{minipage}{17cm}
\begin{center}
\caption{Summary of all possible traps}
\label{table8}
\begin{tabular}{ccc}
\hline
                          & Locations                 & Conditions                                              \\ \hline
Ice line traps         & equation (\ref{ril_vis})  & equation (\ref{condition_il}) only for $r_{il}< r_{edge}$\\
Dead zone traps        & equation (\ref{eq_rp})    & N/A                                                     \\
Heat transition traps  & equation (\ref{r_transit})&  $r_{ht}> r_{edge}$    \\
\hline
\end{tabular} 
\end{center}
\end{minipage}
\end{table*}

We now integrate the analyses of $\S$ \ref{wo_dz}, \ref{w_dz} and \ref{discus} into a single comprehensive framework for 
understanding planetary system architectures. More specifically, we investigate under what conditions which barriers 
are effective, and identify the location of the active barriers in the entire disc. Thus, we call any 
barrier a planet trap, since it plays a role of halting migration as well as of affecting the subsequent formation of 
planetary systems. We summarise all possible planet traps and the necessary conditions for them in Table \ref{table8}. 
We find that up to three planet traps can exit in one disc model while the minimum case is just a dead zone trap.

Before examining the location and conditions in Table \ref{table8}, we re-derive equations (\ref{ril_vis}) and 
(\ref{r_transit}) by simply substituting $\Sigma_0$ with $\dot{M}$ using equation (\ref{mdot}). Then 
equation (\ref{ril_vis}) becomes 
\begin{equation}
 \frac{r_{il}}{r_0} = \left[ \frac{1}{T_{m,\mbox{H}_2\mbox{O}}^{12}(r_{il})}
                        \frac{27 \bar{\kappa}_0 \mu_g \Omega_0^3} {64 \sigma_{SB} \alpha \gamma k_B}
                        \left( \frac{\dot{M}}{3 \pi} \right)^2
                 \right]^{2/9} \propto \dot{M}^{4/9},
 \label{ril_vis2}
\end{equation}
where $\bar{\kappa}_0= 2\times 10^{16}$, and equation (\ref{r_transit}) becomes 
\begin{eqnarray}
 \label{r_transit2}
 \frac{r_{ht}}{r_0} & = & \left[ \frac{1}{T_{m0}} \left( \frac{r_0}{R_*} \right)^{3/7} 
                          \left( \frac{27 \bar{\kappa}_0  \mu_g \Omega_0^3}{64 \sigma_{SB} \alpha \gamma k_B} 
                               \left( \frac{\dot{M}}{3 \pi} \right)^2  \right)^{1/3} 
                          \right]^{14/15} \\
                   & \propto & \dot{M}^{28/45}, \nonumber
\end{eqnarray}
where $\bar{\kappa}_0= 2 \times 10^{-4}$. Also, equation (\ref{eq_rp}) is rewritten as 
\begin{equation}
 \frac{r_{edge}}{r_0} = \left( \frac{\dot{M}}{3 \pi (\alpha_A+\alpha_D) \Sigma_{A0} H_0^2 \Omega_0} 
                         \right)^{\frac{1}{s_A+t+3/2}},
 \label{eq_rp2}
\end{equation}
where we assume that $s_A$, $t$, and $\Sigma_{A0}$ are all external parameters. This rearrangement of equations 
is very useful for integrating our separate analyses done in $\S$ \ref{wo_dz} and \ref{w_dz} into one picture.

\begin{figure}
\begin{center}
\includegraphics[height=6.95cm]{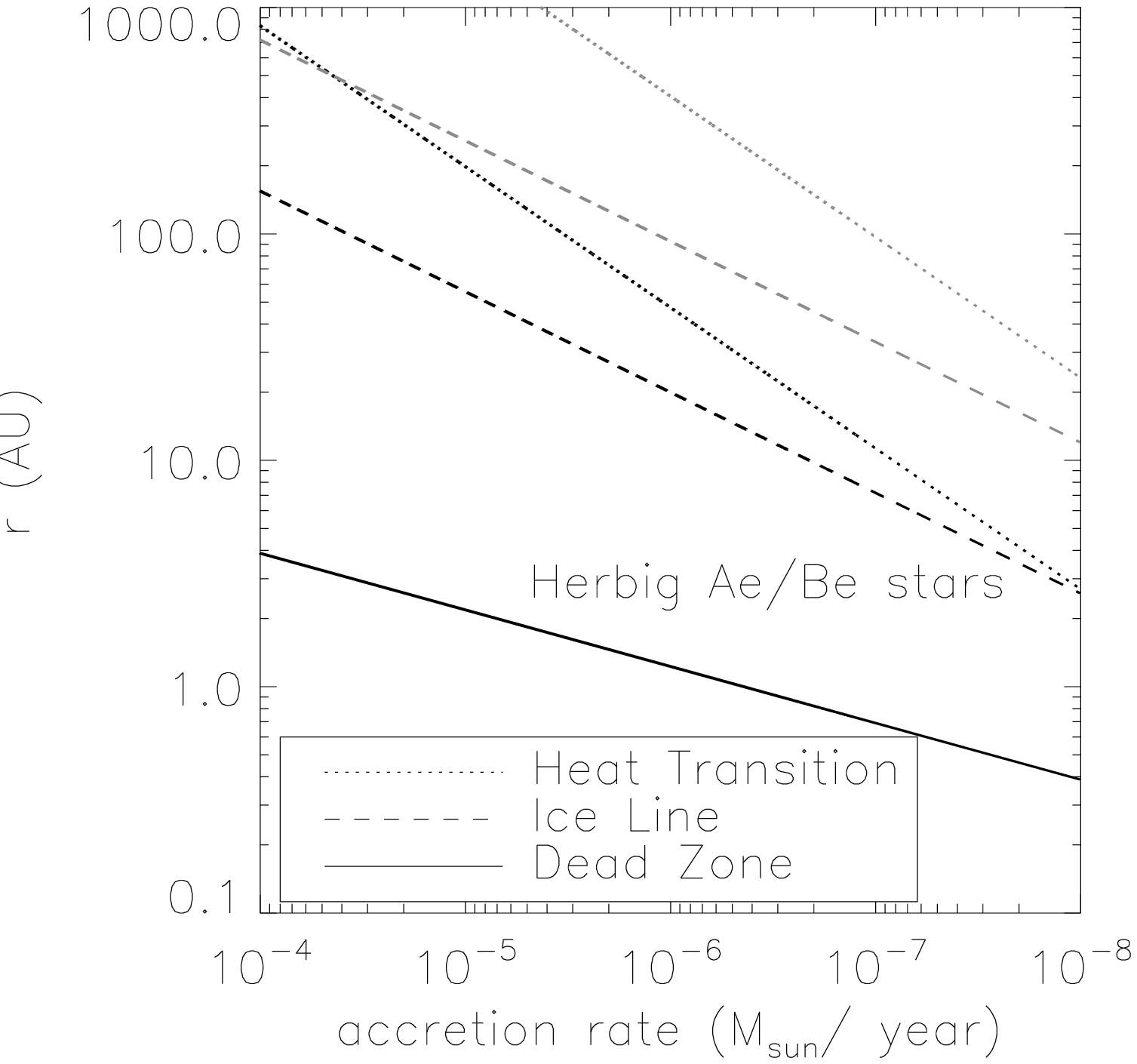}
\includegraphics[height=6.95cm]{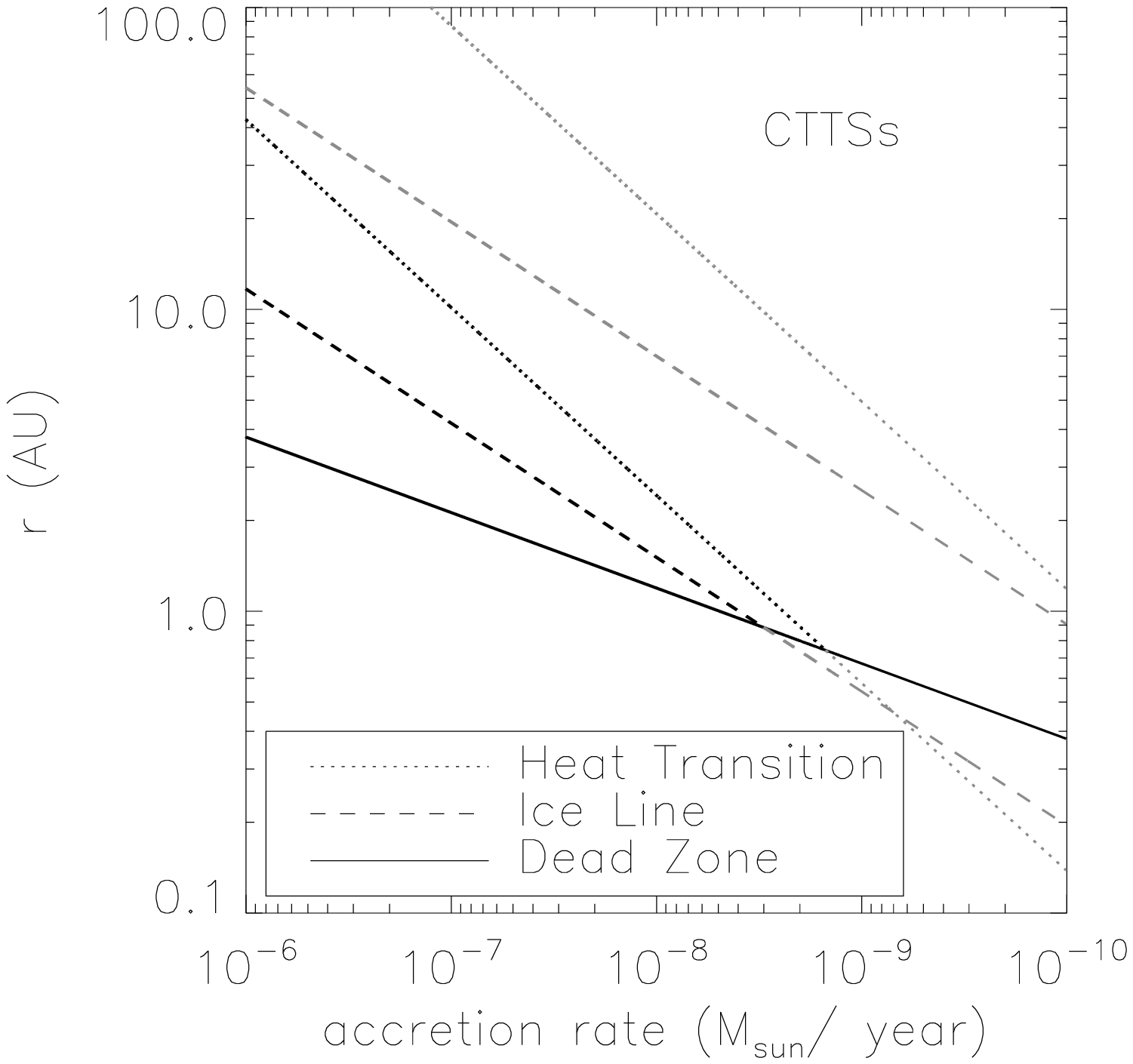}
\includegraphics[height=6.95cm]{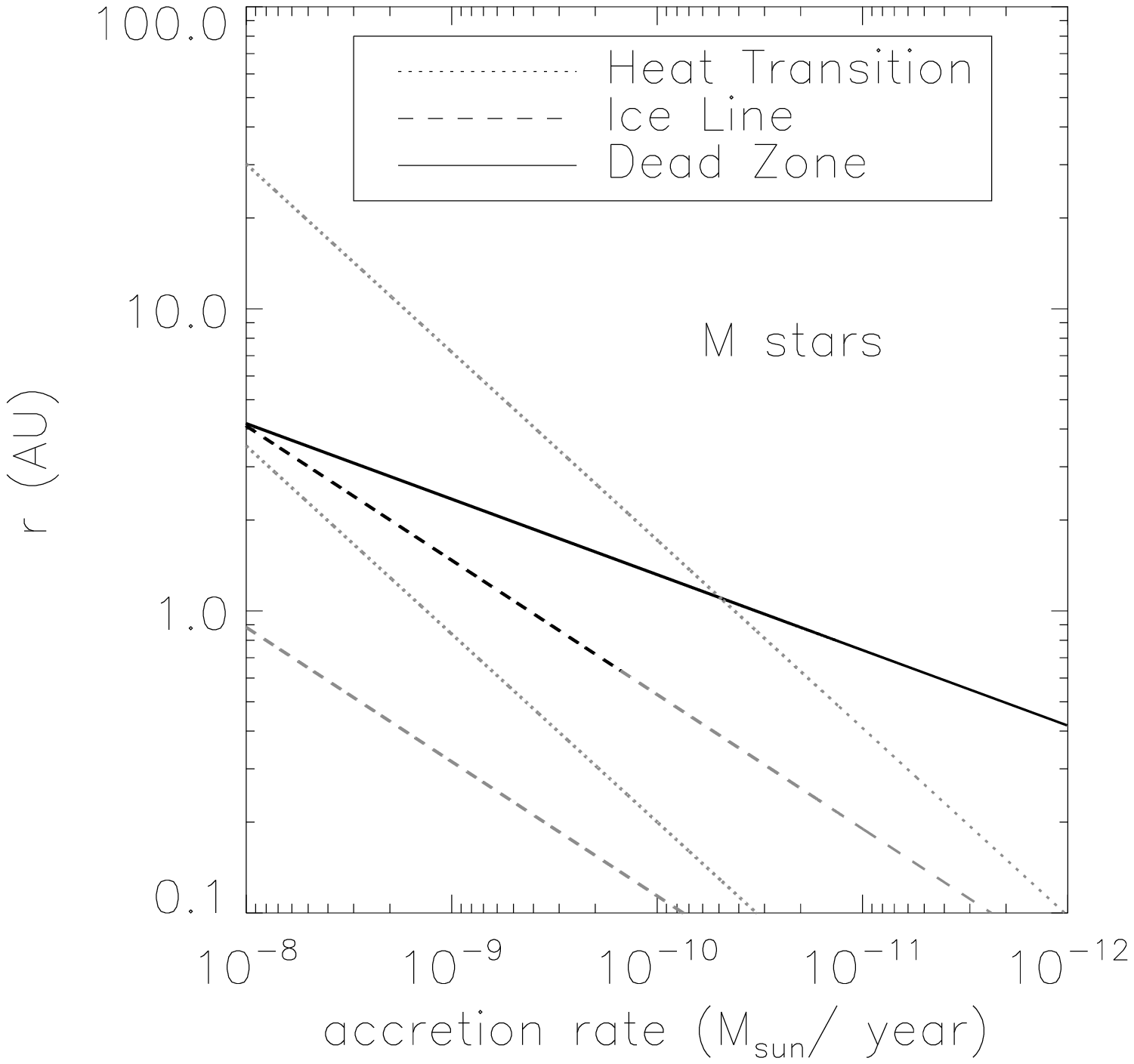}
\caption{The location and evolution of all three planet traps. The heat transition traps with $\alpha=10^{-2}$ and 
$\alpha=10^{-5}$ are denoted by the lower and upper dotted lines, respectively (see equation (\ref{r_transit2})). 
The ice line traps with $\alpha=10^{-2}$ and $\alpha=10^{-5}$ are denoted by the lower and upper dashed lines, 
respectively (see equation (\ref{ril_vis2})). The dead zone traps are shown by the solid line (see equation 
(\ref{eq_rp2})). Taking into account their active conditions (see Table \ref{table8}), all effective traps are 
denoted by the black lines. For the Herbig Ae/Be star case, all three traps are effective for a wide range of 
$\dot{M}$. For the CTTS case, the ice line and heat transition traps merge with the dead zone trap. For the M star 
case, the ice line trap disappear for low values of $\dot{M}$ (see equation (\ref{condition_il})). Thus, planet 
traps interact with each other, depending on the stellar and disc parameters.}
\label{fig6}
\end{center}
\end{figure}

The important result is that the locations of all three traps are mainly controlled by $\dot{M}$, but 
with different power-law dependencies. This suggests that some traps can merge with each other due to time-dependent, 
viscous evolution of discs. This is clearly shown in Fig. \ref{fig6}. In this figure, the radius of all three traps 
is plotted as a function of $\dot{M}$ by setting 
$(s_A,t,\Sigma_{A0})=(3,-1/2,\Sigma_0/100)$ using the value of $\Sigma_0$ in Table \ref{table3}. For stellar 
parameters, we adopt the values of Table \ref{table3}. We calculate $H_0$, assuming stellar irradiation to be dominant. 
As discussed in the above two sections, the heat transition traps can be active for our models with $\alpha =10^{-2}$ 
and their disc radii are represented by equation (\ref{r_transit2}) (see the lower dotted lines). For comparison 
purposes, we also show $r_{ht}$ for models with $\alpha=10^{-5}$ (see the upper dotted line). The ice line traps can 
exist in both active and dead regions. Thus, their disc radii are expressed by equation (\ref{ril_vis2}) with 
$\alpha=10^{-2}$ and $10^{-5}$ for the active and dead regions, respectively (see the lower and upper dashed lines). 
The solid line is for the location of the dead zone trap (see equation (\ref{eq_rp2})). Taking into account the 
conditions for them (see Table \ref{table8}), the locations of all active traps are shown by the black lines.  

\subsection{A picture of forming planetary systems around stars with various masses}

\begin{table*}
\begin{minipage}{17cm}
\begin{center}
\caption{The position of planet traps}
\label{table9}
\begin{tabular}{cccc}
\hline \hline
                                     & Herbig Ae/Be stars                              \\
                                     & $r_{edge}$ (AU) & $r_{il}$ (AU) & $r_{ht}$ (AU) \\ \hline
$\dot{M}=10^{-4}$($M_{\odot}$/year) & 3.9             & 155           & 830           \\
$\dot{M}=10^{-6}$($M_{\odot}$/year) & 1.2             & 20            & 47            \\ 
\hline \hline
                                     & CTTSs                                           \\
                                     & $r_{edge}$ (AU) & $r_{il}$ (AU) & $r_{ht}$ (AU) \\ \hline
$\dot{M}=10^{-6}$($M_{\odot}$/year) & 3.7             & 11.7          & 42.4          \\
$\dot{M}=10^{-8}$($M_{\odot}$/year) & 1.2             & 1.5           & 2.4           \\
\hline \hline
                                     & M stars                                         \\
                                     & $r_{edge}$ (AU) & $r_{il}$ (AU) & $r_{ht}$ (AU) \\ \hline
$\dot{M}=10^{-8}$($M_{\odot}$/year) & 4.2             & N/A           & N/A           \\
$\dot{M}=10^{-9}$($M_{\odot}$/year) & 2.3             & 1.5           & N/A           \\
\hline \hline
\end{tabular} 
\end{center}
\end{minipage}
\end{table*}

We present a picture of how planetary systems are established based on planet traps and discuss how host stars 
affect the size of planetary systems formed around them.

\subsubsection{General picture}

\begin{figure*}
\begin{minipage}{17cm}
\begin{center}
\includegraphics[height=5.5cm]{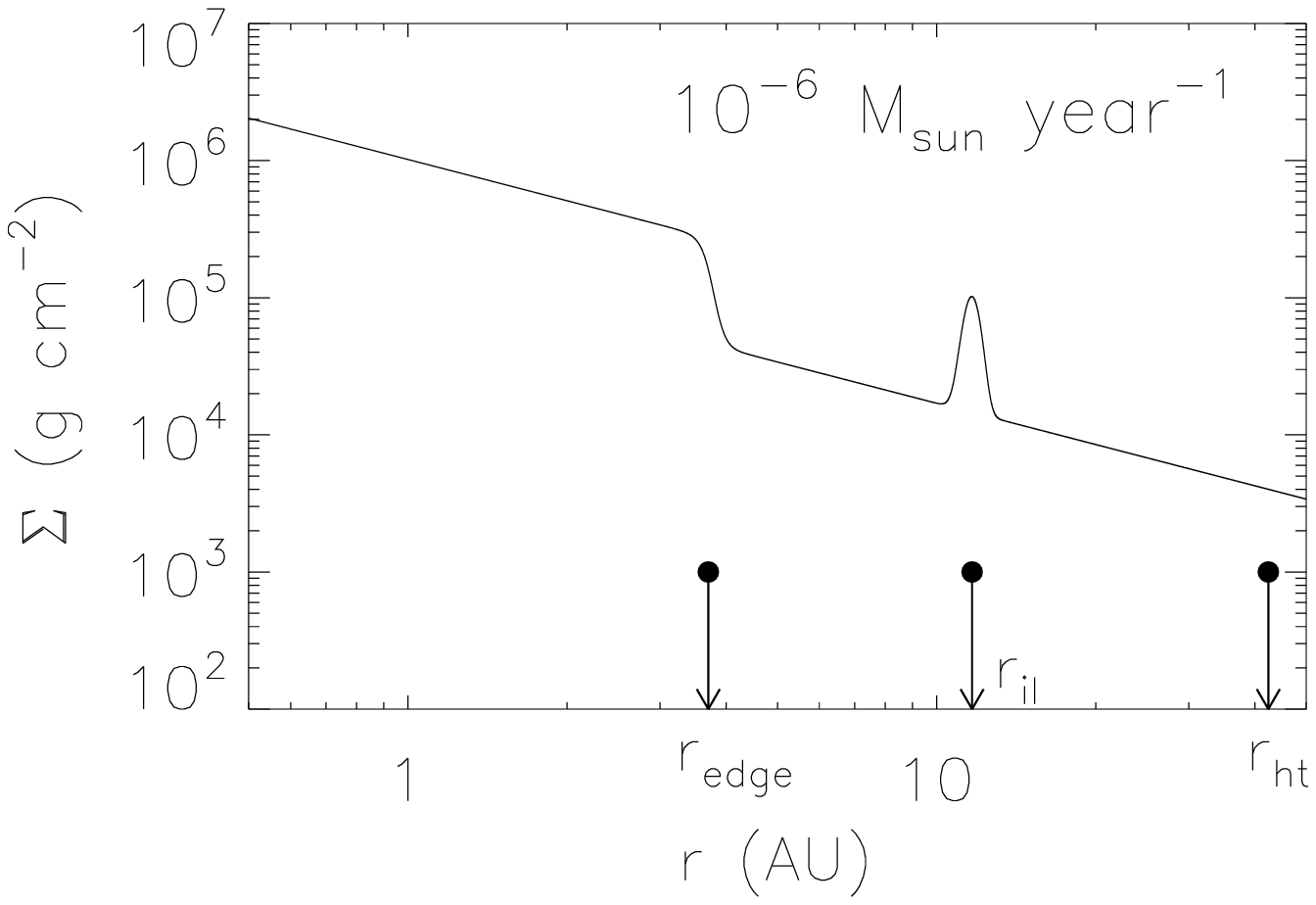}
\includegraphics[height=5.5cm]{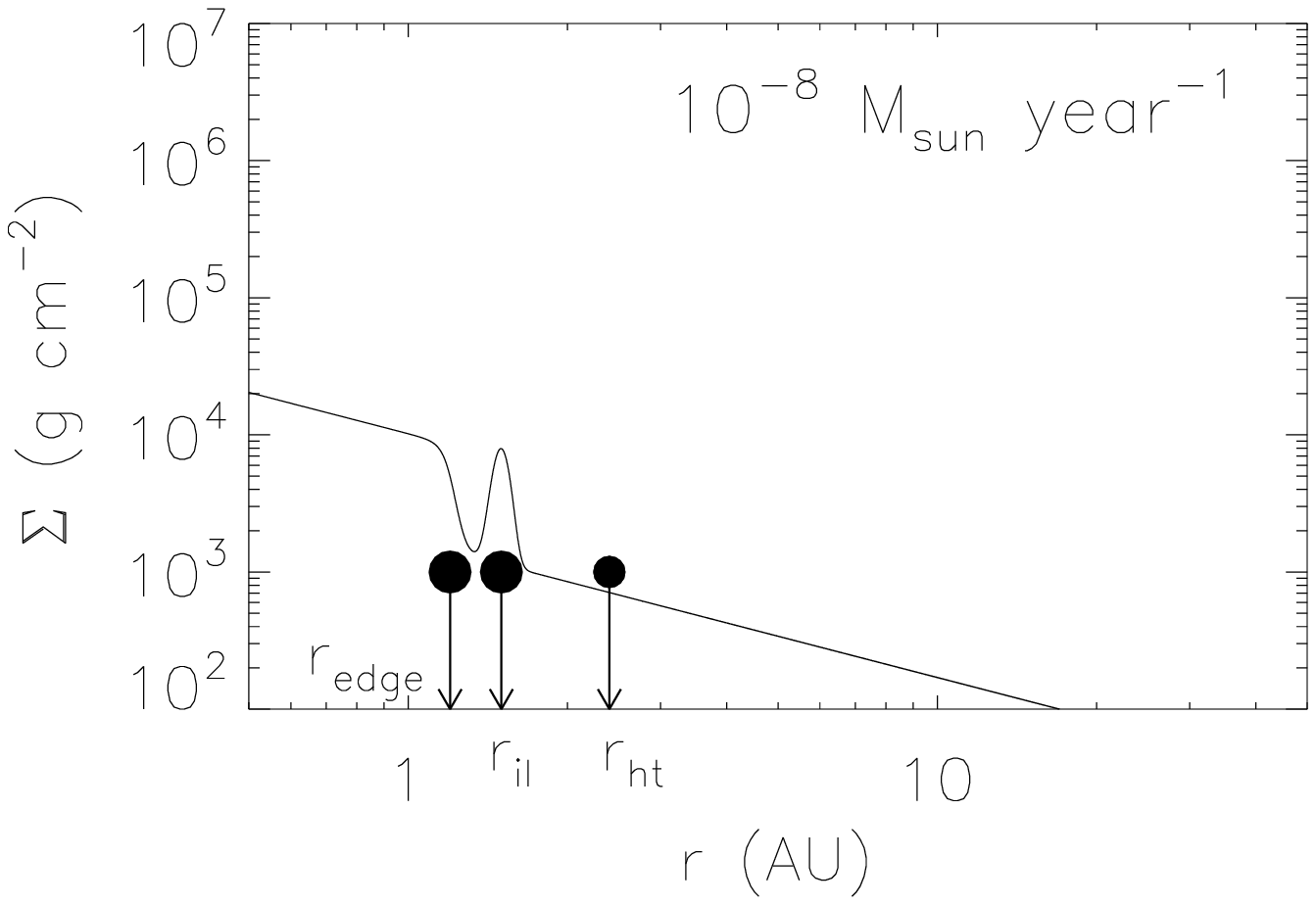}
\caption{Our general picture of the formation of planetary systems. The disc radius of each planet trap comes 
from Fig. \ref{fig6} for the case of CTTSs. The position of each planet trap is summarised in Table \ref{table9}. 
The structure of surface density is based on our disc model discussed in $\S$ \ref{layered_structures}. The dots denotes 
trapped protoplanets, and their size represents their masses. At the early stage of disc evolution (on the left panel), 
the accretion rate and the surface density are higher, and the separation between protoplanets which are captured at 
each trap is larger. Planetary masses are smaller. At the later stage of disc evolution (on the right panel), the 
accretion rate and the surface density become lower, and the separation becomes smaller. The planetary masses are 
now much larger. We assume that solid density at the outer edge of dead zones and ice lines is larger than that 
at the heat transition, based on the higher surface density of gas. It well illustrates how the orbital movement of 
planet traps at different rates affects the formation of planetary systems.}
\label{fig7}
\end{center}
\end{minipage}
\end{figure*}

Fig. \ref{fig7} schematically illustrates this. The disc radius of each planet trap comes from Fig. \ref{fig6} for 
the case of CTTSs (also see Table \ref{table9}). The structure of surface density is plotted based on our disc models 
discussed in $\S$ \ref{layered_structures}. Since we do not model the time dependency of $\dot{M}$ that gives the 
realistic value of surface density, we simply adopt the values of $\Sigma_0$ and $10^2 \times \Sigma_0$ in Table 
\ref{table3} at $\dot{M}=10^{-8} M_{\odot}/year$ and $\dot{M}=10^{-6} M_{\odot}/year$, respectively. These choices 
do not affect our picture. 

At the early stage of disc evolution ($\dot{M}=10^{-6} M_{\odot}/year$), the mass of 
protoplanets (denoted by the size of dots) which are captured in each trap is small, and they are widely separated. 
As the disc accretes onto the host star, the surface density of gas and the accretion rate decrease, and the position 
of the planet trap region with protoplanets captured moves inwards. Concurrently, the growth of these protoplanets 
proceeds. Since the rate at which each planet trap moves differs (see Table \ref{table9}), the separations between 
these trapped planets shrink. Also, the solid surface density in each planet trap probably varies in 
accordance with the variation of the surface density of gas. Therefore, the growth rate at the dead zones and ice 
lines may be higher than that at the heat transition. At some point, the gravitational interaction between trapped 
planets could exceed planetary migration and planet-planet scattering effects begin to dominate. We leave the later 
stage of evolution of planetary systems into our forthcoming paper. However, we can speculate how planetary systems 
evolve as follows. 

If the trapped (proto)planets distribute themselves closely enough, the most massive planet can surely survive the 
planet-planet interaction. On the other hand, the other two trapped planets could scatter inwards or outwards, or could 
be ejected from the system, depending on the mass of the most massive planet. Thus, our picture can in principle 
predict how planetary systems are formed, and what differentiates the formation of multiple planets versus single 
planet in a system. 

\subsubsection{Dependence on stellar masses and accretion rates}

We discuss the effects of stellar masses and accretion rates on the scale of planetary systems based on our picture. 
Table \ref{table9} summarises the position of each planet traps at two different values of $\dot{M}$ for stars with 
various masses in Fig. \ref{fig6}. One immediately observes that the scale of orbital radii of planets captured 
in planet traps is a function of stellar masses and disc accretion rates. For Herbig Ae/Be stars, planetary systems 
can extend from the order of au to $\sim 10^3$ au. For CTTSs, the size of planetary systems shrinks to $\sim$ 10 au. 
For M stars, planetary systems may be well confined within a few au. We discuss more on the dependence on the stellar 
mass by comparing with the observations below. 

\subsection{Implications for observations}

We discuss the validity of our picture by comparing with the observations. Fig. \ref{fig6} shows that the location of 
the dead zone traps for all cases is on order of 1 au (see the CTTS case). This is in accord with the observations 
which show statistically that the most gas giants are found around 1 au \citep[e.g.][]{wum09}. Therefore, our finding 
may imply that the dead zones play a significant role in the formation and location of these planets. 

For the case of Herbig Ae/Be stars (the top panel), there are three traps for high accretion rates. It is interesting 
that the range of disc radii for the heat transition and ice line traps covers the (large) range of orbital radii 
($\sim 10^{2}-10^{3}$ au) of massive planets around massive stars such as A stars which are observed by the direct 
imaging method \citep[e.g.][]{mmbz08}. For a lower accretion rate, the heat transition and ice line traps merge. 
However, the position of this merger is well separated from that of the dead zone trap. Thus, at least two planets may 
survive the planet-planet interaction. This multiplicity trend for massive stars is confirmed again by the direct 
imaging method (although the detected semi-major axes are considerably larger than our prediction.)

For the CTTSs case (the middle panel), three traps exist at first, and then as the disc accretion rate drops, the ice 
line and heat transition traps merge into the dead zone traps. As mentioned above, the interaction of three active 
traps arises around 1 au. Based on our picture, planets in multiple systems are less massive than single systems, since 
more massive planets provide more destructive effects on planetary systems. This prediction well explains the 
observation trend that (apparently) single planets statistically have larger masses than planets in multiple systems 
\citep{wum09}. More recently, the Kepler mission also confirmed the same trend \citep{l11}. Thus, the interaction of 
planets in their traps is essential for a systematic understanding of how planetary systems are formed. 

For the case of M stars (the bottom panel), only the dead zone trap is active at first. The location of this trap can 
explain the observed Super-Earth around M stars \citep{bbfw06}. As the accretion rate decreases, the ice line 
trap appears and then disappears due to the constraint derived from equation (\ref{condition_il}) 
(see Table \ref{table8}). In this case, it is more sensitive to when and what mass of planets are captured at each trap. 

\subsection{Parameter study}

\begin{figure*}
\begin{minipage}{17cm}
\begin{center}
\includegraphics[height=5.5cm]{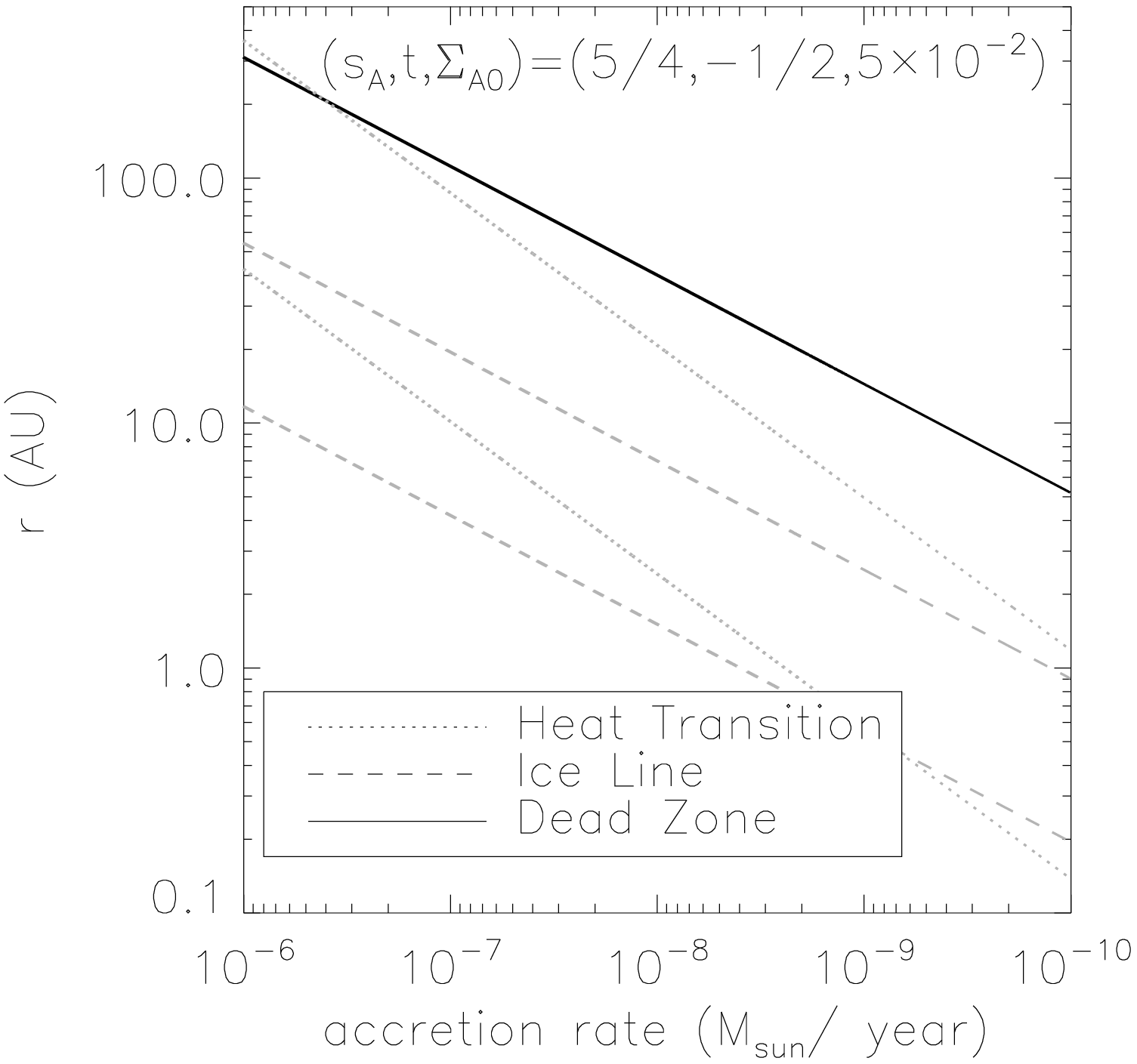}
\includegraphics[height=5.5cm]{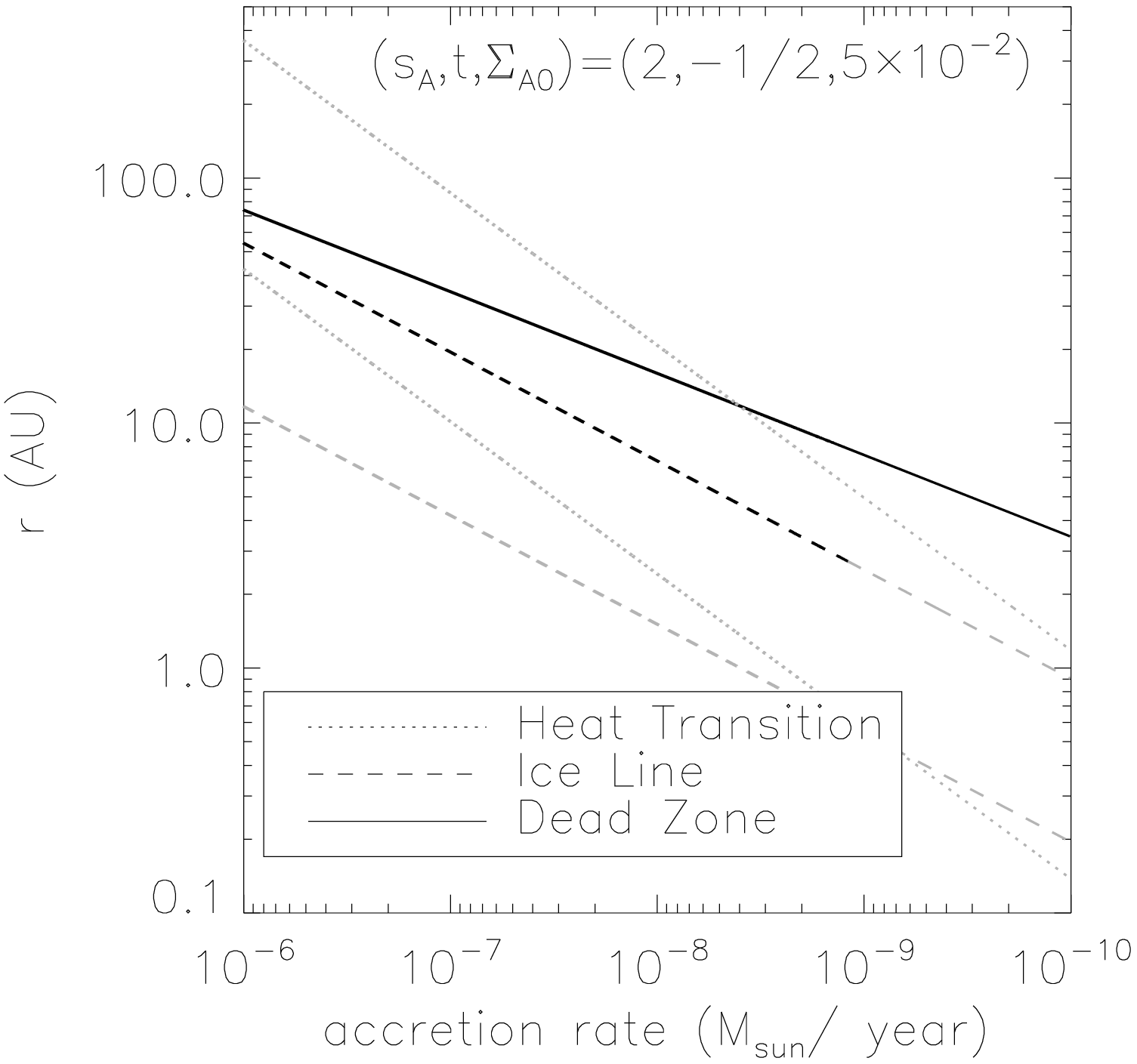}
\includegraphics[height=5.5cm]{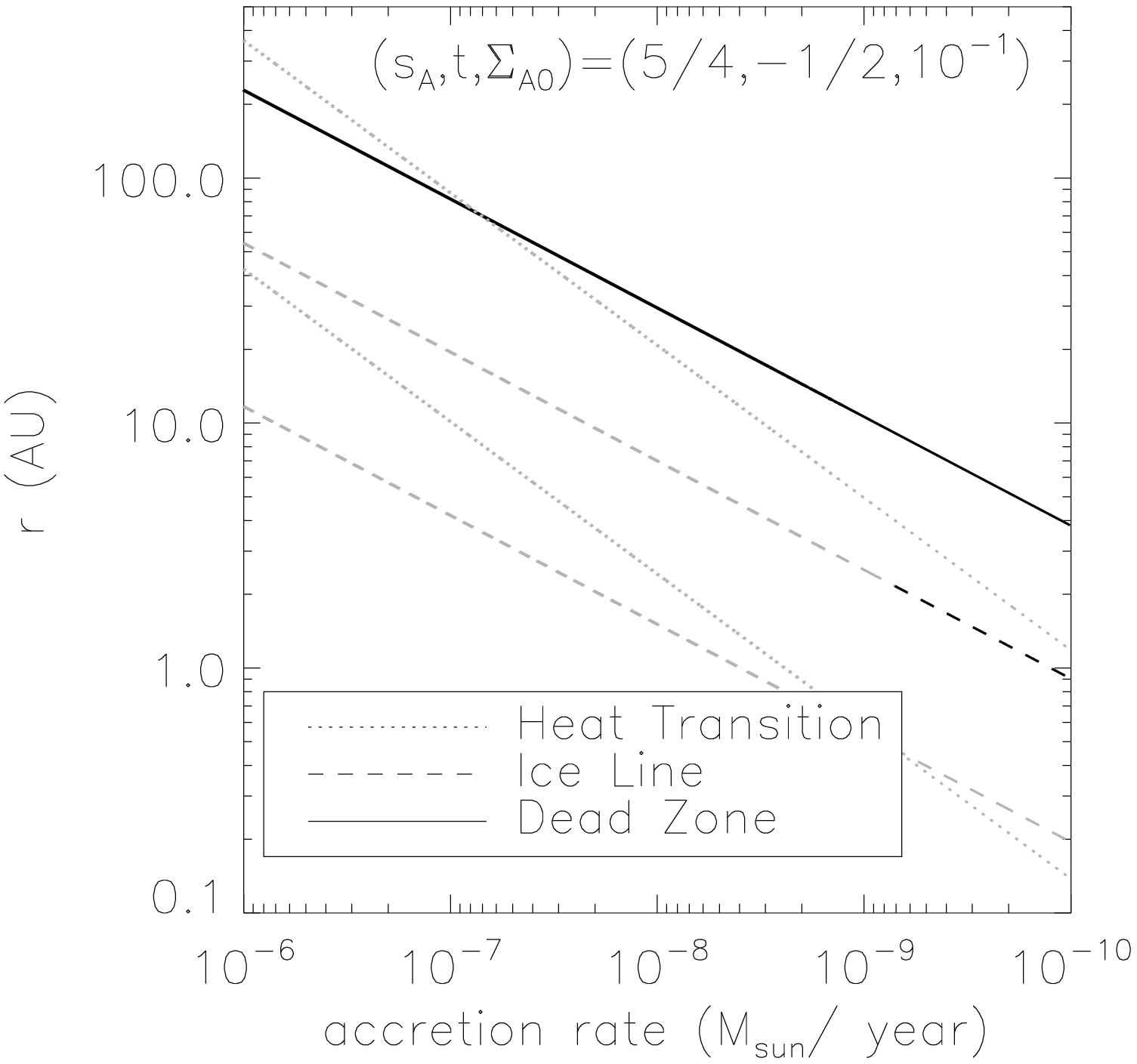}
\includegraphics[height=5.5cm]{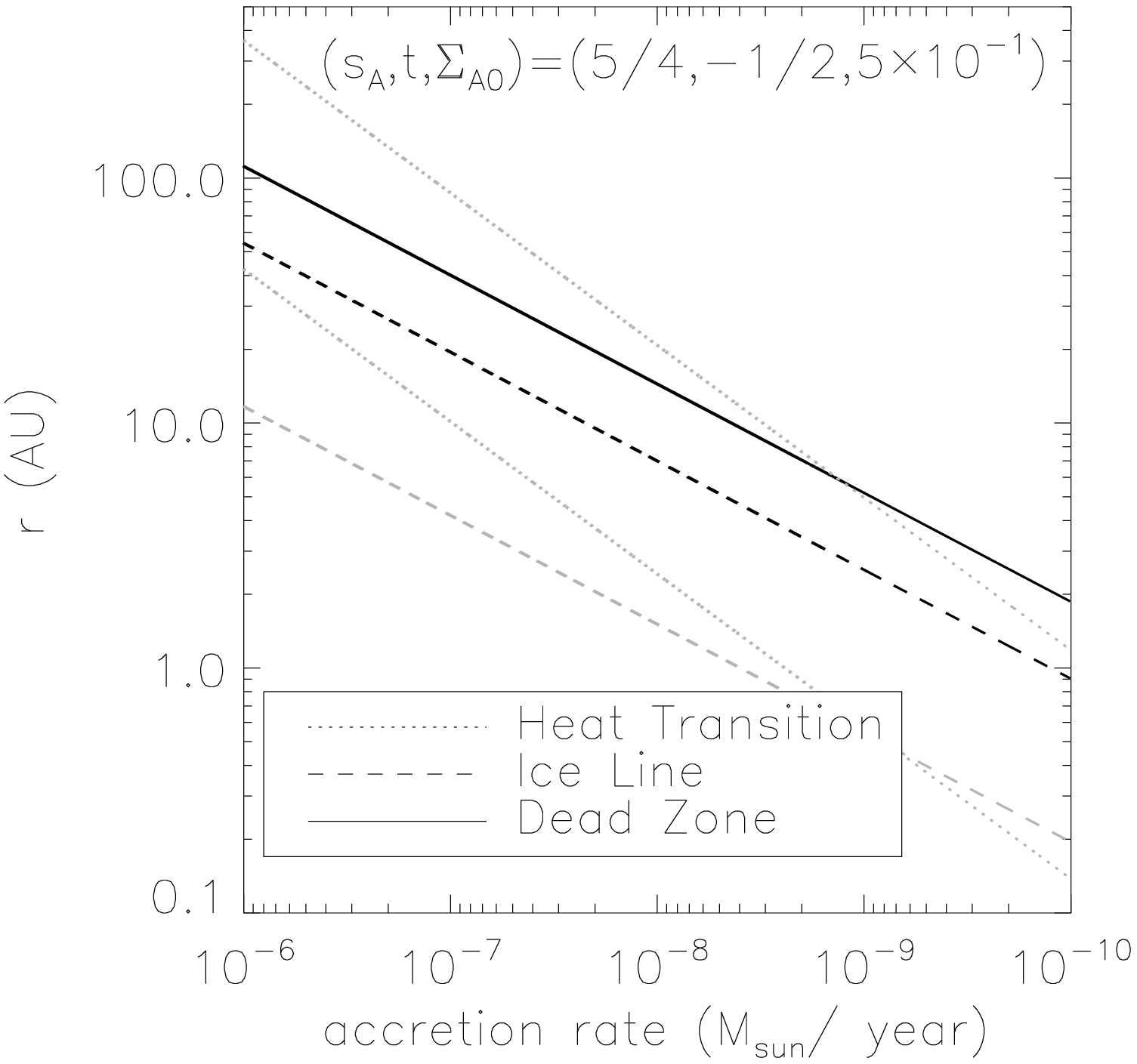}
\includegraphics[height=5.5cm]{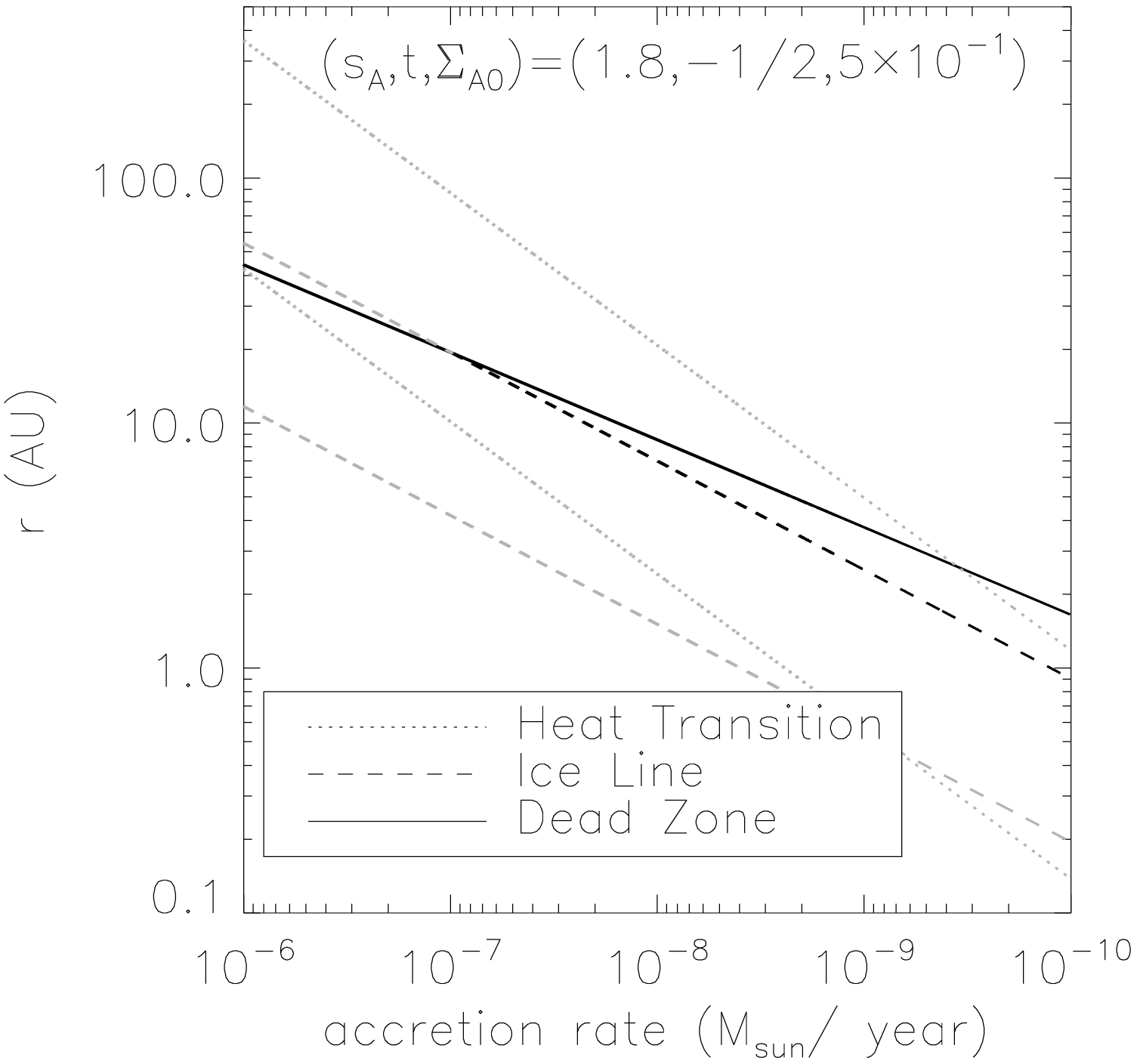}
\includegraphics[height=5.5cm]{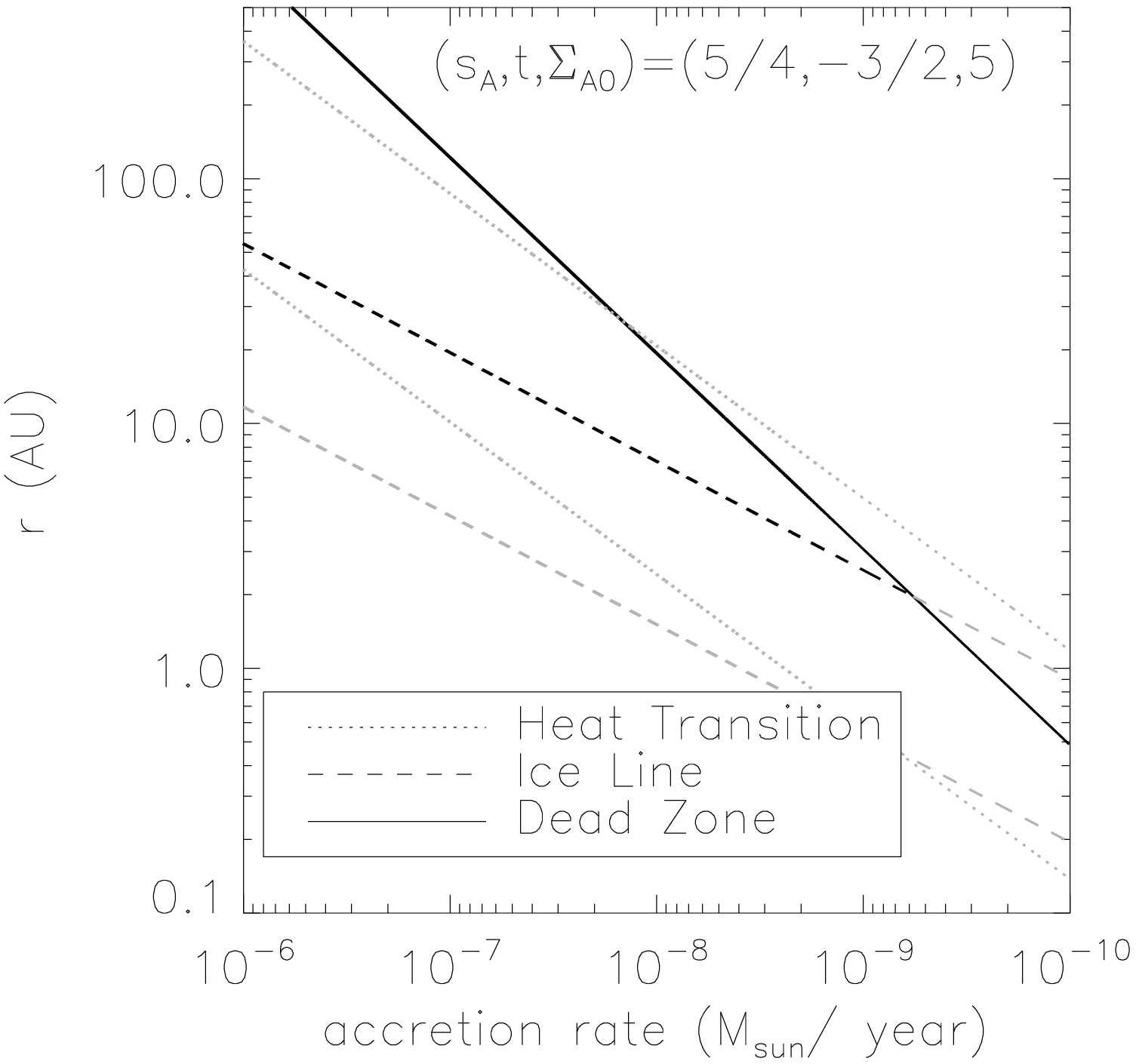}
\includegraphics[height=5.5cm]{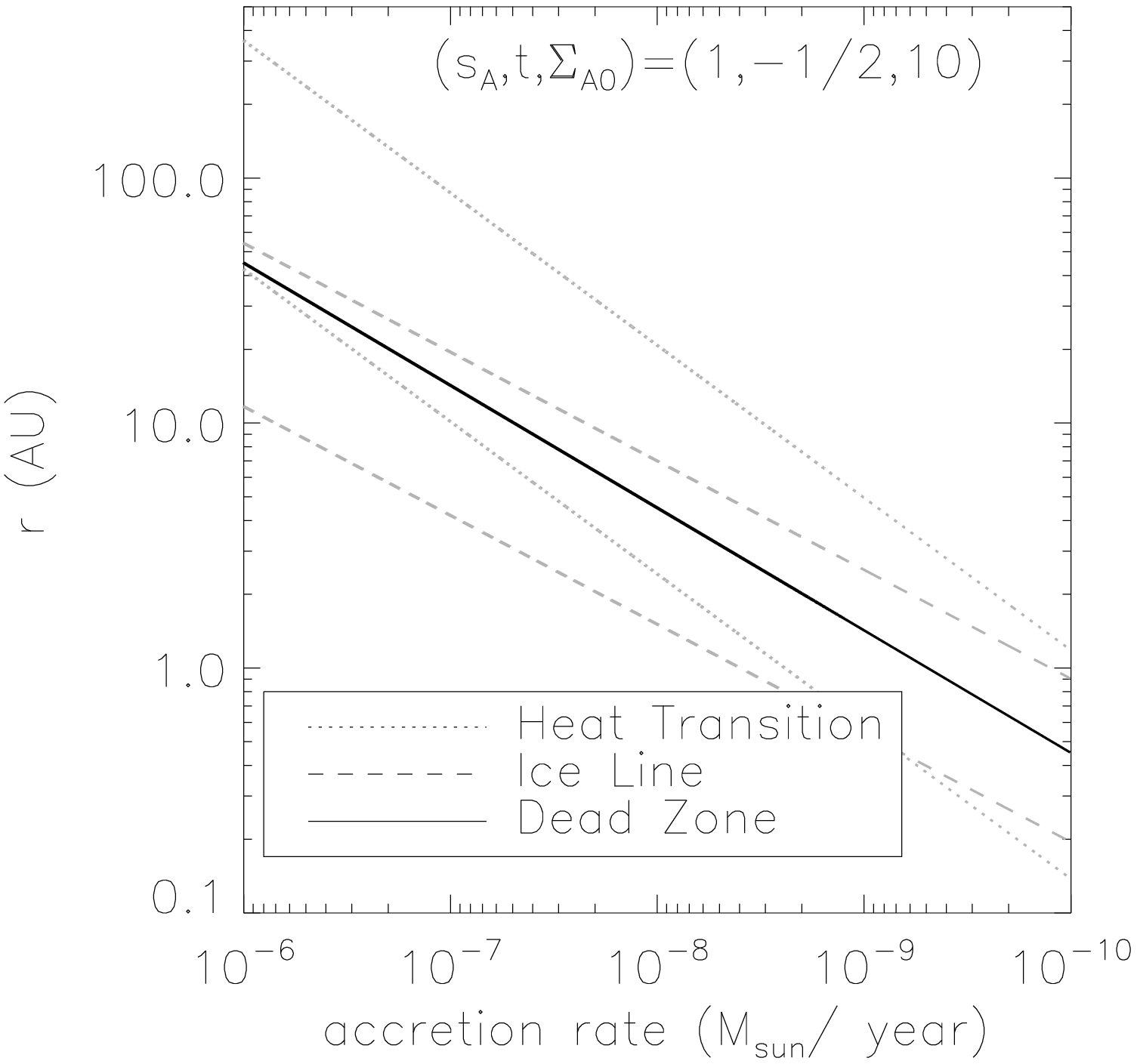}
\includegraphics[height=5.5cm]{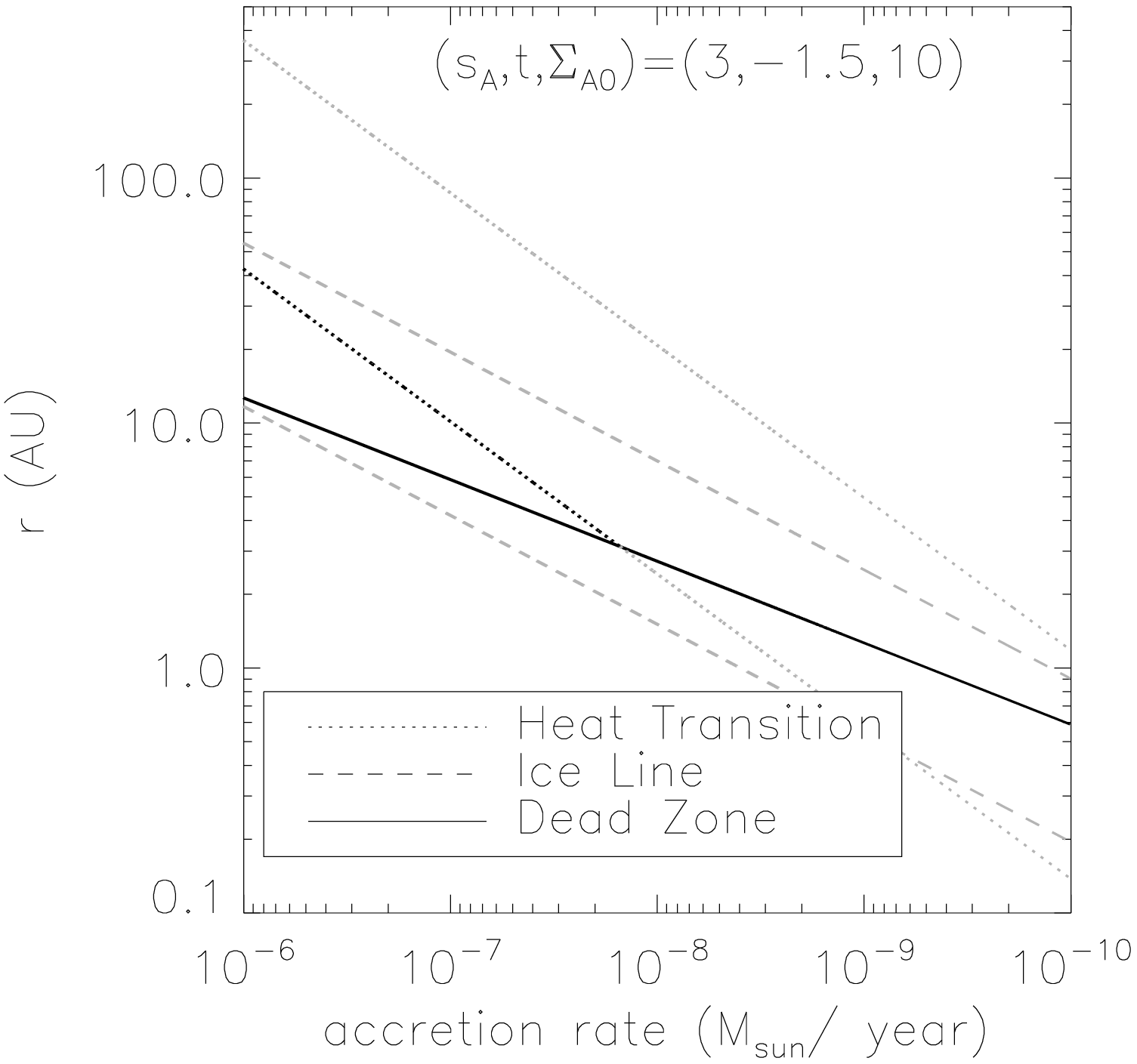}
\includegraphics[height=5.5cm]{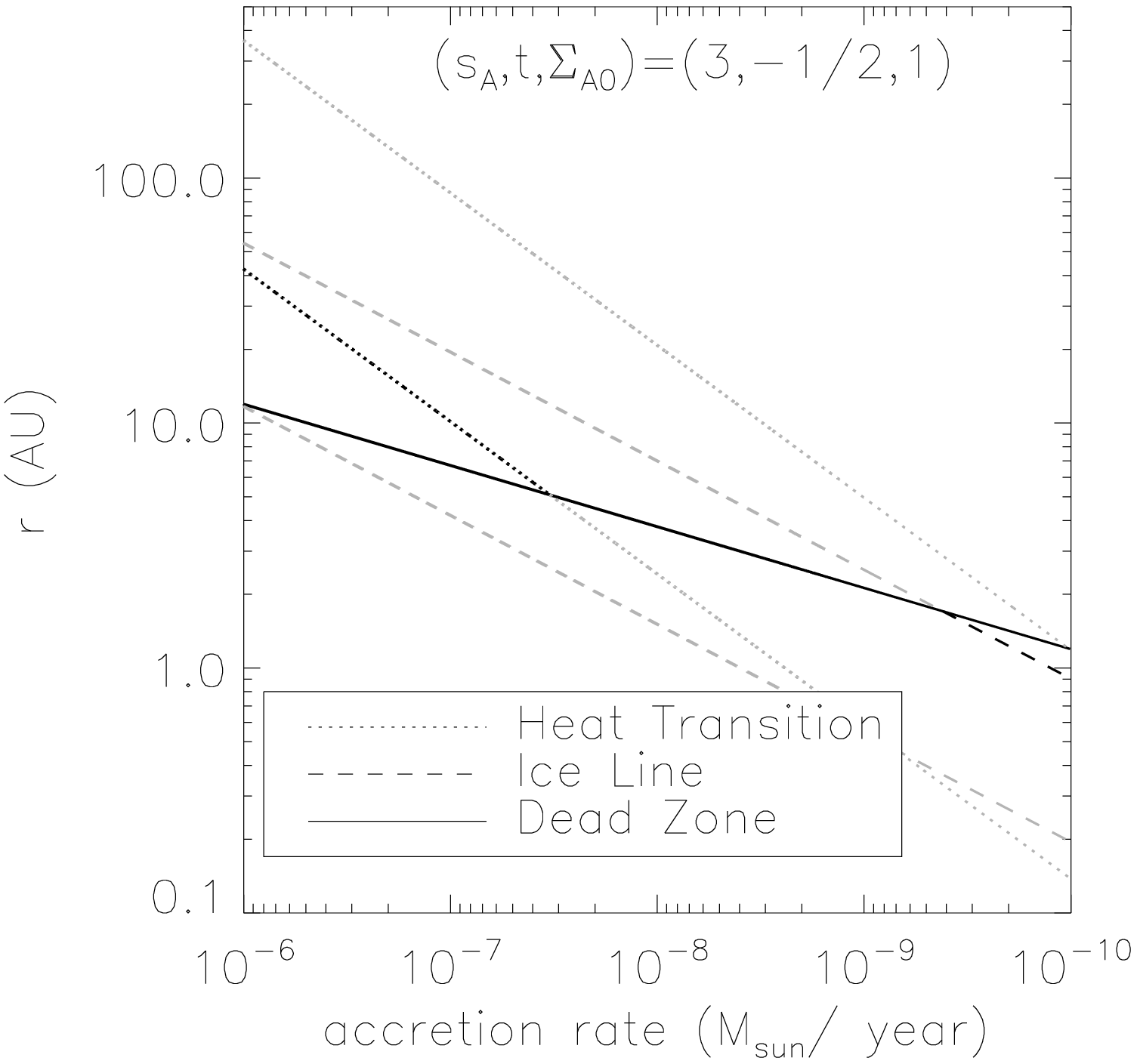}
\includegraphics[height=5.5cm]{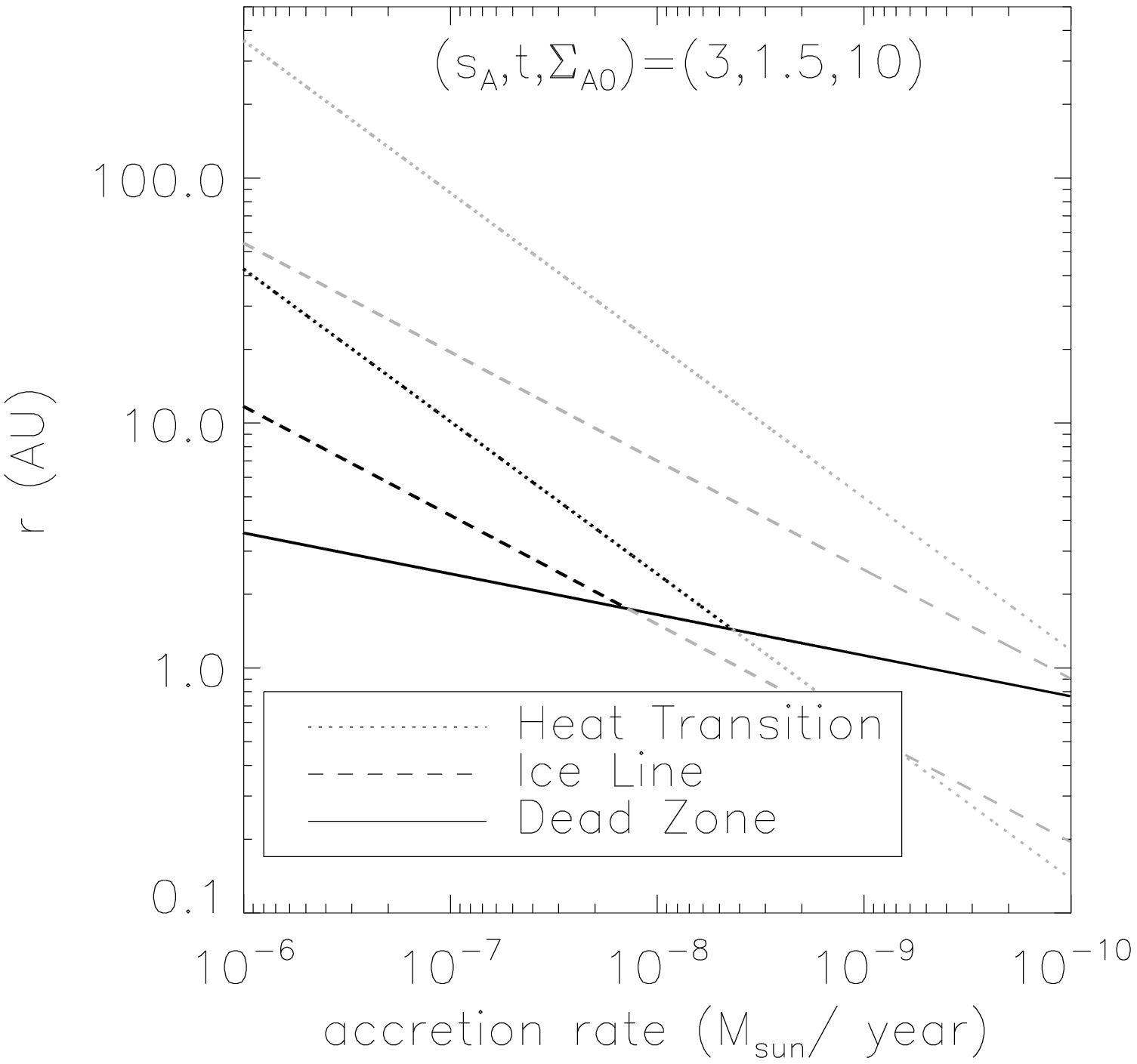}
\includegraphics[height=5.5cm]{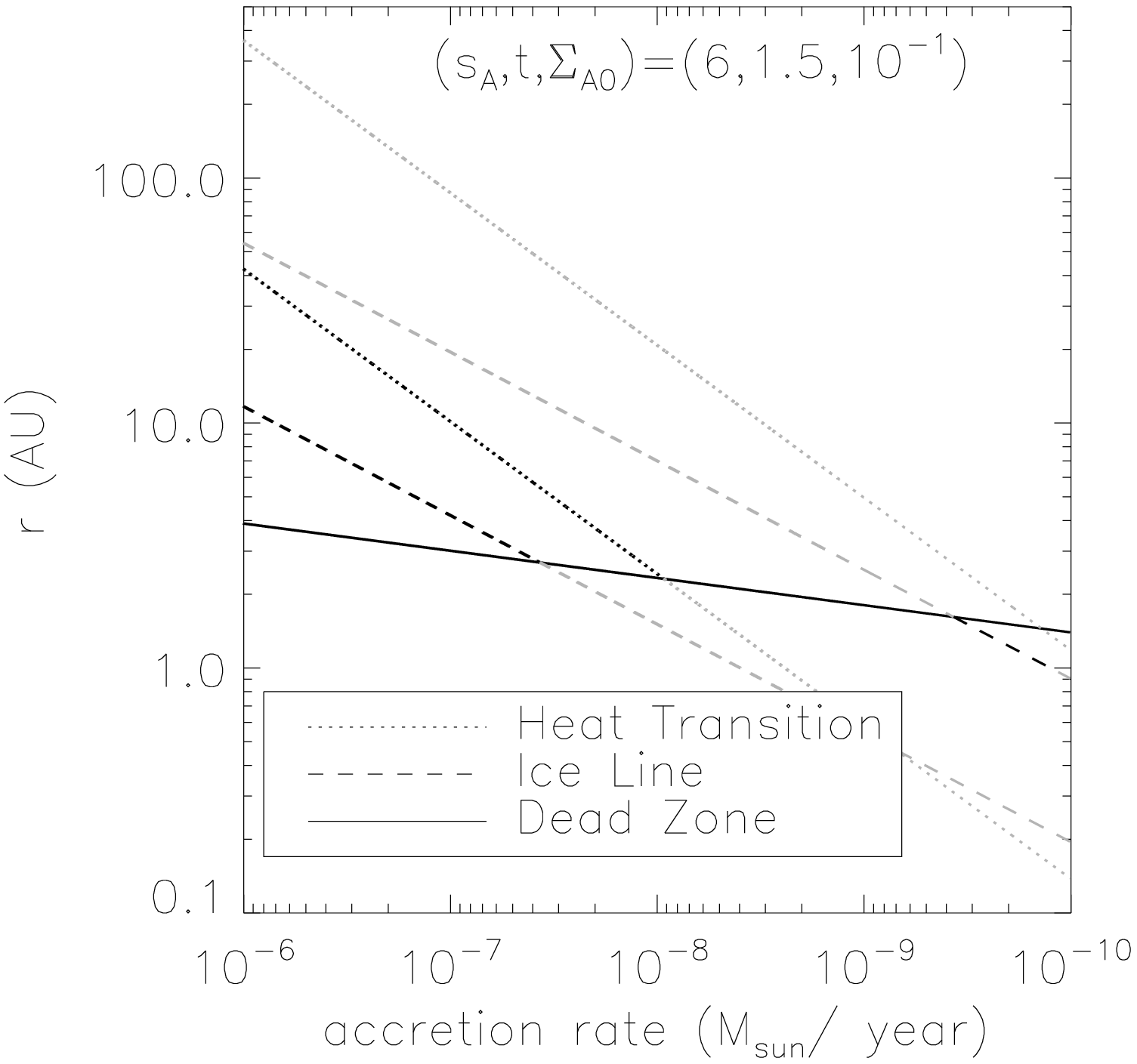}
\includegraphics[height=5.5cm]{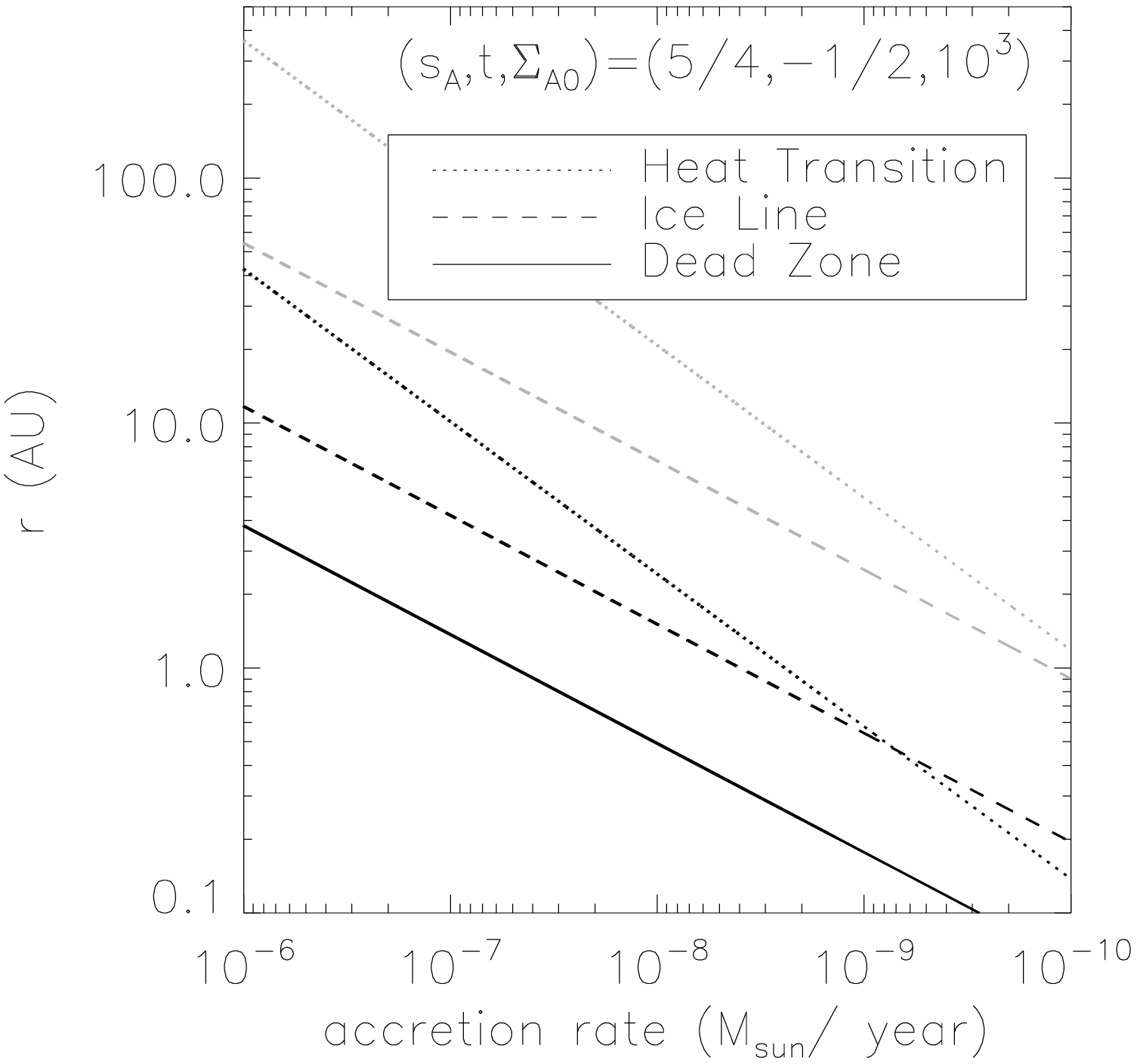}
\caption{Parameter study on the location and evolution of three planet traps around CTTSs. The definitions of all 
lines are the same as Fig. \ref{fig6}. All possible cases are shown. Our choices of parameters are shown in each panel. 
The presence of multiple planet traps in a single disc is more common. Therefore, the 
interaction of the planet traps is likely essential.}
\label{fig8}
\end{center}
\end{minipage}
\end{figure*}

We perform a parameter study on $s_A$, $t$, and $\Sigma_{A0}$ (see Fig. \ref{fig8}). We adopt the values of the CTTSs 
case. Fig. \ref{fig8} shows all the possible cases. The locations of $r_{il}$ and $r_{ht}$ are the same on all panels 
(since they are independent of these parameters, see equations (\ref{ril_vis2}) and (\ref{r_transit2})). The position 
of $r_{edge}$ is scaled by $\Sigma_{A0}$ while the dependence of $\dot{M}$ on $r_{edge}$ is controlled by the sum of 
$s_A$ and $t$ (see equation (\ref{eq_rp2})). One immediately observes that the presence of multiple traps in a single 
disc is more likely than only a single planet trap over a wide range of $\dot{M}$. Hence, the interaction of planets 
which are growing and lodged in their traps can be considered as essential. As discussed above, the growth rate of 
trapped protoplanets which controls when the planet-planet interaction initiates is crucial. We will address this issue 
in the forthcoming paper, by taking into account the growth of protoplanet as well as disc evolution.   

\section{Application to our Solar system} \label{nice}

In this section, we apply our picture to the origin of the Solar system - in particular to the Nice model as 
developed by \citet{tgm05,mlt05,glt05}. In this model, the dynamics of outer planets is investigated by simulating 
planetary migration in a debris disc and the subsequent planet-planet interaction. Thus, the model assumes that all 
gas in the disc is depleted and planetary migration is caused by interactions with planetesimals. However, these 
N-body based simulations depend on their initial conditions which are not justified. More recently, 
\citet{mtc10} investigated planetary migration in dissipating gaseous discs and showed that the semi-major axes are 
established by planetary migration in gaseous discs. Once gas is depleted, other orbital quantities such as 
eccentricities are determined by the planet-planet interaction. Therefore, it is crucial to examine the initial 
conditions of the Nice model in discs where gas is still present, in order to validate their results.

\begin{figure}
\begin{center}
\includegraphics[height=8cm]{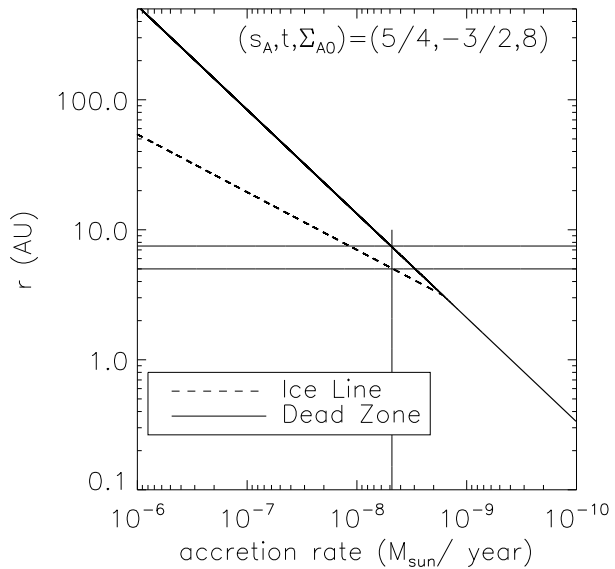}
\caption{Application to the Nice model. Two planet traps exist in a single disc (as Fig. \ref{fig6}). The horizontal 
solid lines denote the initial condition of the semi-major axes for Jupiter ($r=5.45$ au, see the lower line) and 
Saturn ($r\approx$ 8.2 au, see the upper line), respectively in the Nice model. The core of Jupiter is captured at 
the ice line trap while that of Saturn is at the dead zone trap. At $\dot{M}=5.5 \times 10^{-9} M_{\odot}/year$, 
the trapped cores reach the initial condition adopted in the Nice model.}
\label{fig9}
\end{center}
\end{figure}

Fig. \ref{fig9} shows the most plausible case for the initial condition of the Nice model in which Jupiter is 
placed at 5.45 au while Saturn at $\sim$ 8.2 au. We find it by varying the parameters ($s_A, t, \Sigma_{A0}$). 
In this figure, we show the two expected traps: the ice line and dead zone traps. These two planet traps may be the 
building sites of the two gas giants (Jupiter and Saturn). The formation of the icy giants (Uranus and Neptune) 
may originate from scattering processes \citep{tdl99}. This idea is also supported by the Nice model. Thus, the 
core of the assumption in the Nice model is the initial semi-major axes of two gas giants (Jupiter and Saturn; 
see the horizontal solid lines). Fig. \ref{fig9} indicates that these two traps deliver the cores of two gas giants 
to the initial semi-major axes for the Nice model towards the end of the gas disc's lifetime, when the gas accretion 
rate have fallen to $\dot{M}=5.5 \times 10^{-9} M_{\odot}/ year$ (see the vertical solid line). At this moment, 
the disc still have considerable amount of gas ($\sim$ 1 per cent of $\Sigma_0$), and hence the initial condition 
adopted in the Nice model may be too simplified.

Another interesting point is that Jupiter hovers around the ice line trap (see Fig. \ref{fig9}). Since the 
dust density there is significantly enhanced, the formation of the trapped core of Jupiter speeds up considerably. 
This may be a reason why the mass of Jupiter is larger than that of Saturn. Recently, the significant improvements 
of the Nice model were achieved by incorporating planetary migration driven by the gas discs \citep[references herein]{m10}. 
These studies showed that a key is the mass ratio of Jupiter to Saturn. If Jupiter is more massive than Saturn, the 
initial conditions adopted in the original Nice model are readily achieved. This is possible if there is outward 
migration of these two gas giants, which according to the model is consequence of capturing them into the 2/3 mean 
motion resonances (MMRs). If such a mass ratio of these two gas giants is not attained, both planets migrate 
inwards too rapidly to survive. Although these trends of migration happen without any planet trap, no one can 
currently explain such a mass ratio. Thus, our analyses are likely important for giving at least one explanation 
for deriving the initial conditions for the Nice model. We will address more comprehensive roles of planet traps 
in the Nice model in a future publication.  

As the separation of these two traps decreases with decreasing accretion rates, the two trapped cores can induce 
scattering of any other cores formed around them. Thus, our approach is very useful to investigate the formation of 
planetary system more self-consistently. In the subsequent paper, we will include the growth of planets in viscously 
evolving discs.

\section{Conclusions} \label{conc}

We have systematically investigated how inhomogeneities arise in protoplanetary discs and how they create planet traps 
where the direction of planetary migration changes sign: from inwards to outwards. We analytically investigate the 
various mechanisms of planet traps activated by both corotation and Lindblad torques. For the traps related to the 
corotation torque, we have simply examined the disc structures that lead to outward migration. For the traps related to 
the Lindblad torque, we have derived the relations which predict when the torque reversal occurs. 

Our analytical relations derived from the Lindblad torque succeed in reproducing the results of the previous 
numerical studies done for the dead zones: the density and thermal barriers ($\S$ \ref{dead_zone}). This confirms 
that our assumptions and treatments are valid for capturing the detailed, complex physics occurring in the simulations. 
Furthermore, we have shown that the ice lines can serve as barriers due to the Lindblad rather than corotation torque. 
In discs with MRI induced turbulence, the ice lines create the layered structure, so that the strength of turbulence 
($\alpha$) reduces to the value of the dead zones. Thus, the ice lines can be considered to be self-regulated, 
localised dead zones. We have also discussed the possibility that a single disc can have several ice line barriers 
which become effective due to different species such as water-ice and CO-ice. A more detailed understanding of the 
structure of ice lines is required for addressing these features. Finally, we have estimated the mass range over 
which our analyses can apply. In either laminar or turbulent discs, our modeling can be useful for understanding 
almost observed exoplanets.  

We summarise our major findings below.

\begin{enumerate}

\item We have shown that the heat transition of protoplanetary discs from viscosity to stellar irradiation dominated 
heating, activates a new barrier, which we call the heat transition barrier (see $\S$ \ref{steirra}). This barrier 
arises due to the entropy-related horseshoe drag and is caused by the temperature transition there. At disc radii 
inside the heat transition radius, $r_{ht}$, the temperature slope determined by viscous heating is very steep, 
resulting in outward migration. On the other hand, a shallower temperature profile controlled by stellar irradiation 
dominates at larger radii beyond $r_{ht}$. Such a region drives inward migration. Thus, planets which migrate toward 
the location of the heat transition are captured there.

\item We have also shown that a single disc can have up to three planet traps: the dead zone, ice line, and heat 
transition traps (see $\S$ \ref{synthesis}). This multiplicity of planet traps is likely quite common. We have derived 
fully analytical relations for their disc radii and their effective conditions. These relations imply that dead zone 
traps may be the most important in order to explain why so many gas giants are observed around 1 au. 
Also, the presence of the heat transition and ice line traps may be the origin of massive planets with (very) large 
orbital radii around massive stars which are detected by the direct imaging method.

\item We have demonstrated that these planet traps can interact with each other, as they follow the time-dependent, 
viscous evolution of the disc (see Fig. \ref{fig6}). More specifically, we have found that their characteristic radii 
evolve with different dependence on $\dot{M}$ as the disc accretion rate $\dot{M}$ drops. This result shows that 
different planet traps move inwards differently, as discs accrete onto the central stars and star formation is 
terminated. (Equivalently, this evolution results in decreasing the surface density of gas ($\Sigma_0$) and accretion 
rates ($\dot{M}$).) Our analyses are important for comprehensively understanding the formation of planetary 
systems around stars with various masses (see Fig. \ref{fig7}). Since trapped planets can grow quickly, the interaction 
of active traps can initiate planet-planet interaction, which is controlled by the most massive, trapped planet. Thus, 
it is important to investigate how trapped planets gain their masses in viscously evolving discs. We will address this 
issue in a forthcoming paper.

\item We have applied our results to planet traps in discs around massive (Herbig Ae/Be), intermediate (CTTS), and 
low (M) mass stars. In all cases, the heat transition trap is the outermost trap in the disc while the dead zone or the  
ice line trap is the innermost. At low accretion rates ($\dot{M} \sim 10^{-8}M_{\odot}/year$) in CTTS system as an 
example, the traps are located at $r_{edge}=1.2$ au, $r_{il}=1.5$ au, and $r_{ht}=2.4$ au, respectively 
(see Table \ref{table9}). 

\item We have shown that the position of planet traps strongly depends on stellar masses and disc accretion rates. 
This indicates that host stars dictate a preferred scale of planetary systems formed around them 
(see Table \ref{table9}). For Herbig Ae/Be stars, the size of planetary systems may expand from the order of a few au 
to $\sim 10^3$ au. For CTTSs, it shrinks to a few ten au. For M stars, any planetary systems formed around them may be 
well confined within the order of au. 

\item We have applied our analyses to the Nice model for Solar system evolution. By choosing the appropriate 
parameters ($s_A,t,\Sigma_{A0}$), we have shown that the ice line trap can capture the core of Jupiter while the 
dead zone trap Saturn. The locations of these trapped cores match those of the Nice model at 
$\dot{M}=5.5 \times 10^{-9} M_{\odot}/year$ which is achieved towards the end of disc evolution. Furthermore, our 
analyses suggest why Jupiter is more massive than Saturn, which is one of the most crucial assumptions adopted in 
the improved Nice model. 

\end{enumerate}

In a subsequent paper, we will investigate the growth of planets which are captured at the planet traps during 
the time-dependent, viscous evolution of the disc. Inclusion of planetary growth will allow us to address when the 
planet-planet interaction dominates planetary migration. 
  
\section*{Acknowledgments}

The authors thank Shigeru Ida and Charles Lineweaver for stimulating discussions, especially about ice lines, and
an anonymous referee for useful comments on our manuscript. YH is 
supported by McMaster University, as well as by Graduate Fellowships from SHARCNET and the Canadian Astrobiology 
Training Program (CATP). REP is supported by a Discovery Grant from the Natural Sciences and Engineering Research 
Council (NSERC) of Canada.

\bibliographystyle{mn2e}

\bibliography{mn-jour,adsbibliography}

\appendix

\section{Extension of Paardekooper's torque formula} \label{app1}

Here, we discuss the extension of Paardekooper's torque formula that is valid exclusively in pure power-law discs 
(see equation (\ref{corotation_barrier})). Since entropy gradients that scale an entropy-related corotation torque can 
be straightforwardly obtained from the disc profiles, we focus on the extension of Lindbald and vortensity-related 
corotation torques. For disc models, we adopt the same one in $\S$ \ref{cc}, that is, general profiles for the 
surface density (see equation (\ref{sigma_density_jump}) ) and power-law profiles for the disc temperature 
($T\propto r^{t}$). We only examine the case of $s=-1$ for $\Sigma_{int}$ (see equation (\ref{sigma_density_jump})), 
because the results for the MMSN case are qualitatively similar. We also discuss the effects of the disc temperature 
that has more general profiles. 

For the Lindblad torque, we indeed derive an analytical relation that controls the direction of migration for 
the above disc model in $\S$ \ref{dead_zone} (see equation (\ref{density_barrier})). Exploiting the analysis, we estimate 
the critical value of $F_{g,crit}$ above which the Lindblad torque dictates outward migration:
\begin{equation}
 F_{g,crit} \equiv \frac{\tanh(1/c)+h_p(s-t/2+7/4)}{\tanh(1/c)-h_p(s-t/2+7/4)},
\end{equation}
where $h_p$ is the disc aspect ratio ($H/r$) at the position of a planet (see $\S$ \ref{dz_dj} for a full 
derivation). Adopting $h_p=0.1$, we find that $F_{g,crit}=1.5$ for $c=1$ while $F_{g,crit}=2.7$ for $c=3$. 
As discussed in 
$\S$ \ref{dz_dj}, the direction of migration is determined only by the Lindblad torque if the 
density modification is larger than $F_{g,crit}$. This is because the Lindblad resonances distribute much further 
way from a planet than the horseshoe region. Based on the results of MG04 (see their fig. 1), the density distortion 
that is produced by the opacity and heat transitions is likely less than $F_{g,crit}$. In the following discussion, 
we adopt these critical values of $F_{g,crit}$. Thus, Lindblad torques surely become weaker around the opacity and heat 
transitions, but never give the dominant contribution to the direction of migration. Therefore, we adopt the original 
term derived in \citet{pbck09} for consistency in the usage of the torque formulation. 
 
\begin{figure}
\begin{center}
\includegraphics[width=8.5cm]{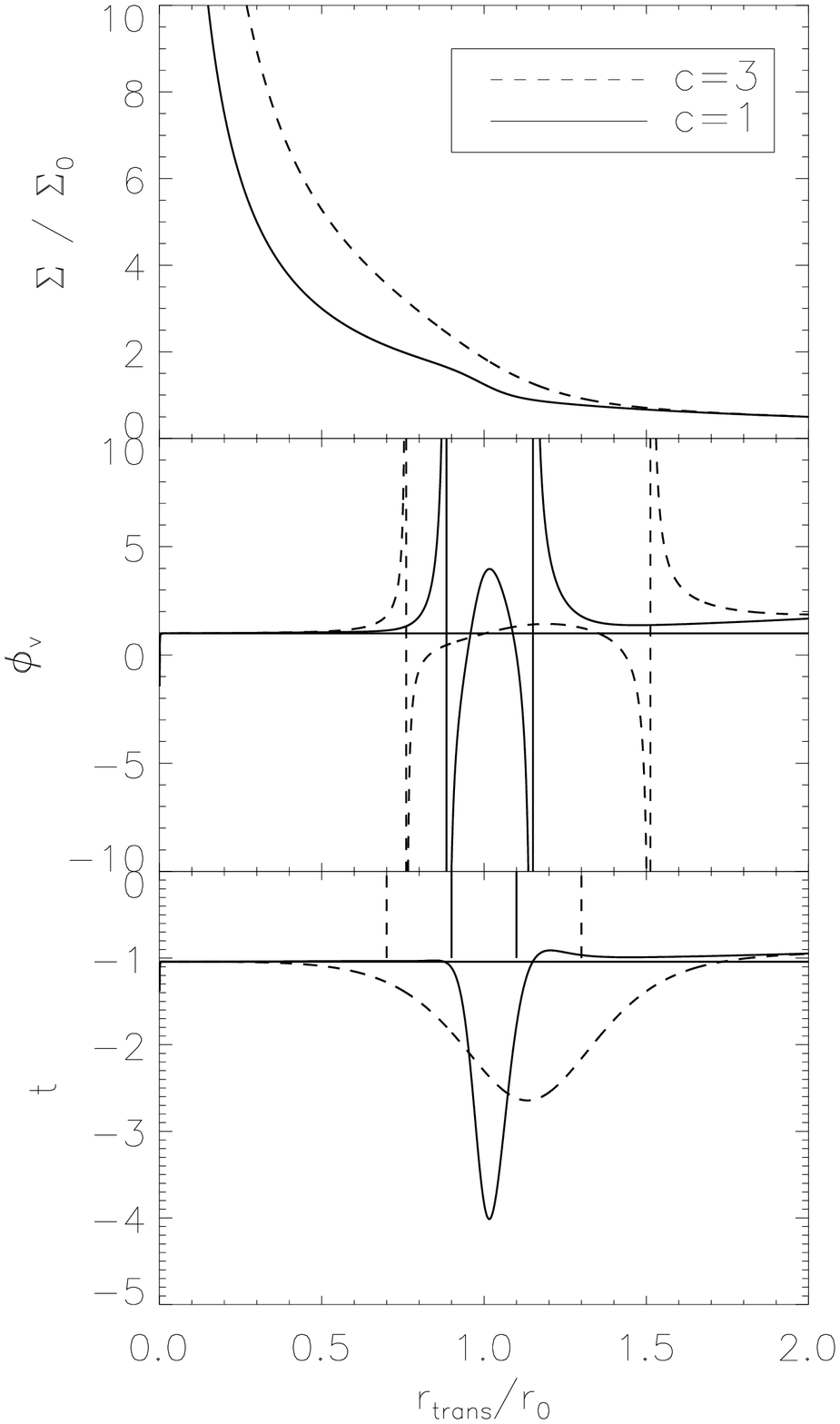}
\caption{The structure of the vortensity correction factor $\phi_v$ for discs with general profiles. On the top panel, 
the surface density structures are plotted (see equation (\ref{sigma_density_jump})). On the middle, the resultant 
behaviours of $\phi_v$ are shown. On the bottom, the critical values of $t$ affected by $\phi_v$ are plotted. We 
stress that the choice of $c=1$ is more appropriate for opacity and heat transitions. For the purpose of a parameter 
study, we also consider $c=3$. The results of pure power-law discs are denoted by the horizontal lines on the middle 
and bottom panels. The crucial effects of $\phi_v$ arise from Gaussian-like functions that are well confined with the 
transition width $(\omega=cH)$ (also see Fig. \ref{figA2}). Inclusion of $\phi_v$ significantly lowers the required 
value of $t$ around $r=r_{trans}$. This indicates that the vortensity-related corotation torque becomes considerably 
larger there and makes planets migrate inwards. Thus, proper treatments of the vortensity-related corotation torque 
deviates the results of the pure power-law discs, but the deviations occur only in the local region centered at 
$r=r_{trans}$.}
\label{figA1}
\end{center}
\end{figure}

For the vortensity-related corotation torque, we add a correction factor $\phi_v$ in the original formulation 
(see equations (\ref{required_t}) and (\ref{phi_v})). This is because gradients of vortensity that are the core of 
the vortensity-related corotation torque, are very sensitive to the disc structure. Fig. \ref{figA1} shows the surface 
density structure, the resultant behaviour of $\phi_v$, and the critical value of $t$ below which planets migrate 
outwards (see equation (\ref{required_t})) on the top, middle, and bottom panel, respectively. Since the vortensity 
gradients depend on the disc temperature through $\Omega$ (see equations (\ref{Boort}) and (\ref{omega})), 
self-consistent treatments are required to derive $t$. For simplicity, however, we fix $t=-1.2$ for determining 
$\phi_v$ in this figure. This profile is established for viscously heated, optically thick discs (see $\S$ \ref{vh}). 
We also discuss the effects of the disc temperature with more general profiles below.  
   
\begin{figure}
\begin{center}
\includegraphics[width=8.5cm]{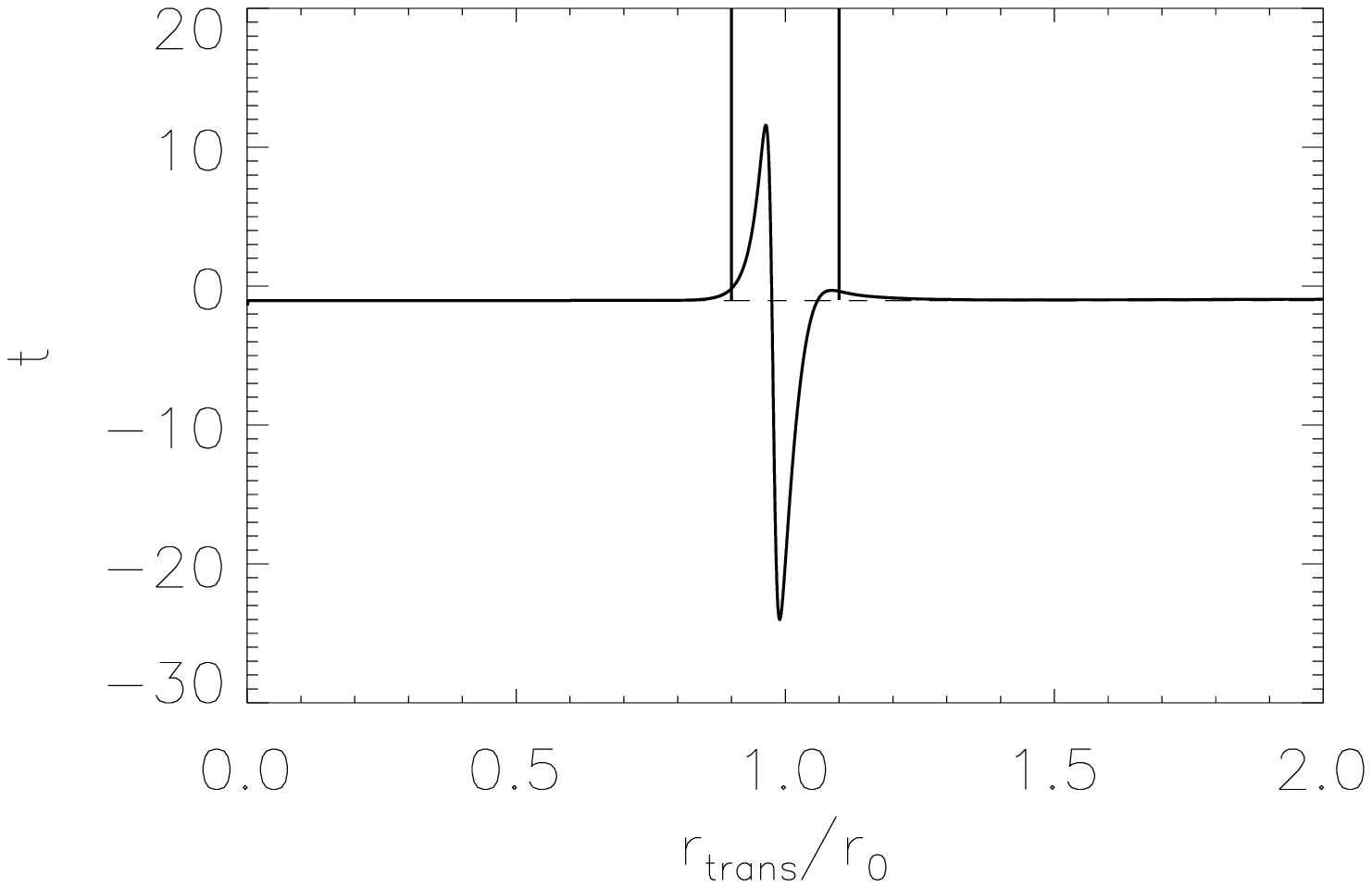}
\caption{The critical value of $t$ for the disc temperature with general profiles. We adopt an expression similar to 
equation (\ref{sigma_density_jump}) for the disc temperature. Also, we consider the case of $c=1$. By comparing Fig. 
\ref{figA1}, the extent wherein $\phi_v$ plays a crucial role is strongly diminished. This supports the usage of the 
pure power-law approximation.}
\label{figA2}
\end{center}
\end{figure}

Fig. \ref{figA1} shows very complicated behaviour for $\phi_v$. It is obvious that the choice of $c=1$ (the solid line) 
is more appropriate for the opacity and heat transitions based on the results of MG04. However, we also consider 
$c=3$ (the dashed line) for a parameter study. By comparing the middle and bottom panels, one can observe immediately 
that the crucial effects of $\phi_v$ on $t$ arise from Gaussian-like functions that distribute around $r_{trans}$. 
It is important that the extent of the functions is well characterised by the transition width ($\omega=cH$) (see the 
vertical lines on the bottom panel). For the case of $c=3$, the function expands beyond the transition width. However, 
this is a result of our choice of the temperature profile. Adopting more general profiles similar to equation 
(\ref{sigma_density_jump}) for the disc temperature, the extent of this function strongly diminishes 
(see Fig. \ref{figA2}). Thus, it is reasonable to conclude that the effects of $\phi_v$ are very local and well 
confined with the transition width. Another important point is that the required value of $t$ is strongly reduced by 
$\phi_v$. Such small values of $t$ are never accomplished in protoplanetary discs. This means that the 
vortensity-related corotation torque becomes significantly large and results in inward migration. Therefore, inclusion 
of $\phi_v$ reduces the possible outward migration region which is a consequence of viscous heating and 
entropy-related corotation torques.  

In summary, proper treatments of the vortensity-related corotation torque are important for discs with general profiles. 
However, the deviation from the power-law treatments is well confined in the vicinity of the position of disc 
inhomogeneities, and therefore power-law approximation provides reliable results in our level of extension.  

\section{Taylor expansion of the tidal torque} \label{app2}

Here, we derive the Taylor expansion of quantities related with the tidal torque in terms of $1/m \ll 1$ 
since the torque takes the maximum value at the wavenumber $m\approx 10$ \citep{ward97}. 
\begin{eqnarray}
\frac{\Sigma}{H} \cdot \alpha_r^2 & =       & \frac{\Sigma}{H}(r_p)\alpha_r^{s-t/2+1/2} \nonumber \\
                                  & \approx & \frac{\Sigma}{H}(r_p) 
         \left[ 1+\frac{2\epsilon }{3m}\sqrt{1+\xi^2} \left( s - \frac{t}{2} + \frac{1}{2}\right)  \right],
\end{eqnarray}  
and 
\begin{eqnarray}
\psi & \approx & \left [ K_1(\Lambda _0)+2\sqrt{1+\xi^2}K_0(\Lambda _0) \right] + \nonumber \\
     &  &        \frac{\epsilon }{m} \left[ \left(\frac{5}{6} + \frac{2\xi^2}{3} \right)
                    \sqrt{1+\xi^2}K_1(\Lambda _0) 
                  + \left(\frac{1}{6} - \frac{\xi^2}{3}\right)K_0(\Lambda _0) \right] \nonumber \\
     & \equiv  & \psi_0 + \frac{\epsilon }{m}\Delta \psi
\end{eqnarray}
with
\begin{equation}
\Lambda \approx \frac{2}{3}\sqrt{1+\xi^2} - \frac{\epsilon }{m}\frac{1+\xi^2}{3} 
                        \equiv \Lambda _0 +\frac{\epsilon }{m}\Delta \Lambda, 
\end{equation}
and
\begin{eqnarray}
K_n(z)             & = & K_{n-2}(z) + \frac{2(n-1)}{z}K_{n-1}(z), \\
\frac{dK_n(z)}{dz} & = & -\frac{1}{2} \left( K_{n-1}(z) + K_{n+1}(z) \right).
\end{eqnarray}
Thus, 
\begin{equation}
  \psi^2 \approx \psi_0^2[1+2(\epsilon /m)(\Delta \psi/\psi_0 )].
\end{equation}
Taking the limit $\xi\rightarrow 0$ (since the effect of gas pressure becomes negligible in the Keplerian 
discs),
\begin{equation}
2\frac{\Delta \psi}{\psi_0} \approx \frac{1}{3}\frac{5K_1(2/3)+K_0(2/3)}{K_1(2/3)+2K_0(2/3)}\sim \frac{5}{6}.
\end{equation}

\section{Ice lines of other molecules} \label{app3}

Here, we investigate the minimum abundance of a molecule that is required for its ice line to be a barrier, by 
using the analyses done in $\S$ \ref{layered_structures}. Assuming the dust density to increase 
by the amount of ice which is formed at the ice line radius of a specific molecule, $r$, the required dust density 
enhancement is (using equations (\ref{iceline_barrier1}) and (\ref{eq_rp})), for $r/r_{edge}<$1,
\begin{equation}
 f_{d1} > \frac{K_1-K_2}{K_3} =f(\alpha_A, \alpha_D, s_A, t, h, r, f_{d2}, c), 
\label{eq_in}
\end{equation}
where
\begin{equation}
 K_1 = h_p \left( -\frac{3}{2}t + \frac{1}{4} \right) 
         \frac{\alpha_A + \alpha_D}{\alpha_A - \alpha_D} 
         \left( \frac{r_{edge}}{r} \right)^{s_A+t+3/2},
\end{equation}
\begin{equation}
 K_2 = h \left[ B \left( s_A -\frac{t}{2} + \frac{7}{4} \right) 
                     - (B-1)\frac{t+3}{c^2} \left( -\frac{t}{2} + \frac{7}{4}  \right)
                                          \right],
\end{equation}
and
\begin{equation}
 K_3  =  \frac{\tanh(1/c)}{2} \left( B - (B-1)\frac{t+3}{c^2} \right)  + \frac{K_2}{2},
\end{equation}
For $r/r_{edge}>$1, the dust density enhancement needed is given as
\begin{equation}
 g_{d} > \frac{-3t/2+1/4}{3t/2-1/4+\sqrt{\ln 2}/ch},
 \label{eq_out_temp}
\end{equation}
where equation (\ref{iceline_barrier2}) is used. In order to gain a unified condition for the entire of discs, 
we find the relation between $f_{d1}$ and $g_d$ which is given as
\begin{eqnarray}
  g_d & = & \frac{\alpha_A - \alpha_D}{\alpha_D} \left[ 1 - \frac{\alpha_A}{\alpha_A+\alpha_D} \right. \\ \nonumber
      & \times & \left. \left( \frac{r}{r_{edge}} \right)^{s_A+t+3/2} 
        \left( 1+ \frac{f_{d1}}{2} \right)^{-1} \left( 1+ f_{d2}\right)^{-1}  \right].
\end{eqnarray}
As a result, equation (\ref{eq_out_temp}) becomes 
\begin{equation}
 f_{d1} > \frac{2(P_1+P_2-1)}{1-P_2},
 \label{eq_out}
\end{equation}
where 
\begin{equation}
 P_1 = \frac{\alpha_A}{\alpha_A+\alpha_D} \left( \frac{r}{r_{edge}} \right)^{s_A+t+3/2} \left( 1+ f_{d2}\right)^{-1}, 
\end{equation}
and 
\begin{equation}
 P_2 = \frac{\alpha_D}{\alpha_A - \alpha_D}\frac{-3t/2+1/4}{3t/2-1/4-\sqrt{\ln 2}/ch}.
\end{equation}

\begin{figure}
\begin{center}
\includegraphics[width=8.26cm]{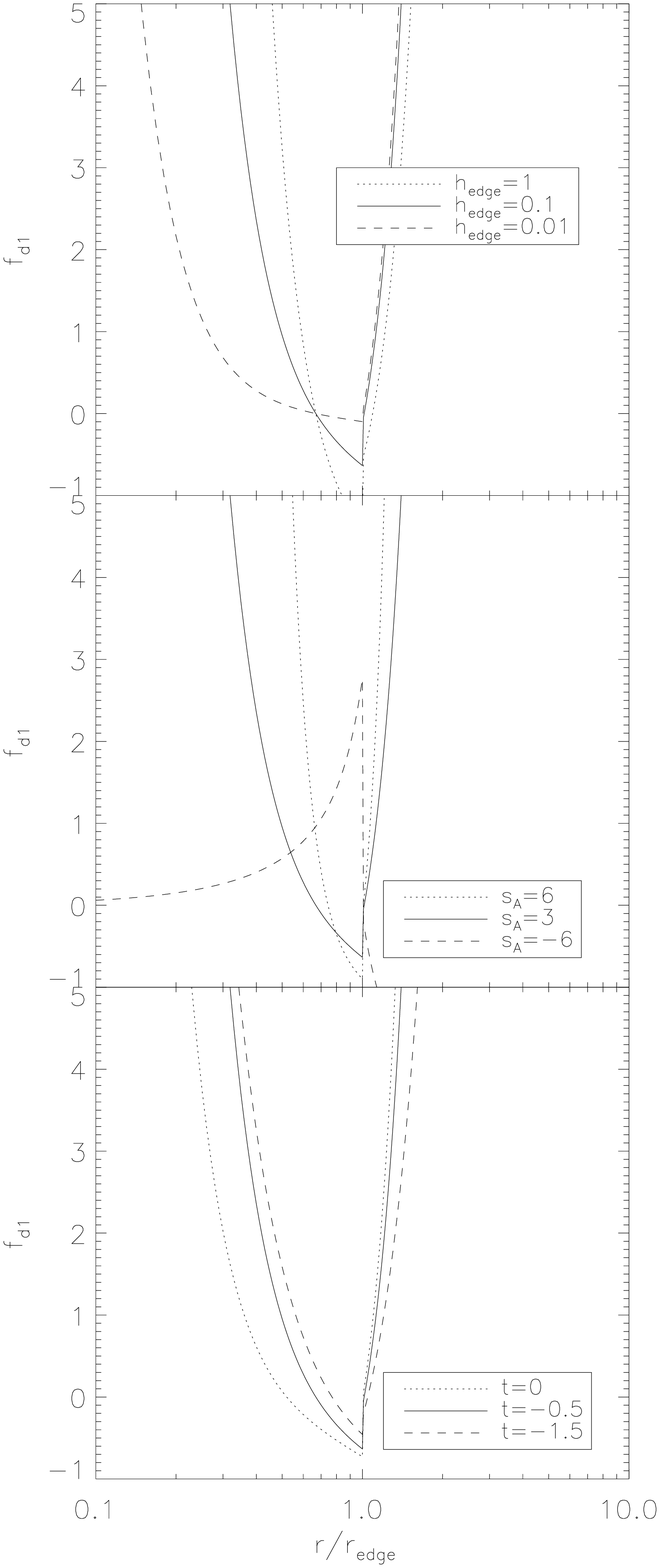}
\caption{The enhancement $f_{d1}$ of the dust density required to produce ice line barriers. For $r/r_{edge}<$1, 
equation (\ref{eq_in}) is adopted while equation (\ref{eq_out}) is used for $r/r_{edge}>$1. We define a fiducial 
model as ($s_A,t,h_{edge},f_{d2},c$)=(3,-1/2,0.1,0,1). We vary $h_{edge}$, $s_A$, and $t$ on the top, middle, and 
bottom panel, respectively. A large value of $f_{d1}$ is required for ice lines to be a barrier, if 
$s_A+t+3/2$ is positive. On the other hand, the presence of ice line barriers becomes insensitive to $f_{d1}$ if 
$s_A+t+3/2$ is negative (see the dashed line on the middle panel). It implies that the minimum value of $f_{d1}$ 
strongly depends on the detail structure of ice lines.}
\label{figC1}
\end{center}
\end{figure}

Fig. \ref{figC1} shows the threshold value of $f_{d1}$ as a function of $r/r_{edge}$. We define a fiducial model as 
($s_A,t,h_{edge},f_{d2},c$)=(3,-1/2,0.1,0,1). The required value of $f_{d1}$ are above the lines. We performed 
parameter studies on $h_{edge}$, $s_A$, and $t$ and the results are shown in the top, middle, and bottom panels, 
respectively. 

For $r/r_{edge}<$1, the dominant term in equation (\ref{eq_in}) is $( r_{edge}/r)^{s_A+t+3/2}$ in $K_1$. We see that a 
large value of $f_{d1}$ is required for outward migration, if the value of $s_A+t+3/2$ is positive. Fig. \ref{figC1} 
shows that small separations between the ice line and dead zone barriers are again preferred. If $s_A+t+3/2<$0, 
a tiny amount of ice is enough to provide a barrier (see the dashed line on the middle panel). This indicates that 
ice lines of relatively abundant molecules can work as barriers. For $r/r_{edge}>$1, the dominant term in 
equation (\ref{eq_out}) is $( r/r_{edge})^{s_A+t+3/2}$ in $P_1$. Consequently, a significantly large value of 
$f_{d1}$ is needed for outward migration, if the value of $s_A+t+3/2$ is positive. In fact, it may not be possible 
for ice lines of any molecule to be a barrier in this case. On the other hand, if $s_A+t+3/2$ is negative, the amount 
of ice does not matter. (see the dashed line on the middle panel). This implies that ice lines of any species can act 
as barriers. Thus, the required $f_{d1}$ strongly depends on the sum of $s_A$ and $t$.

What value of $s_A$ is most appropriate for realistic discs? For $r/r_{edge} <$1, $s_A$ should be 
positive. Otherwise, dead zones cannot exist. This indicates that larger values of $f_{d1}$ are required as 
$r/r_{edge}$ decreases, and hence only specific, abundant species in discs can work as a barrier. For $r/r_{edge} >$1, 
$s_A$ should be negative, since $\Sigma_A \approx \Sigma$. Thus, it is likely that the abundance of molecules may not 
be important for providing a barrier. As a result, several ice line barriers which are excited by different species 
such as water-ice and CO-ice may be present in a single disc. However, the value of $s_A$ strongly depends 
on how icy dust absorbs free electrons and how these distributions evolve with time.

\bsp

\label{lastpage}

\end{document}